\documentclass{aa}  
\usepackage{natbib}
\usepackage{graphicx}
\usepackage{pdflscape}
\usepackage{xcolor}
\usepackage{subfig}
\usepackage{threeparttable}
\usepackage{txfonts}

\newcommand{\hi}{H{\sc i}}

\newcommand{\hii}{H{\sc ii}}

\newcommand{\msol}{\mbox{M$_{\odot}$}\,}

\newcommand{\asec}{$^{\prime\prime}$\,}
\newcommand{\amin}{$^{\prime}$\,}
\newcommand{\adeg}{$^{\circ}$\,}
\newcommand{\fdeg}{.\!\!^\circ}
\newcommand{\fasec}{.\!\!$^{\prime\prime}$}

\newcommand{\TBF}{\color{black}}
\usepackage[colorlinks=true, allcolors=blue]{hyperref}

\begin{document} 

\title{A global view on star formation: The GLOSTAR Galactic plane survey}
\subtitle{IV. Radio continuum detections of young stellar objects in the Galactic Centre region}
\titlerunning{GLOSTAR: YSOs in the Galactic Centre}

   \author{H.\,Nguyen
          \inst{1}\fnmsep\thanks{Member of the International Max Planck Research School (IMPRS) for Astronomy and Astrophysics at the Universities of Bonn and Cologne.}
          \and
          M.\,R.\,Rugel\inst{1}
          \and
          K.\,M.\,Menten\inst{1}
          \and
          A.\,Brunthaler\inst{1}
          \and
          S.\,A.\,Dzib\inst{1}
          \and
          A.\,Y.\,Yang\inst{1}          
          \and
          J.\,Kauffmann\inst{2}
          \and
          T.\,Pillai\inst{3,1}
          \and
          G.\,Nandakumar\inst{4,5}          
          \and
          M.\,Schultheis\inst{6}
          \and
          J.\,S.\,Urquhart \inst{7}
          \and
          R.\,Dokara \inst{1,\star}
          \and
          Y.\,Gong \inst{1}
          \and
          S-N.\,X.\,Medina \inst{1}
          \and
          G.\,N.\,Ortiz-Le{\'o}n \inst{1}
          \and
          W.\,Reich \inst{1}
          \and
          F. Wyrowski \inst{1}
          \and
          H.\,Beuther \inst{8}
          \and
          W.\,D.\,Cotton \inst{9,10}
          \and
          T.\,Csengeri \inst{11}
          \and
          J.\,D.\,Pandian \inst{12}
          \and
          N.\,Roy \inst{13}
          }

   \institute{Max-Planck-Institut f\"ur Radioastronomie, Auf dem H\"ugel 69, 53121 Bonn, Germany\\
            \email{hnguyen@mpifr-bonn.mpg.de}
            \and
            Haystack Observatory, Massachusetts Institute of Technology, 99 Millstone Road, Westford, MA 01886, USA
            \and
            Institute for Astrophysical Research, Boston University, Boston, MA 02215, USA\and
            Research School of Astronomy \& Astrophysics, Australian National University, ACT 2611, Australia
            \and
            ARC Centre of Excellence for All Sky Astrophysics in Three Dimensions (ASTRO-3D)
            \and
            Universit\'e C\^ote d'Azur, CNRS, laboratoire Lagrange, Blvd de L'Observatoire, F-06304 Nice
            \and
            Centre for Astrophysics and Planetary Science, University of Kent, Ingram Building, Canterbury, Kent CT2 7NH, UK
            \and
            Max Planck Institute for Astronomy, Königstuhl 17, D-69117 Heidelberg, Germany
            \and
            National Radio Astronomy Observatory, 520 Edgemont Road, Charlottesville, VA 22903, USA
            \and
            South African Radio Astronomy Observatory, 2 Fir St, Black River Park, Observatory 7925, South Africa
            \and
            Laboratoire d'astrophysique de Bordeaux, Univ. Bordeaux, CNRS, B18N, all\'ee Geoffroy Saint-Hilaire, 33615 Pessac, France
            \and
            Department of Earth \& Space Sciences, Indian Institute of Space Science and Technology, Trivandrum 695547, India
            \and
            Department of Physics, Indian Institute of Science, Bengaluru 560012, India
            }

   \date{Received 12 March 2021 / Accepted 13 April 2021}

 
  \abstract
   {The Central Molecular Zone (CMZ), a $\sim$200\,pc sized region around the Galactic Centre, is peculiar in that it shows a star formation rate (SFR) that is suppressed with respect to the available dense gas. To study the SFR in the CMZ, young stellar objects (YSOs) can be investigated. Here we present radio observations of 334 2.2\,$\mu$m infrared sources that have been identified as YSO candidates.}
   {Our goal is to investigate the presence of centimetre wavelength radio continuum counterparts to this sample of YSO candidates which we use to constrain the current SFR in the CMZ.}
   {As part of the GLObal view on STAR formation (GLOSTAR) survey, D-configuration Very Large Array (VLA) data were obtained for the Galactic Centre, covering $-2^{\circ}$\,$<l<2^{\circ}$\, and $-1^{\circ}$\,$<b<1^{\circ}$\, with a frequency coverage of 4--8\,GHz. We matched YSOs with radio continuum sources based on selection criteria and classified these radio sources as potential \ion{H}{ii}~regions and determined their physical properties.}
   {Of the 334 YSO candidates, we found 35 with radio continuum counterparts. We find that 94 YSOs are associated with dense dust condensations identified in the 870\,$\mu$m ATLASGAL survey, of which 14 have a GLOSTAR counterpart. Of the 35 YSOs with radio counterparts, 11 are confirmed as \ion{H}{ii}~regions based on their spectral indices and the literature. We estimated their Lyman continuum photon flux in order to estimate the mass of the ionising star. Combining these with known sources, the present-day SFR in the CMZ is calculated to be $\sim$0.068\,\mbox{M$_{\odot}$}\,yr$^{-1}$, which is $\sim$6.8$\%$ of the Galactic SFR.
   Candidate YSOs that lack radio counterparts may not have yet evolved to the stage of exhibiting an \ion{H}{ii}~region or, conversely, are older and have dispersed their natal clouds. Since many lack dust emission, the latter is more likely. 
   Our SFR estimate in the CMZ is in agreement with previous estimates in the literature.}
   {}

   \maketitle

\section{Introduction}
\label{sect:intro}
The study of high-mass stars is vital to the understanding of the 
evolution of star formation in galaxies. They directly influence
their surrounding environments by feeding energy through various 
feedback processes back into the interstellar medium (ISM).
This can alter the efficiency of the remaining gas to form new
stars and thus directly impact the evolution of their host galaxies.
It is therefore important to understand the formation of high-mass
stars themselves. Observations of star forming sites in our own galaxy, the Milky Way, 
are easier to resolve due to their proximity. Their study allows us to extend our understanding of high-mass star formation (HMSF) to those in other galaxies as well \citep{KennicuttEvans2012}.

The term massive young stellar object (MYSO) has been used for sources in a wide range of evolutionary stages. They start as objects that are still deeply embedded in their parental dense molecular cloud core and that are powered by accretion, often forming in clusters. 
Once nucleosynthesis commences, they start to ionise their surroundings (e.g. \citealt{zinn2007}; \citealt{hoare2007}; \citealt{breen2010}), develop into hyper- and later ultra-compact \hii~regions that further evolve into compact \hii~regions (\TBF{such as} the Orion Nebula). The most luminous O-type stars therein clear their surroundings of obscuring dust and eventually make them and the much more numerous lower-mass members of the young stellar clusters, whose centres they occupy, visible (predominantly) in nearby parts of the Galaxy that do not suffer heavy line of sight and local visual extinction. The 1--2 million year old Orion Nebula Cluster (ONC; \citealt{orion1989}) at a distance of only $\sim 400$\,pc \citep{mentenONC, kounkel2017} is a nearby prominent example. Observations of the earliest stages of development are difficult because of the embedded nature of YSOs, as well as by the comparatively short lifetime of massive stars and their short formation timescales ($\sim$10$^{5}$\,years). The \hii~region phase, however,
gives a clear indication that, \TBF{in particular, `high-mass'} star formation has recently occurred \citep{woodChurchwell1989}. 

High-mass star formation occurs in dense clumps within giant molecular clouds (GMCs). Thus, one would expect a high concentration in the so-called Central Molecular Zone (CMZ), which contains about 3--10\%\ of the molecular material in our Galaxy (e.g. \citealt{gusten1989}; \citealt{rodriguez2004}). The CMZ \citep{morrisSerabyn1996} is a roughly $\sim 200$\,pc sized region that covers a range of $-0\fdeg7<l<1\fdeg8$ and \mbox{$-$0\fdg3 < $b$ < 0\fdg2} in Galactic coordinates at a distance of 8.2\,kpc \citep{gravity2019}\footnote{GRAVITY determines a geometric distance of $8178\pm26$\,pc to the central super-massive black hole Sgr~A*.}. The CMZ's physical conditions 
are extreme in comparison to other GMCs 
in the Milky Way as the gas temperature, the pressure, and magnetic field
strengths are a few to several orders of magnitude higher \citep{morrisSerabyn1996}. It is clear that the question of present-day star formation in the CMZ is an important one, in particular given 
the presence of a few massive star clusters, the Arches and the Quintuplet clusters, and the central cluster in the immediate vicinity of the super-massive black hole Sgr~A$^*$ in the centre of the Galaxy (e.g. \citealt{cotera1996}, \citealt{kobayashi1983}). Theses clusters have ages of 2--4\,Myr, while the massive \mbox{`(mini-)starburst}' region Sagittarius B2 (Sgr~B2), which has a mass of 8$\times10^{6}$\,\msol, is a prominent active star factory (\citealt{figer2002}, \citealt{schneider2014}, \citealt{lis1990}, \citealt{schmiedeke2016}). On the other hand, the infrared dark cloud M025+0.11, termed \TBF{`the Brick'}, contains a comparable, if not somewhat lower, mass to Sgr~B2 ($\sim$10$^{5}$\,\msol), but it shows few signs of active star formation (e.g. \citealt{LisMenten1998, Henshaw_2019} and references therein). Recent observations by \cite{walker2021} show unambiguous signs of low- to intermediate-mass star formation and potential evidence for future high-mass star formation, however, to a much lesser degree than Sgr~B2.

The star formation rate (SFR) of galaxies has been shown empirically to follow
the Kennicutt--Schmidt relation (\citealt{schmidt1959}, \citealt{Kennicutt1998}), which infers a power law relation between the SFR per unit area and the total gas mass. One can further correlate the SFR with the amount of dense ($n>10^{4}$\,cm$^{-3}$) molecular gas in our Galaxy to also show a linear relation (e.g. \citealt{lada2010}, \citealt{lada2012}). Despite the amount of dense gas available, the SFR is a factor of 10--100 lower than expected in the CMZ 
and it does not follow the Kennicutt--Schmidt relation (e.g. \citealt{longmore2013a}, \citealt{csengeri2016}), although it may have done so in the past \citep{kruijssen2014}.
Various investigations of the SFR in the CMZ have been performed using different
methods such as YSO counting (e.g.\citealt{yusef2009}, \citealt{an2011}, \citealt{immerCMZ}, \citealt{nandakumar2018}), free-free emission \citep{longmore2013a}, and infrared luminosity \citep{barnes2017}.
Systematic uncertainties in various methods used for determining the SFR as the cause for the much lower value 
were ruled out by \cite{barnes2017} as they obtained similar average SFRs by comparison of the above YSO counting and free-free emission
measurements. Low SFRs have also been found in specific high-density clouds in the CMZ (e.g. \citealt{kauffmann2017a}, \citealt{lu2019}). While not applicable to clouds that already show traces of star formation at later stages, as in Sgr~B2, some theoretical models suggest that the lower SFR is due to these clouds being in an early evolutionary stage where active star formation is not yet observable (see also, e.g. \citealt{kruijssen2014}; \citealt{krumholz2015}; \citealt{krumholz2017}).
In stark contrast, the mini-starburst region Sgr~B2 is one of the most prolific star formation factories in the Galaxy \citep{ginsburg2018}.

Various hypotheses exist for this difference in SFRs between the CMZ and typical star forming environments:
On the one hand, the formation of high-mass stars may require a higher critical density threshold than that of low-mass stars \citep{krumholzMckee2008}; on
the other hand, the turbulent environment of the CMZ itself is increasing
this density threshold (e.g. \citealt{kruijssen2014}; \citealt{rathborne2014}).
Further studies of HMSF in the CMZ and the physical processes therein are therefore crucial for our understanding of star formation in our Galaxy as well as other galaxies. \TBF{The question of Why star formation in the CMZ is so unevenly distributed and why it is absent in so much of its volume is still open to this day.}
Obtaining a census of and characterising YSOs in the CMZ will help to address these points and is the focus of this paper.

Searching for YSOs is a direct way of identifying on-going and recent star formation and to determine the current SFR. The low number of YSOs identified in some parts of the Galactic Centre 
motivates new searches. Finding YSOs in the CMZ 
requires studies in multiple wavelengths.
Very recently formed YSOs are surrounded
by dense envelopes of gas and dust \citep{zinn2007} and
as they begin to heat this nearby dust, the energy is re-emitted in the 
infrared regime and as such, most  studies of YSOs have been
performed with infrared photometry (\citealt{schuller2006}, \citealt{yusef2009}).
However, the large and spatially variable extinction in the CMZ
($A_{\rm V}$ = 20 -- 40\,mag; \citealt{schultheis2009}) can cause confusion regarding the
identification of YSOs. Furthermore, the extinction in dense regions can be of the order of $\sim$\,100\,mag making proper counts of forming stars impossible in these clusters, thus missing a large fraction of the stars currently forming. Studies classifying YSOs with near-infrared (NIR) photometry cannot identify YSOs uniquely since AGB stars, red giants, and even super giants can have similar photometric colour signatures \TBF{similar to} YSOs due to foreground
extinction \citep{schultheis2003}.
To distinguish them, spectroscopic observations are required, 
with ambiguities in the classification schemes remaining even with this method
(e.g. \citealt{an2011}, \citealt{immerCMZ}).

\begin{figure*}[]
    \begin{tabular}{c}
    \centering
    \includegraphics[scale=0.6]{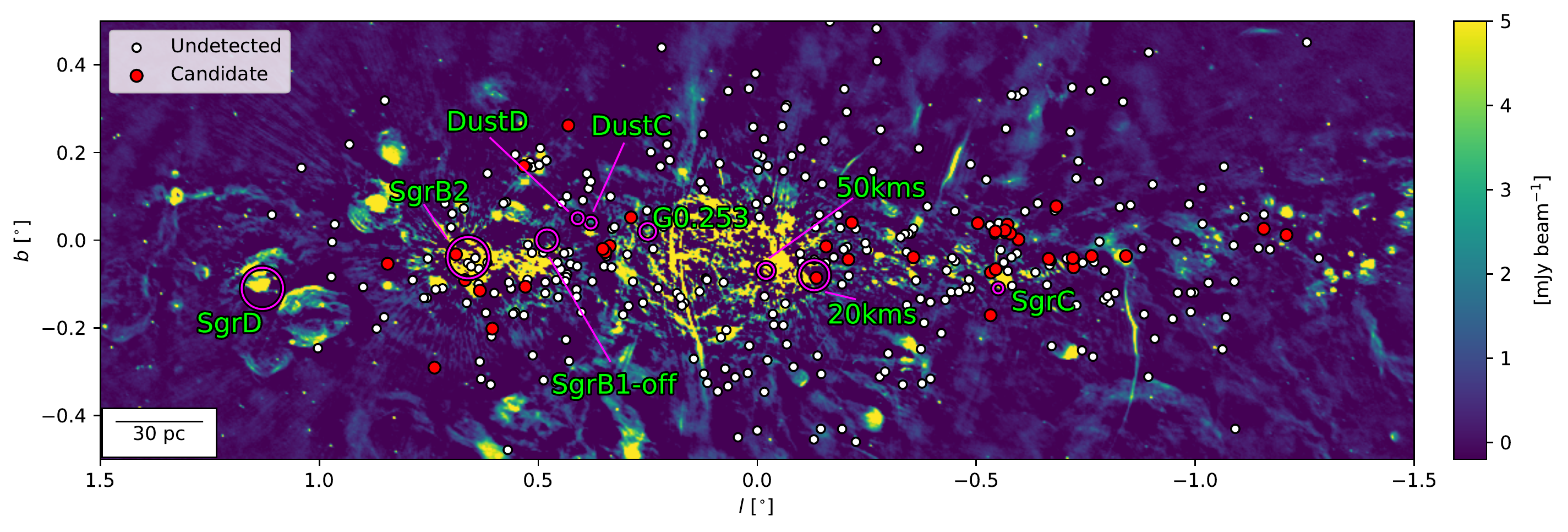}

    \end{tabular}
    \caption{GLOSTAR 5.8\,GHz detections towards YSOs from \citet[][red circles]{nandakumar2018}. Non-detections are shown as white dots. The background image shows the GLOSTAR 5.8\,GHz D-configuration continuum image restricted to the region studied by \cite{nandakumar2018}. The image has been clipped with minimum and maximum limits of $-$0.1\,mJy\,beam$^{-1}$ and 5\,mJy\,beam$^{-1}$, respectively, to better emphasise visibility of the low intensity radio features. The main GMCs presented in the CMZ are highlighted with purple circles with effective radii from \cite{kauffmann2017a} and the references therein.}
    \label{fig:yso_overlay_glostar}
\end{figure*}
Recently, \cite{nandakumar2018} presented a study aimed at identifying YSOs in the CMZ. 
They conducted K-band ($2.2\,\mu$m) spectroscopic NIR observations
of photometrically identified YSOs in the CMZ and detected 91 viable sources that they used to develop a new photometric YSO classification scheme that tries to eliminate
contamination from late-type and evolved stars as those revealed by their spectroscopic observations. To estimate the SFR, a larger sample was needed and thus, in combining the photometric catalogue of YSO candidates from SIRIUS (J(1.25\,$\mu$m), H(1.63\,$\mu$m), and K$_{\rm{S}}$(2.14\,$\mu$m) filters; \citealt{nishiyama2006}) and the point-source catalogue of the Spitzer IRAC survey (3.6--8.0\,$\mu$m;  \citealt{ramirez2008}), they produced a final sample of 334 YSO candidates using their new classification scheme.
The estimated masses from spectral energy distribution (SED) fitting models obtained from \cite{robitaille2017} range from 2.7 to 35\,\msol and furthermore peaks at $\sim 8$\,\msol, suggesting that $>50\,\%$ of these sources are already in the high-mass regime. To further investigate if these sources are indeed sites of HMSF, we used sub-millimetre and radio wavelengths to constrain the evolutionary stage.

In this paper, we used the sample of 334 YSO candidates from \cite{nandakumar2018} to search for radio continuum counterparts at \mbox{4--8\,GHz} obtained with the Karl\,G.\,Jansky Very Large Array (VLA) as part of the GLObal view of STAR formation (GLOSTAR; \citealt{sac2019}, Brunthaler et al. submitted) survey. Radio continuum sources can be signposts of free-free emission from \hii~regions. 
As of yet, it is unknown how many of the NIR-identified YSOs trace \hii~regions and so a census of the association between YSOs and \hii~regions would shed light on this.  
We produced a catalogue of radio continuum counterparts and investigated their nature by calculating their spectral indices. We used spectral indices to distinguish source types depending on whether the emission is thermal or non-thermal which helps to classify sources as bona fide \hii~regions or otherwise. We further determined the fraction of YSOs that have counterparts and physical reasons for the absence of an \hii~region. We also cross-matched these YSO candidates with the APEX Telescope Large
Area Survey of the Galaxy (ATLASGAL; \citealt{atlasgal}) 870\,$\mu$m dust emission data, which trace the early natal environments of high-mass stars. If the YSO is still very young, we would expect to see it embedded in a compact cold dust envelope traced by sub-millimetre wavelengths.
In this way, we investigated if there is any clear association of these YSOs with either the earliest or the latest stages of massive star formation in order to shed light on the complete spread of evolutionary stages of this census of YSOs. Lastly, we infer the SFR from the free-free
emission of the \hii~regions detected in our field.

We structure this paper as follows: In Section~\ref{sect:obs} we give a short summary of the data used in this paper. Section~\ref{sect:results} details
our source selection criteria in finding radio continuum counterparts and the
determination of their physical parameters needed to calculate the SFR.
Section~\ref{sect:discussion} discusses our comparison with other surveys and other prominent regions in the Galactic Centre and the properties of the YSO-sample with radio continuum counterparts as well as the SFR in the CMZ. We present
the conclusions and summary in Section~\ref{sect:summary}.



\section{Observations}\label{sect:obs}

The GLOSTAR survey (\citealt{sac2019}, Brunthaler et al. submitted) is an on-going survey with the VLA and the Effelsberg 100\,m telescope between 4--8 GHz of the Galactic mid-plane from
$-2^{\circ}<\textit{l}<60^{\circ}$ and $|\textit{b}|<1^{\circ}$ as well as
the Cygnus-X star-forming complex. VLA observations were mainly conducted in D- and B-configurations whose angular resolutions correspond to 18\asec and 1.5\arcsec at 5.8\,GHz, 
respectively, to detect various tracers of different stages of early star formation using methanol, formaldehyde, and radio recombination lines as well as radio continuum to describe the stellar evolution process of massive stars.

This work is a targeted search for continuum sources towards YSOs identified with NIR photometry \citep{nandakumar2018} using only
the continuum data obtained from the VLA in D-configuration for
the Galactic Centre ($|\textit{l}|<1.5^{\circ}$ and $|\textit{b}|<1^{\circ}$). We briefly summarise the data properties (for details see, e.g. \citealt{sac2019}, Brunthaler et al., submitted, Dzib et al., in prep.). Observing in the C-band (4--8\,GHz), the
correlator setup consists of two 1-GHz-wide sub-bands centred at 4.7 and
6.9\,GHz. Each sub-band was further divided into eight intermediate frequency windows of 128 MHz with each window consisting of
64 channels with widths of 2\,MHz. Approximately 2520 pointings were used to 
cover an area of $2^{\circ}\times4^{\circ}$.
Flux calibration was done using 3C286 which was used as the band-pass calibrator and J1820-2528 as the phase calibrator. We used the Obit package \citep{obit} for the calibration as well as the imaging of the continuum data. The data were rearranged into nine different frequency bands (spectral windows) of a similar fractional bandwidth. Each pointing was first cleaned individually and then combined into a large mosaic for each frequency band. The final mosaic at the reference frequency was created by combining the individual, primary beam corrected images of each of the frequency bands.

The effective frequency of the averaged image is 5.8\,GHz
with a FWHM of 18\asec. The average noise level increases from
$\sim$0.07\,mJy\,beam$^{-1}$ to $\sim$1\,mJy\,beam$^{-1}$ as one moves closer to the Galactic mid-plane, which is as expected since the majority of emission is in the plane of the disk such as the black hole in the centre of our galaxy.
In comparison to other regions studied in GLOSTAR, emission-free regions typically have noise levels of around $\sim$0.06\,mJy\,beam$^{-1}$, but they can steeply increase to $\sim$0.45\,mJy\,beam$^{-1}$ towards the Galactic mid-plane \citep{sac2019}. We note that the VLA B-configuration and Effelsberg data have not yet been imaged and will be analysed in future works.


\section{Results}\label{sect:results}
\subsection{2.2\,$\mu$m sources as equivalent ONCs in the CMZ}\label{sect:COBE}
We would like to understand the nature of the 2.2\,$\mu$m sources investigated in \cite{nandakumar2018} but have the problem that they are far away and suffer from heavy extinction. As such, we make a comparison to a nearby known 2.2\,$\mu$m star cluster.
Due to its close distance of $\sim$400\,pc \citep{mentenONC, kounkel2017,onc_grossschedk}, the ONC \citep{orion1989} is often
used as the template environment for studying HMSF. It contains multiple massive stars that are easily detectable at this distance and also shows bright 2.2\,$\mu$m and radio emission. The YSOs that we use as targets in our investigation
are, however, located at a much farther distance of $\sim$8.2\,kpc \citep{gravity2019} in the Galactic Centre.
Therefore, to put the environment into perspective and to see how the YSO candidates compare to a cluster of massive stars similar to the ONC, we discuss what the ONC would look like
if it was placed in the CMZ at similar infrared wavelengths. 

\begin{figure}[!h]
    \centering
    \includegraphics[width=0.5\textwidth]{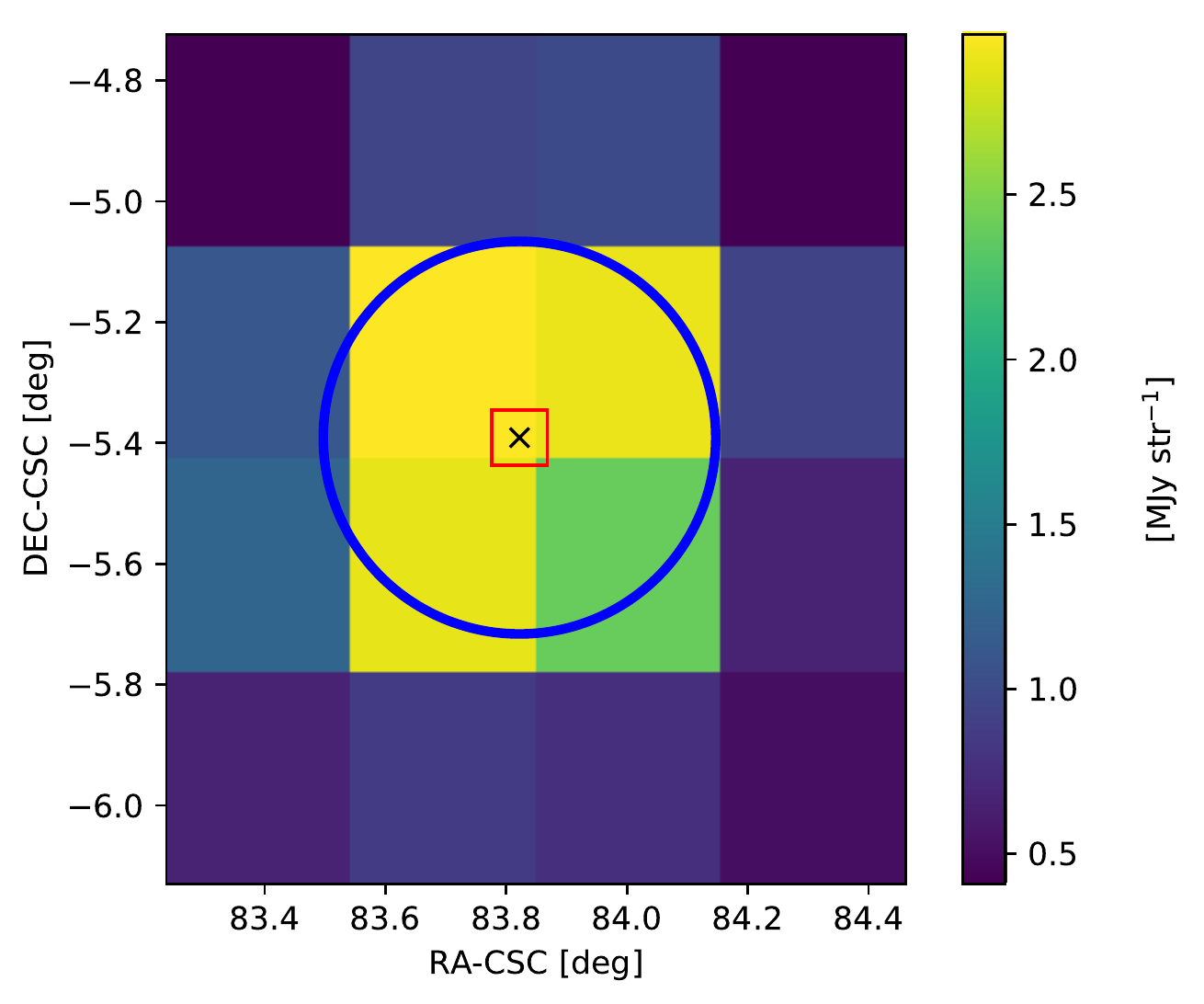}
    \caption{COBE 2.2$\mu$m data of the ONC plotted in equatorial coordinates (CSC; COBE quadrilateralised spherical cube projection). The \text{x} marker denotes the position of the ONC with the surrounding red square signifying a 5\amin width, which is the estimated angular size of the ONC. The size of a DIRBE pixel is 0.32\adeg. The blue circle highlights the four pixels selected for the flux determination.}
    \label{fig:cobe_orion}
\end{figure}

The Cosmic Background Explorer\footnote{The National Aeronautics and Space Administration/Goddard Space Flight Center (NASA/GSFC)} (COBE) space-based mission \citep{COBE} was developed to measure the diffuse infrared and microwave radiation from the early universe.
The Diffuse Infrared Background Experiment's (DIRBE) objective is to search for Cosmic Infrared Background by making absolute brightness measurements of the diffuse infrared radiation in ten photometric bands from 1 to 300\,$\mu$m. We obtained the map corresponding to 2.2\,$\mu$m
and plotted the intensities for each pixel near the ONC in Galactic coordinates (see Fig.~\ref{fig:cobe_orion}). Given the size of a DIRBE pixel (0.32\adeg) and the size of the 
ONC ($\sim$5\amin), it is unclear which pixel is correctly associated. We carried out calculations for
both the nearest singular pixel as well as the sum of all four surrounding pixels. The resultant
peak flux and integrated flux density are 93.18\,mJy and 630.22\,mJy, respectively. Now if we were  to place the ONC at a distance of 8.2\,kpc, this woud scale the fluxes to 0.22\,mJy and 1.49\,mJy respectively, diminishing them by $\sim 400$~times.

To compare to the YSO sample, we converted this photometric flux into photometric magnitude units
given that $F_{v}=F_{0}\times10^{-m/2.5}$
where $m$ is the magnitude, $F_{v}$ is the flux in Jy,
and $F_{0}$ is the zero point for a given filter system (for 2.159\,$\mu$m, $F_{0}=666.7$ where this 
corresponds to the K$_{\rm{S}}$ filter system used for the SIRIUS catalogue containing the YSOs). Using the conversion tool\footnote{\url{https://irsa.ipac.caltech.edu/data/SPITZER/docs/spitzermission/missionoverview/spitzertelescopehandbook/19/}}$^{,}$\footnote{\url{http://ssc.spitzer.caltech.edu/warmmission/propkit/pet/magtojy/index.html}} for photometry, the resultant apparent magnitudes for the peak and integrated flux densities are 16.2\,mag and 14.1\,mag, respectively. When comparing with only the 
YSOs selected from spectroscopic KMOS observations (see Table~2 in \citealt{nandakumar2018}),
the magnitude values are at the lower range but are still possible to be observed, suggesting that KMOS would be able to observe ONC-like sources at the distance of the Galactic Centre.
\subsection{GLOSTAR source selection}\label{sect:source} 

\cite{nandakumar2018} provide a catalogue of 334 YSO candidates made with the new colour-colour diagram selection criteria aimed at disentangling YSOs from late-type stars. Using spectral line data, they first separated NIR sources as YSOs or cool, late-type stars through the absence of $^{12}$CO\,(2,0) absorption and the presence of Br$\gamma$ emission. They found that in a H-K$_{\rm{S}}$ versus H-[8.0] colour-colour diagram, they could distinctly separate the two groups of stellar objects. This gives us a sample with very few contaminants even if not all of them have been spectroscopically
identified as YSOs. Given this list of YSOs, we searched the GLOSTAR data to see if 
there are associations with 5.8\,GHz continuum emission. The region that we study in this work is shown in Fig.~\ref{fig:yso_overlay_glostar}, where we have also overlaid the
complete sample of YSO candidates investigated and highlighted the sources coinciding with radio emission. One can already see that there is a statistical bias present from the YSOs as they are not evenly distributed across the Galactic Centre. We see that in some of the known massive star forming regions such as Sgr~D, we do not find a large number of YSOs from \cite{nandakumar2018}. As such, we do not provide a complete census of all \hii~regions in this work. 

We first investigated if there is a spatial coincidence between the YSO sample
and the GLOSTAR continuum emission by using a radius for the angular separation of $\sim$10\asec,
which is approximately equal to half of the GLOSTAR continuum data synthesised beamsize, corresponding to $\sim$0.4\,pc at a distance of 8.2\,kpc (\citealt{gravity2019}). 
We find close to $\sim$100 YSOs which have potential 5.8\,GHz counterparts
that match this spatial selection criterion. We note that for certain extended regions (e.g. Sgr~B2), it can be difficult to associate a single YSO to them. We used a sensitivity threshold to select the closest possible YSO as its counterpart, which is described below.

We examined the validity
of these candidate radio sources by looking at their intensity, shape, 
and the likelihood of being a radio artefact. Taking the 
root-mean-square (rms) of a nearby emission-free region as the noise level, we considered
sources that have peak pixel intensities of at least $5\sigma$. To confirm the detection
of the radio source, we used eight of the nine spectral windows
from the GLOSTAR-VLA data since the ninth spectral window has a much higher 
noise level. We have kept sources that
have clear and consistent structure in at least half of their
spectral windows. The data can be strongly influenced by the strong emission from powerful
sources in the Galactic Centre such as from Sgr~A${}^{\ast}$ that can lead to strong sidelobe effects. 
This can manifest as a false detection or a source appearing variable in intensity and shape
across the multiple spectral windows. In general, the observed field is known to be extremely crowded
and one needs to be wary of the surrounding environment of each source.
The rms ranges from 0.351\,mJy\,beam$^{-1}$ to 11.914\,mJy\,beam$^{-1}$ with
median and mean values of 0.709\,mJy\,beam$^{-1}$ and 1.558\,mJy\,beam$^{-1}$,
respectively. The lower noise value corresponds to regions which are located in emission-free regions far offset from strong emission sources and can act as
a lower limit, while the higher end corresponds to the average environment of a strong
emission source. For our investigation, we calculated the rms at each individual source separately.

\begin{figure}
    \centering
    \includegraphics[width=0.5\textwidth]{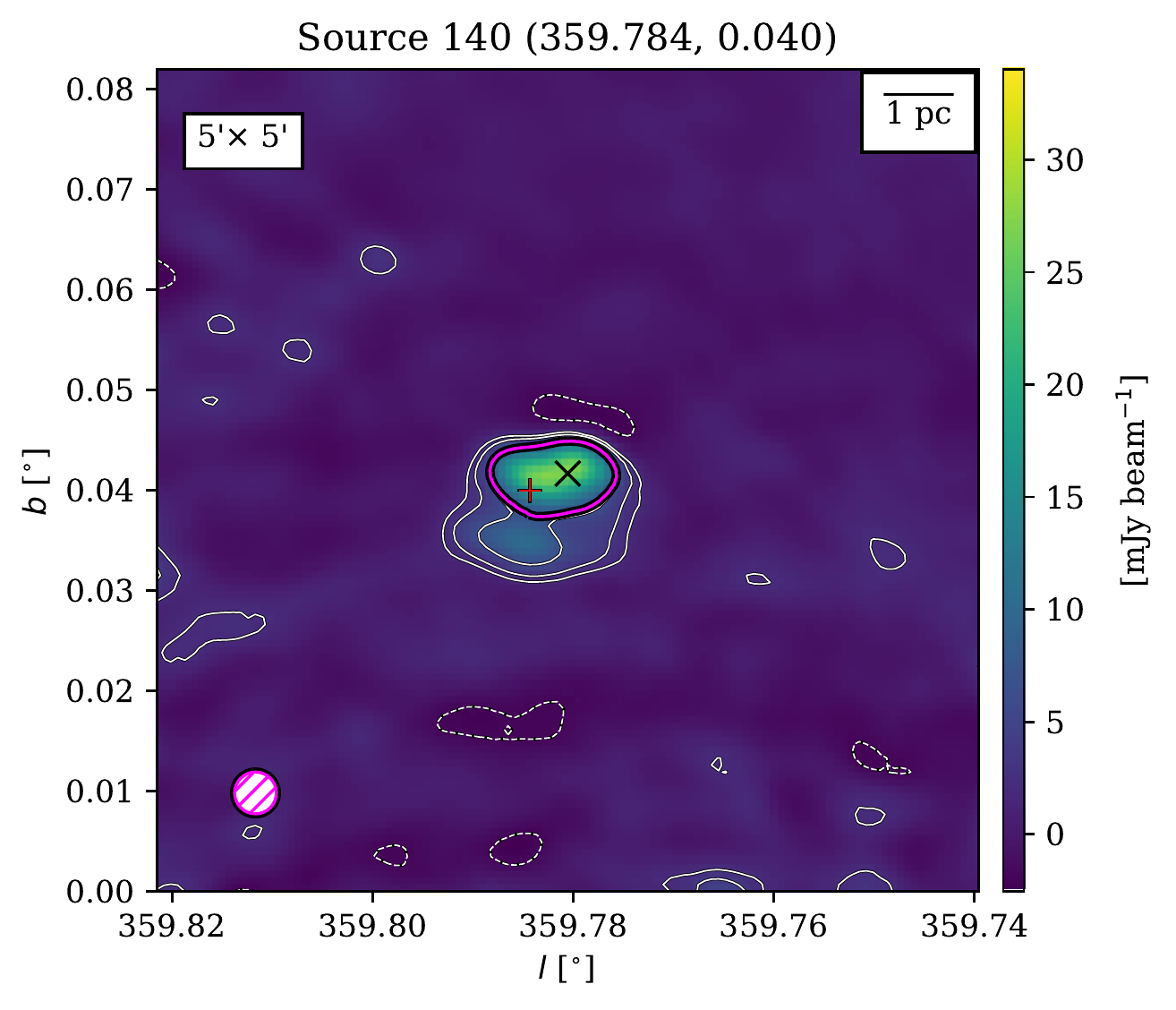}
    \caption{Detection of a GLOSTAR 5.8\,GHz counterpart for source 140 from \citet{nandakumar2018}. The GLOSTAR 5.8\,GHz continuum map is shown in colour with positions of the peak radio emission and the YSO marked with a black X and red cross, respectively. White contours correspond
    to 3, 5, and 10\,$\sigma$ ($\sigma=0.634$\,mJy\,beam$^{-1}$). The magenta contour outlines the area from which the flux density was calculated. 
    The colour bar maximum was chosen to be 1.5$\times$ the peak intensity of the 
    coinciding 5.8\,GHz source. The beamsize is shown in magenta in the bottom left corner.}
    \label{fig:example_source}
\end{figure}

With these criteria for a continuum detection,
we removed roughly two-thirds of the candidates from our consideration as they do not
meet our intensity detection threshold (continuum emission >$5\sigma$), are spatially separated with more than half a beam, have an unclear association with an extended continuum structure, or are most likely affected by
sidelobes. The final list of 35 sources that we investigate further is given in Table~\ref{tab:sources1}. For each radio source, we detail the
position of the YSO and the position of peak intensity of the associated
radio source as well as the peak intensity, $S_{p}$. We calculated the integrated
flux density, $S_{int}$, using 5$\sigma$ contours or covering an area of at least a GLOSTAR beam
size. This also serves to estimate an effective
source diameter, $D_{\text{eff}}$, where we assume the \hii~region is
spherically symmetric. The rms used to define the detection limit for each
source is also given in the table. We note that for sources 88, 230, 241, and 311, we made a manual integration contour. This was done as a 5$\sigma$ contour does not perfectly capture just the emission of the local compact source we are interested in. In these cases, our flux estimation acts as a lower limit. The general shape of the radio
sources are classified as compact (C; $\sim$66\%), extended (E; $\sim$28\%), or
extended and complex (EC; $\sim$6\%). An example source of our final sample is shown in 
Fig.~\ref{fig:example_source} (and the rest in Figs.~\ref{fig:manySource1}-~\ref{fig:manySource6}). This illustrates an example association of a YSO candidate with an extended radio feature as it satisfies our criteria of being within 10\asec of the continuum radio and within a 5$\sigma$ contour. However, this is not always easily discernible for all of our sources. For example, sources 307 and 311 were previously observed at 10.7\,GHz with Effelsberg and listed as sources 40 and 41, respectively \citep{seiradakis1989}. These single dish observations report higher flux densities of 28.5\,Jy and 4.3\,Jy, respectively, compared to our $\sim$0.2\,Jy and $\sim$1.2\,Jy. Furthermore, the sizes are of the order of 2 arcminutes compared to our arcsecond scale. The fact that these properties are so different from our higher angular resolution data indicates that for these sources in particular, they are sub-components of a larger and complex multi-component source. The Effelsberg data from \citep{seiradakis1989} capture the more extended emission that is resolved out at the angular resolution of our GLOSTAR-VLA data. This can be seen for other sources in our final sample as well, where the radio source we associate with our YSO candidate may be a part of a much more complicated multi-component source. For example, source 3 (see \ref{fig:manySource1}) shows a continuum feature at 5$\sigma$ above the local noise and is consistent in all the frequency bands of the GLOSTAR continuum data. However, it might not be a completely isolated and compact source as it is not distinctly 5$\sigma$ above the local extended emission. In these cases, we recognise that the \hii~region associated with the YSO might actually be larger in some cases and would require higher resolution data to resolve the source.

We note that the distance to these radio counterparts is assumed to be the same as the YSOs placing them in the CMZ. Association with distance tracers such as \hi~absorption or the methanol maser to these radio sources as well as detected molecular lines for the YSOs would be needed to clearly determine their association along the line of sight.

\begin{sidewaystable*}
\caption{Positions and integrated flux densities of the candidate YSO--5.8\,GHz radio continuum
associations. Grouped here are sources with a continuum detection and angular separation within a GLOSTAR beam.}
\label{tab:sources1}
\centering
\begin{threeparttable}

\begin{tabular}{r| r r r r r r r r r r r r r r r} 
\hline\hline 
Source ID & \multicolumn{2}{c}{YSO} & \multicolumn{2}{c}{Cont$_{5.8\,\text{GHz}}$} & $S_{p}$ & $S_{int}$ & $\delta S_{i}$ & $\alpha(S\propto\nu^{\alpha})$ &$D_{\text{eff}}$ & Separation & rms  & Q$_{shape}$ &  Catalogues \\ 
       & $l[^{\circ}]$ & $b[^{\circ}]$ & $l[^{\circ}]$ & $b[^{\circ}]$ & [mJy beam$^{-1}$] & [mJy] & [mJy] & & [\asec]& [\asec] & [mJy beam$^{-1}$]& & & &  \\
\hline
3& 358.7918& 0.0117& 358.7920& 0.0117& 2.1& 4.6& 0.2&-&33.3& 0.6& 0.4&   C&  WC\\ 
5& 358.8437& 0.0259& 358.8444& 0.0257& 19.5& 27.6& 0.3&$-0.2\pm0.03$&37.5& 2.9& 0.8&   C&  WC,N\\ 
34& 359.1581& -0.0363& 359.1594& -0.0365& 28.0& 109.5& 1.1& $-0.3\pm0.03$ &56.9& 4.8& 0.9&   E&  WK, N\\ 
44& 359.2365& -0.0364& 359.2389& -0.0361& 11.0& 11.2& 0.1&$-0.1\pm0.12$&24.3& 8.8& 1.3&   E&  WC\\ 
51& 359.2777& -0.0618& 359.2785& -0.0625& 0.9& 0.9& 0.0& -&<18.0& 3.6& 0.3&   C&  WC\\ 
54& 359.2796& -0.0407& 359.2806& -0.0417& 7.7& 10.1& 0.2& $-0.2\pm0.2$&33.9& 4.8& 2.3&   E&  WQ,A\\ 
64& 359.3347& -0.0425& 359.3361& -0.0438& 8.7& 9.1& 0.1&$-0.3\pm0.2$&26.6& 6.6& 0.8&   C&  WC\\ 
66& 359.3173& 0.0771& 359.3187& 0.0764& 2.7& 3.4& 0.1&$0.5\pm0.5$&26.6& 5.6& 0.5&    C&  -\\ 
78& 359.4288& 0.0353& 359.4292& 0.0347& 6.7& 8.0& 0.1&$1.3\pm0.3$&29.0& 2.5& 0.9&   EC&  WC\\ 
80& 359.4037& 0.0016& 359.4049& -0.0007& 13.5& 19.2& 0.2&$-0.4\pm0.2$&34.7& 9.3& 0.9&   C&  WC,A\\ 
82& 359.4225& 0.0152& 359.4236& 0.0146& 7.8& 11.1& 0.2&$-0.02\pm0.36$&33.5& 4.7& 0.8&   C&  WC\\ 
83& 359.4348& 0.0225& 359.4319& 0.0222& 10.6& 13.6& 0.2&$-0.5\pm0.2$&31.0& 10.4$^{a}$& 0.9&   C&  WC\\ 
87& 359.4673& -0.1713& 359.4687& -0.1715& 53.2& 169.7& 1.8&$0.3\pm0.05$&72.1& 5.4& 0.8&  EC&  WK,A\\ 
88& 359.4665& -0.0735& 359.4678& -0.0727& 7.1& 7.3& 0.1&$-2.3\pm1.8$&26.5& 5.4& 0.4&  C&  WC\\
89& 359.4554& -0.0663& 359.4576& -0.0667& 9.5& 10.1& 0.2&$0.5\pm0.3$&29.9& 8.2& 1.4&  C&  WC,A\\ 
91& 359.4563& 0.0199& 359.4569& 0.0194& 4.3& 5.2& 0.1&-&27.6& 2.9& 0.5&  C&  WC,A\\
93& 359.4964& 0.0387& 359.4962& 0.0378& 4.0& 5.9& 0.2&$0.05\pm0.31$&32.5& 3.2& 0.9&  C&  WC\\ 
115& 359.6436& -0.0389& 359.6437& -0.0396& 16.1& 17.8& 0.2&$0.2\pm0.2$&30.8& 2.4& 1.6&  C&  WC\\ 
135& 359.7919& -0.0439& 359.7917& -0.0458& 9.9& 10.6& 0.1&$-0.1\pm0.5$&28.1& 7.0& 1.4&  C&  WC\\ 
140& 359.7843& 0.0400& 359.7806& 0.0417& 39.5& 71.0& 0.7&$0.2\pm0.1$&40.8& 14.8$^{a}$& 0.9&  E&  WK\\ 
147& 359.8428& -0.0144& 359.8444& -0.0139& 14.6& 22.6& 0.2&$-0.5\pm0.8$&30.8& 6.4& 1.6&  C&  WC\\ 
157& 359.8651& -0.0860& 359.8667& -0.0868& 107.8& 115.2& 1.0& $0.3\pm0.2$&34.3& 6.3& 1.3&  C& A,$^{b}$\\
230& 0.2883& 0.0519& 0.2896& 0.0514& 9.7& 15.9& 0.1& $-0.2\pm0.2$&19.7& 5.1& 2.1&  E&  A\\ 
234& 0.3368& -0.0133& 0.3358& -0.0108& 89.2& 330.0& 3.0&$-0.2\pm0.05$&63.0& 9.9& 1.9&  E&  WK,A\\ 
235& 0.3466& -0.0271& 0.3472& -0.0278& 63.0& 73.2& 0.6&$-0.1\pm0.02$&27.5& 3.4& 2.0&  C&  WC,A\\ 
241& 0.3527& -0.0199& 0.3554& -0.0201& 12.5& 14.0& 0.1&$-0.6\pm0.23$&25.4& 9.8& 1.7&  C&  WC,A\\ 
262& 0.4314& 0.2617& 0.4319& 0.2611& 32.1& 34.8& 0.3&$-0.2\pm0.05$&33.3& 2.8& 0.4&  C&  B\\ 
277& 0.5294& -0.1061& 0.5299& -0.1083& 37.6& 48.8& 0.5&$-0.3\pm0.3$&32.7& 8.1& 6.9&  E&  WC\\ 
284& 0.5332& 0.1687& 0.5347& 0.1694& 31.2& 44.6& 0.5&$-0.2\pm0.1$&39.0& 6.2& 0.9&  C&  WG,A\\ 
296& 0.6050& -0.2017& 0.6056& -0.2021& 22.8& 41.3& 0.6&$-0.4\pm0.16$&50.1& 2.4& 0.5&  C&  WC,N\\ 
299& 0.6331& -0.1152& 0.6354& -0.1139& 97.1& 298.2& 2.6&$-0.2\pm0.2$&55.5& 9.4& 3.4&  E&  WK\\ 
307& 0.6673& -0.0911& 0.6681& -0.0910& 53.1& 192.5& 2.1&$0.04\pm0.19$&55.3& 2.6& 4.9&  E&  A\\ 
311& 0.6879& -0.0325& 0.6896& -0.0343& 909.2&1175.1 & 8.1&$0.3\pm0.04$2&28.2& 8.9& 11.9&  E&  WG,A\\ 
315& 0.7369& -0.2908& 0.7375& -0.2917& 7.5&8.3 & 0.1&$0.3\pm0.2$&26.8& 3.7& 0.4&  C&  -\\ 
323& 0.8439& -0.0538& 0.8458& -0.0549& 22.6& 38.9& 0.4&$0.1\pm0.3$&37.1& 7.8& 1.2&  E&  -\\

\hline
\end{tabular}
\begin{tablenotes}
\scriptsize
\item \textbf{Notes.} {\it From left to right:} Source number and YSO Galactic coordinates from \cite{nandakumar2018}. The Galactic coordinates of the pixel with the peak intensity of the 5.8\,GHz continuum (this work), the peak intensity ($S_{p}$), the integrated flux density ($S_{int}$) and error ($\delta S_{i}$; \citealt{cornish_source}), and the spectral index ($\alpha$). The effective diameter of the source ($D_{\text{eff}}$) was obtained by assuming a circle with an equivalent area to the area enclosed by the contour used for flux determination. The separation indicates the angular distance between the reported YSO position in \cite{nandakumar2018} and the peak intensity position of the radio continuum source where we are constrained by the pixel size (2.5\asec). The root-mean-square (rms) indicates the noise level that was obtained from a nearby emission-free patch. The general morphology (Q$_{shape}$) of the source is classified as compact (C), extended (E), and extended \& complicated (EC). Previous classifications and detections (catalogues) include WISE candidate (WC), radio quiet (WQ), group (WG), or known \hii~regions for nearby sources (WK; all \citealt{Anderson2014}), as well as sources with counterparts in the NVSS survey \citep{NVSS}, \cite{becker1994}, or ATLASGAL \citep{contreras2017}.
\item \textbf{Footnotes.} (a) Despite the slightly larger separation, the YSO lies well within the main 5.8\,GHz emission peak. (b) Associated to the `20\,km\,$^{-1}$ cloud'.
\end{tablenotes}

\end{threeparttable}
\end{sidewaystable*}


\subsection{Spectral index}\label{sect:specIndex}
 
\begin{figure}
    \centering
    \includegraphics[width=0.5\textwidth]{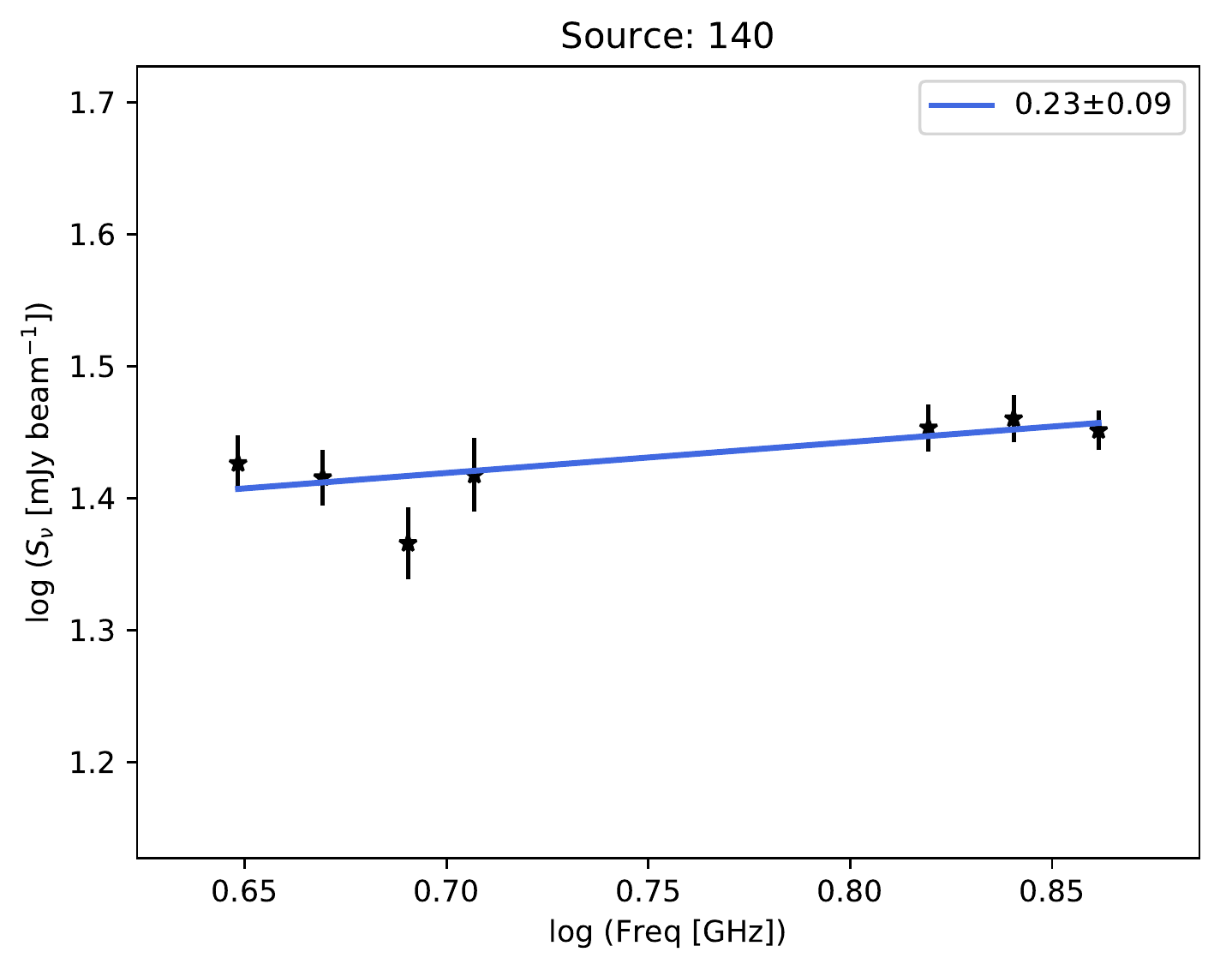}
    \caption{Determination of the spectral index for source 140 (G359.784+0.040). Only frequency bands that
    meet the minimum threshold of 3$\sigma$ are considered. The solid blue line shows the best-fit model determined by the linear regression. The fitted spectral index is indicated in the upper right corner. The frequencies range from 4.45\,GHz to 7.2\,GHz. The spectral indices of the full sample are shown in Appendix~\ref{app:specIndex}.}
    \label{fig:specIndex_fit}
\end{figure}

We used the multiple spectral windows of the GLOSTAR observations at 
different frequencies to perform an estimate of the
spectral index of each continuum source. Following the same procedure for spectral index calculations as \cite{bihr2016}, we extracted the peak intensity from each individual frequency plane at the same
position, using the peak pixel of the continuum emission in the averaged
GLOSTAR image (see Fig.~\ref{fig:specIndex_example}).
We used the intensity at the peak pixel, since the integrated emission may be more
heavily impacted by the frequency-dependent spatial filtering of the
interferometer. It should be noted that for some sources, the shape of the source is inconsistent in every spectral window. Some frequency planes show more
extended features or disappear entirely. We therefore limit the number of
frequency planes used for further analysis to those that have
a peak intensity greater than 3$\sigma$, which we note is lower than the 5$\sigma$ limit used for integrated flux calculations as each spectral window, from which we extracted the peak intensity, does not benefit from the decreased noise from combining all the frequency planes. We used this lower threshold
for detection as each of the individual frequency planes has a lower
signal-to-noise ratio (S/N) compared to the averaged image. From this, we extracted a set
of peak intensities from which to estimate the spectral
index. The possible missing flux, which is inherent to interferometric observations, may affect spectral indices of extended sources by frequency-dependant filtering. This is not further addressed here but will be in the future with the inclusion of GLOSTAR-Effelsberg observations. Assuming that the relationship between the flux and frequency
is $S_{\nu}\propto\nu^{\alpha}$ where $\alpha$ is the spectral index
and $S_{\nu}$ is the frequency-dependent intensity at the associated frequency $\nu$, we used
scipy's \texttt{curve\_fit} to perform a linear fit of the data in
log-space in order to obtain the slope, $\alpha$ (see Fig.~\ref{fig:specIndex_fit}), where the measured errors are only from the fitting procedure.

Using the spectral indices, we classified the continuum sources as 
\hii~regions depending on whether the emission is thermal or non-thermal.
For indices between $-$0.1 and 2, the emission corresponds to
thermal emission that is associated with \hii~regions or
planetary nebulae (PNe). If it is steeply negative, 
that is $<-0.5$, we consider it to be non-thermal, which means the emission is synchrotron in
nature and could come from supernova remnants (SNR) or extra-galactic sources 
such as active galactic nuclei (AGN) (\citealt{condon1984}, \citealt{rodriguez2012}, \citealt{dzib2013}, \citealt{chakraborty2020}).
We record the values of the spectral indices in Table~\ref{tab:sources1}.
We conclude that we can only use the
spectral indices to propose sources as \hii~region candidates since a larger frequency coverage with greater
accuracy would be needed in order to truly constrain the spectral indices and thus their nature.

For 11 sources, the spectral index was in agreement with thermal emission as defined above.
For three sources of our sample, we could not determine reliable spectral indices, as they had less than three spectral windows with a good enough S/N. Four sources show steeply negative spectral indices of $\sim-0.5$, which we classify as non-thermal from our aforementioned definition. However, the errors are quite large and not sufficient to definitively exclude these candidates. The remaining 16 sources have values in between $-0.5<\alpha<-0.1$ and are retained in the analysis that follows as \hii~region candidates.

\section{Discussion}\label{sect:discussion}

\subsection{YSOs and their counterparts}\label{sect:yso_discuss}
To explore the nature of the YSOs, we discuss the properties of the sources that have a
GLOSTAR-VLA radio continuum counterpart. 
Fig.~\ref{fig:yso_overlay_glostar} shows that while the majority of detections are in the Galactic mid-plane, which is also the case in \cite{sac2019}, there is no clear separation between the
sources that have radio continuum counterparts and those that do not.
We do a full comparison with \hii~region catalogues from \cite{Anderson2014}, radio source catalogues from \cite{becker1994} and \cite{NVSS}, and the MMB methanol maser catalogue from \cite{MMBcaswell2010}. The associations are shown in the last column of Table~\ref{tab:sources1}.

We also performed a cross match for known sources in the literature.
To date, the most complete and comprehensive catalogue of \hii~region candidates is presented in \cite{Anderson2014} by using sources from the all-sky 
\textit{Wide-Field Infrared Survey Explorer (WISE)} satellite and investigating
their mid-infrared (MIR) morphology\footnote{\url{http://astro.phys.wvu.edu/wise/}}, confirming candidates 
with existing literature. Of the $\sim$8000 sources in their catalogue, $\sim$450 are within our region of
study. These were used to confirm the nature of the detected GLOSTAR sources. Of our 35 GLOSTAR sources, only six sources do not have a WISE counterpart and six
are classified as, or are a part of, known \hii~regions by \cite{Anderson2014}. The remaining 23 radio sources correspond to \hii~region candidates
based on their MIR morphology. The strong correlation of the infrared WISE sources to our radio sources can be expected given that we selected these radio sources based on a catalogue of NIR sources. The six radio continuum sources without WISE counterparts have source IDs 66, 157, 230, 307, 315, and 323 as listed in Table~\ref{tab:sources1} and are potentially new \hii~regions due to the lack of associations, except for source 157, as it is likely associated with the `20\,km\,s$^{-1}$ cloud'. However, none of them show any methanol masers from the MMB, which suggests that these \hii~regions are at a later stage of HMSF.

Additionally, we searched for counterparts with the CMZoom survey \citep{battersby2020} as it searches the dust continuum at 1.3~mm for signs of compact substructures known to be sites of high mass star formation. We used the catalogue from \cite{hatchfield2020} and find that of the 334 YSOs, only 22 have potential counterparts within 45\asec, where the angular separation criterion is chosen based on the CMZoom's upper sensitivity to structures on that scale. Of these 22 sources, only two also have a GLOSTAR counterpart, sources 307 and 311 from \cite{nandakumar2018}.

The recent study by \cite{lu2019b} observed a smaller region of the CMZ also with the VLA in C-band, but with the interferometer in B-configuration, yielding a higher spatial resolution ($\sim1$\asec). They detected 104 radio continuum sources of a varying nature. Our radio continuum sources 140 and 241 overlap with the sources C54 and C29 from \cite{lu2019b}, respectively, where C54 is a candidate ultra-compact \hii~region. However, a more in-depth analysis in the future with the upcoming higher resolution GLOSTAR-VLA data will provide a more complete comparison between the two data sets.

We additionally compared \TBF{the YSO catalogue} with the compact sources from ATLASGAL to investigate if the YSO is in the young protostar stage, as a young protostar needs to be embedded in compact dense gas, which is visible at sub-millimetre wavelengths.
The earliest stages of HMSF can be observed in 
massive clumps of dust and gas that emit at the optically thin 
(sub)millimetre regime through their thermal emission. ATLASGAL~(\citealt{atlasgal}) observes 
thermal dust emission at 870\,$\mu$m which aids in the study of the early natal
environment of high-mass stars at their early pre-stellar stages. 
With a resolution of 19\asec (FWHM), this unbiased survey has produced a 
catalogue of $\sim$10~000 dense and massive clumps (\citealt{atlasgalCSC}, \citealt{urquhart2014b}, \citealt{atlasgalGSC}).

\begin{figure}
    \centering
    \includegraphics[width=0.5\textwidth]{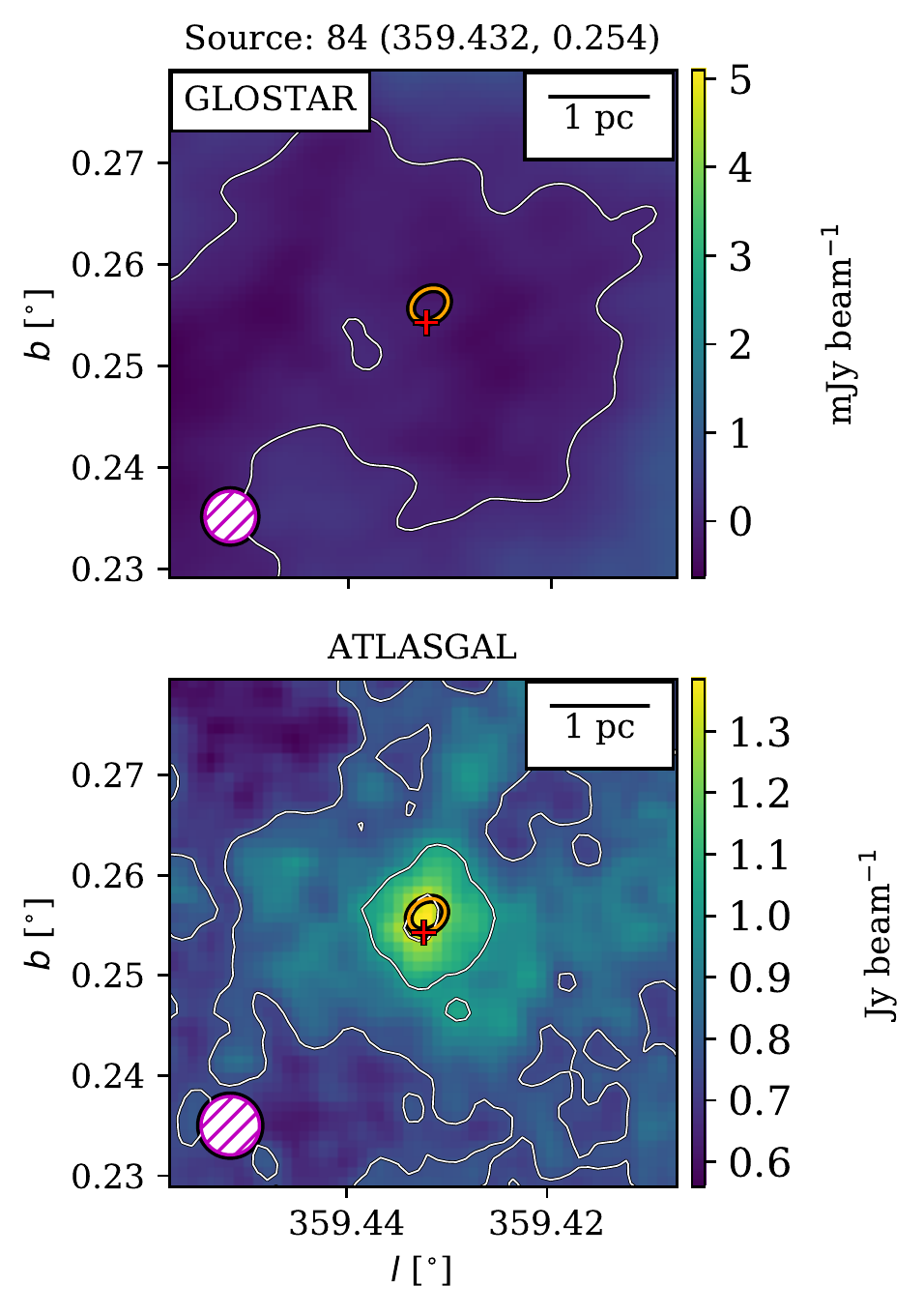}
    \caption{Non-detection of GLOSTAR 5.8\,GHz emission and detection of ATLASGAL 870\,$\mu$m emission towards source 84 from \cite{nandakumar2018}. \textit{Top}: GLOSTAR 5.8\,GHz radio continuum 3\amin$\times$3\amin zoom-in. A red cross denotes the location of the YSO, and the orange ellipse represents
    ATLASGAL sources from \cite{atlasgalCSC}. The GLOSTAR beam size is shown in the bottom left corner. The white contour shows the 0.5\,mJy\,beam$^{-1}$ level, which is the average 5$\sigma$ level for the GLOSTAR data. \textit{Bottom}: Shown is the ATLASGAL map, where the contours are dynamical contours \citep[as formulated by][]{thompson2006}.
    }
    \label{fig:yso_atlasgal}
\end{figure}

\begin{figure}
    \centering
    \includegraphics[width=0.5\textwidth]{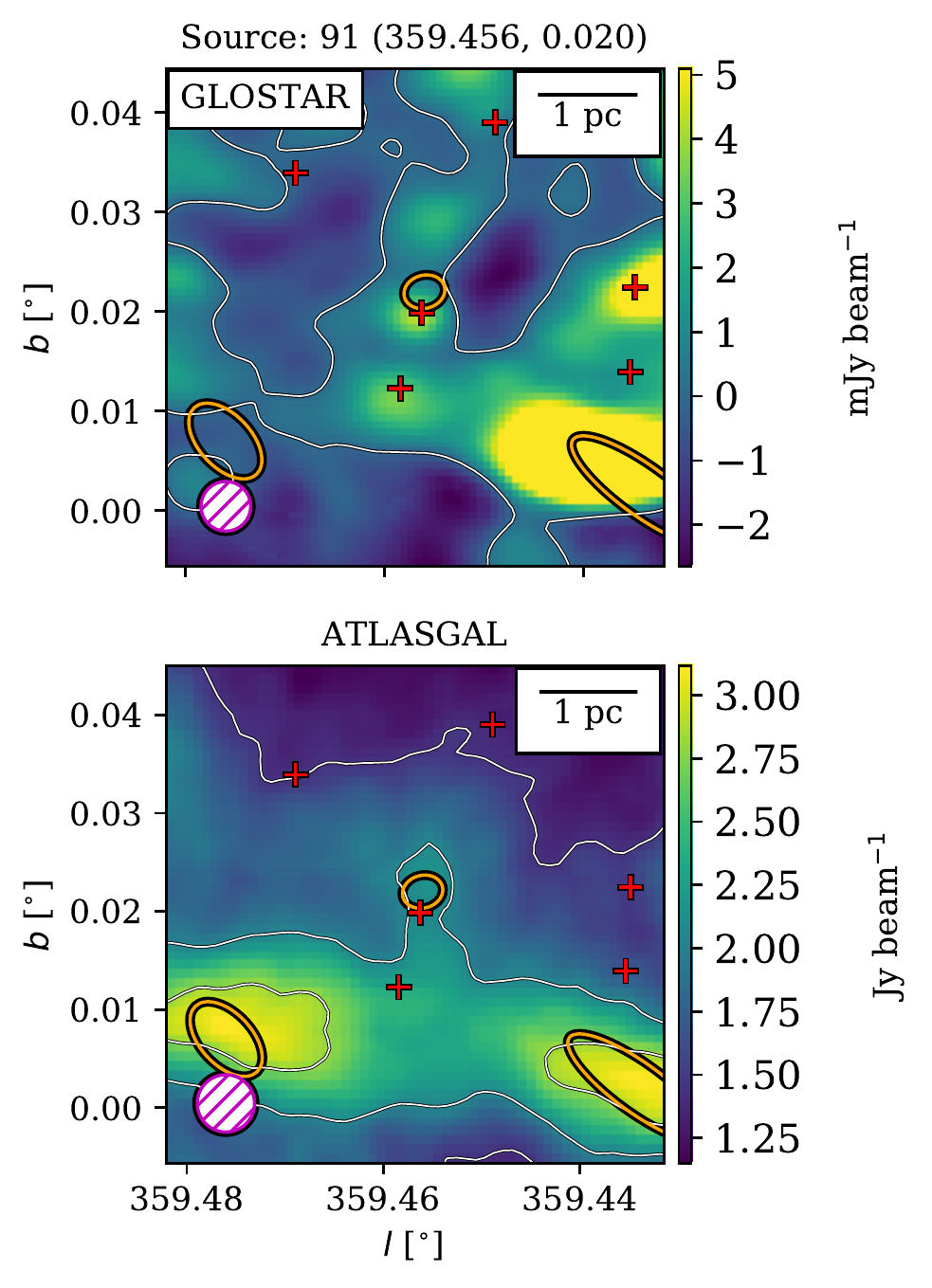}
    \caption{Detections of GLOSTAR and ATLASGAL emission of source 91. Labels and markers are the same as in  Fig.~\ref{fig:yso_atlasgal}. The GLOSTAR image (top) instead has contours at 2.5\,mJy\,beam$^{-1}$. It is the singular YSO candidate with
    both an ATLASGAL and potential GLOSTAR counterpart.
    }
    \label{fig:yso_atlasgal_glostar}
\end{figure}

Comparing the GLOSTAR-VLA data with ATLASGAL can give insight into the nature of
the observed potential \hii~regions as one would expect evolved dust clumps to have a radio continuum counterpart. 
However, of the $\sim$1000 compact sources from ATLASGAL in the CMZ area \citep{atlasgalCSC}, 
only eight have an angular separation of $<10$\asec 
to a YSO candidate from \cite{nandakumar2018} (see Fig.~\ref{fig:yso_atlasgal}). Of these eight associations, YSO candidate 91 has a potential radio continuum 
counterpart at 5.8\,GHz (see Fig.~\ref{fig:yso_atlasgal_glostar}). 
At 870~$\mu$m, it has a peak intensity of $0.54\pm0.23$\,Jy\,beam$^{-1}$, an integrated flux density of $1.89\pm0.5$\,Jy,
and a S/N of 8. While we expect that ultra-compact \hii~regions have both dust and radio emission, this source hardly classifies as an ultra-compact \hii~region as its physical properties (see Table~\ref{tab:res}) do not meet the typical minimum requirements (<0.1\,pc; \citealt{kurtz2002}). 
Furthermore, Fig.~\ref{fig:yso_atlasgal_glostar} shows that the radio continuum source is offset from the main dust clump, suggesting that
the \hii~region is of a more evolved state than the surrounding dense clumps. For the remaining seven sources that do not have a radio continuum counterpart, we suggest that these YSOs are not yet sufficiently evolved as we expect MYSOs to be IR-bright
prior to being able to see the inner and developing \hii~region once it turns on \citep{motte2018}.

This low YSO-ATLASGAL association rate is unexpected. Using the empirical mass-size relation (under the assumption that the sources fill the 19\asec ATLASGAL beam; \citealt{urquhart2018}), it is generally accepted that a cloud mass of 500-1000\,\msol is needed in order for at least one massive star to be formed in the cluster. This corresponds to peak intensities of 0.75-1.5\,Jy\,beam$^{-1}$.
In comparison to the typical noise of ATLASGAL in the CMZ of $\sim$0.2\,Jy\,beam$^{-1}$, we should be able to detect these kinds of clumps.
However, given the extended nature of the dust emission seen in ATLASGAL and the limitations of source extraction, our original $<10$\asec angular separation is likely insufficient to describe the full possible connection between our sample of YSOs and dust clouds. As such, we compare the positions of the full sample of YSO candidates to an emission map of ATLASGAL, which gives a total of 94 YSOs that lie in the dust features at 870\,$\mu$m. Of these YSOs with GLOSTAR counterparts (35), 14 sources would then be classified as having an ATLASGAL association in this way. These sources having coincident associations are consistent with the picture that these YSOs are embedded in their natal dust envelopes. For the remaining sources without radio continuum or dust emission, it may be that they are instead associated with lower-mass dust clumps, suggesting that they may not be high-mass YSOs.
\subsection{Properties of the YSO sub-sample}
\begin{figure*}[!h]
\begin{tabular}{cc}
\includegraphics[width=0.5\textwidth]{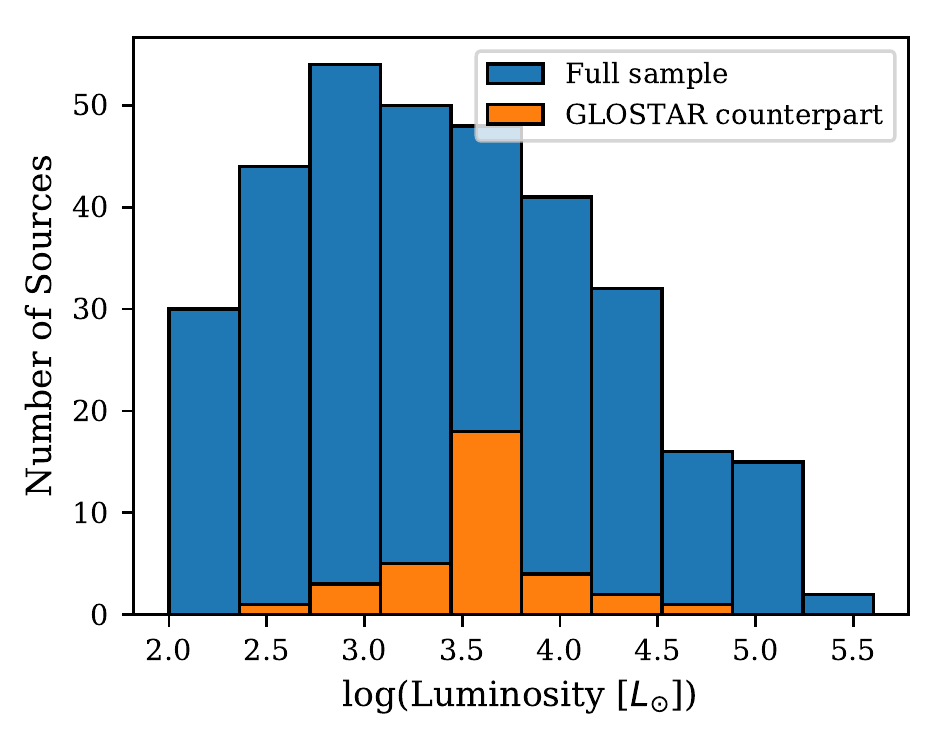}
\includegraphics[width=0.489\textwidth]{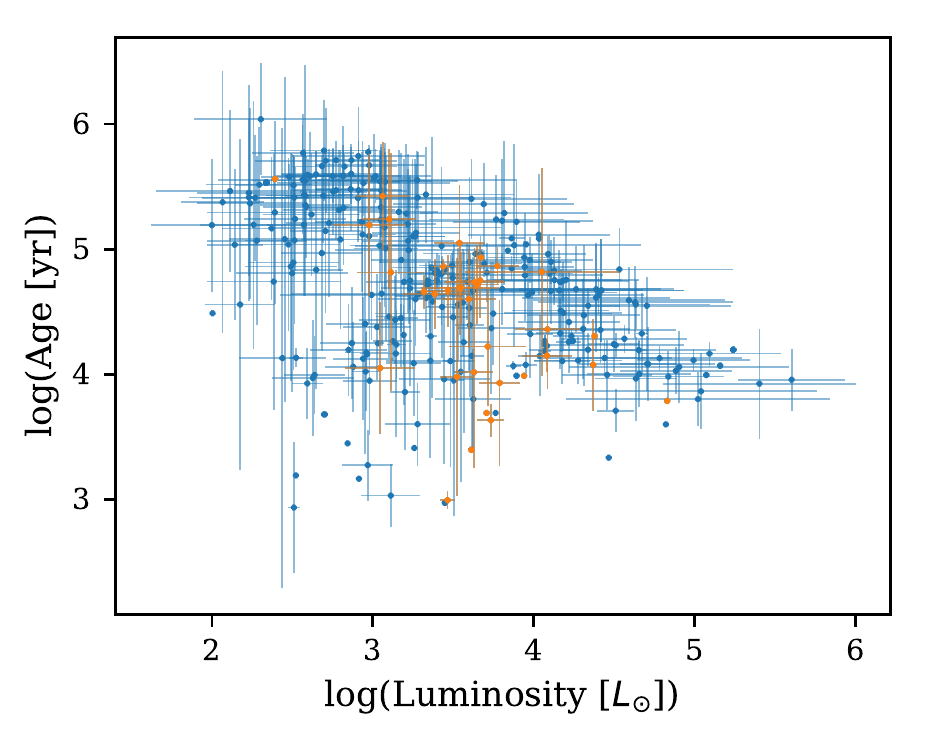}\\
\includegraphics[width=0.489\textwidth]{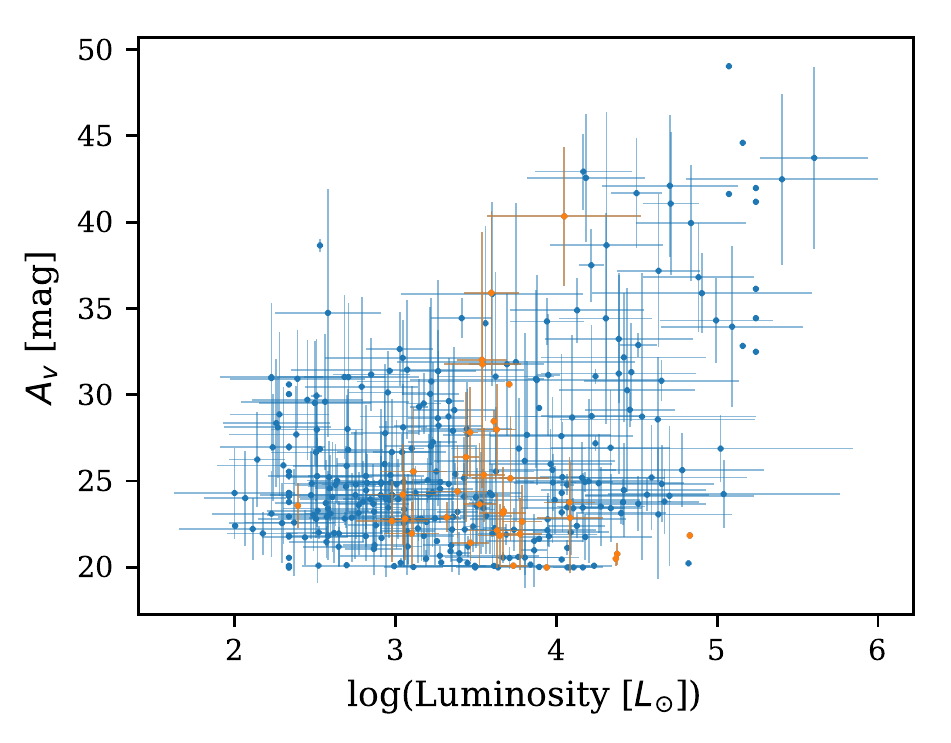}
\includegraphics[width=0.506\textwidth]{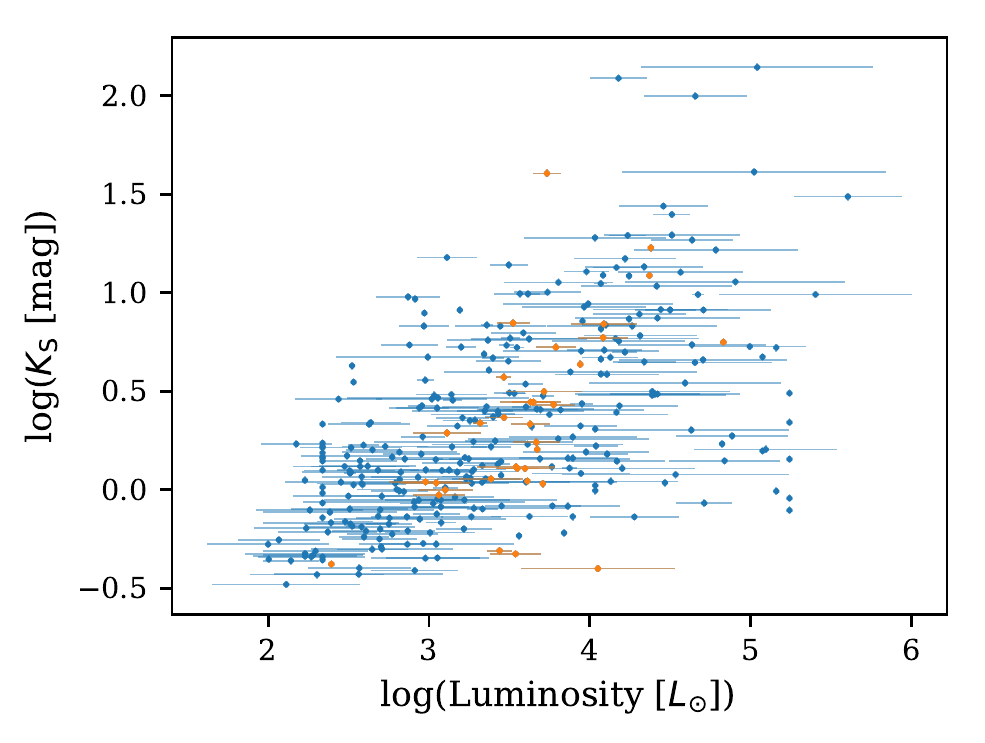}\\
\end{tabular}
\caption{Comparison of the bolometric luminosity of the YSOs to other properties (obtained from SED models \citep{robitaille2017} in the work done by \citealt{nandakumar2018}). \textit{Top left}: Histogram of the YSO luminosity. \textit{Top right}: Estimated age of the YSO obtained from stellar evolution tracks. \textit{Bottom left}: Calculated extinction, $A_{v}$. \textit{Bottom right}: Brightness at photometric band, K$_{\rm{S}}$. Shown in blue are all of the YSO sources which \cite{nandakumar2018} used in their determination of the SFR, while orange denotes only the YSO sources that have a GLOSTAR counterpart.}
\label{fig:yso_properties}
\end{figure*}

Of the 334 YSO sources we investigated to find GLOSTAR counterparts, we found 35 confident
candidates. To determine if there are systematic effects, in Fig.~\ref{fig:yso_properties} we plot a comparison of our
selected sub-sample of YSOs to the full catalogue. We compared the modelled total stellar luminosities from \cite{nandakumar2018} with their derived age, extinction in the V-band, and observed K$_{\rm{S}}$
photometric magnitude to determine if only the most powerful or luminous YSOs have radio-loud \hii~regions. The top left shows the distribution of the luminosities where the median luminosity of the full sample is of the order of $\sim10^{3}$\,L$_{\odot}$, whereas the sample
of sources with GLOSTAR counterparts have a slightly higher median value, but it does not seem to be strongly biased towards either the higher or lower limit of the distribution of the luminosities. We compare the
luminosities with the calculated age of the YSOs in the top right corner of this figure and we see that our sub-sample covers only the young stars, $<10^{5}$\,yr. This could explain the low association rate of the YSOs and GLOSTAR radio continuum data, as the YSOs would still be young and would not have reached the point of developing an \hii~region. However, these ages were calculated from SED model fitting \citep{robitaille2017} where there can be large uncertainties for the age estimation. If they are truly in the early stages, one would expect to have a correlation with ATLASGAL sources. However, as explored in Section~\ref{sect:yso_discuss}, there is limited ATLASGAL correlation, suggesting instead that these sources are either much older or are non-massive YSOs. As explained in the above section, we maintain that the noise level of ATLASGAL does not allow us to use the absence of sub-millimetre sources as a strict constraint on the evolutionary stage in this work. 

Using an assumed distance in the ranges of 7--9\,kpc for their SED models, \cite{nandakumar2018} fitted values for the visual extinction, $A_{\rm V}$, caused by the material along the line of sight from the Sun to the CMZ. We see that most of the sources fall under $A_{\rm V}=36$\,mag and that there are no evident trends in comparing the luminosity with the photometric magnitude. In addition to the large uncertainties, the small sample size inhibits any defining conclusions from these results.

\subsection{Notes on particular sources}\label{sect:gmc}
\begin{figure*}[!h]
\begin{tabular}{ccc}
\includegraphics[width=1\textwidth]{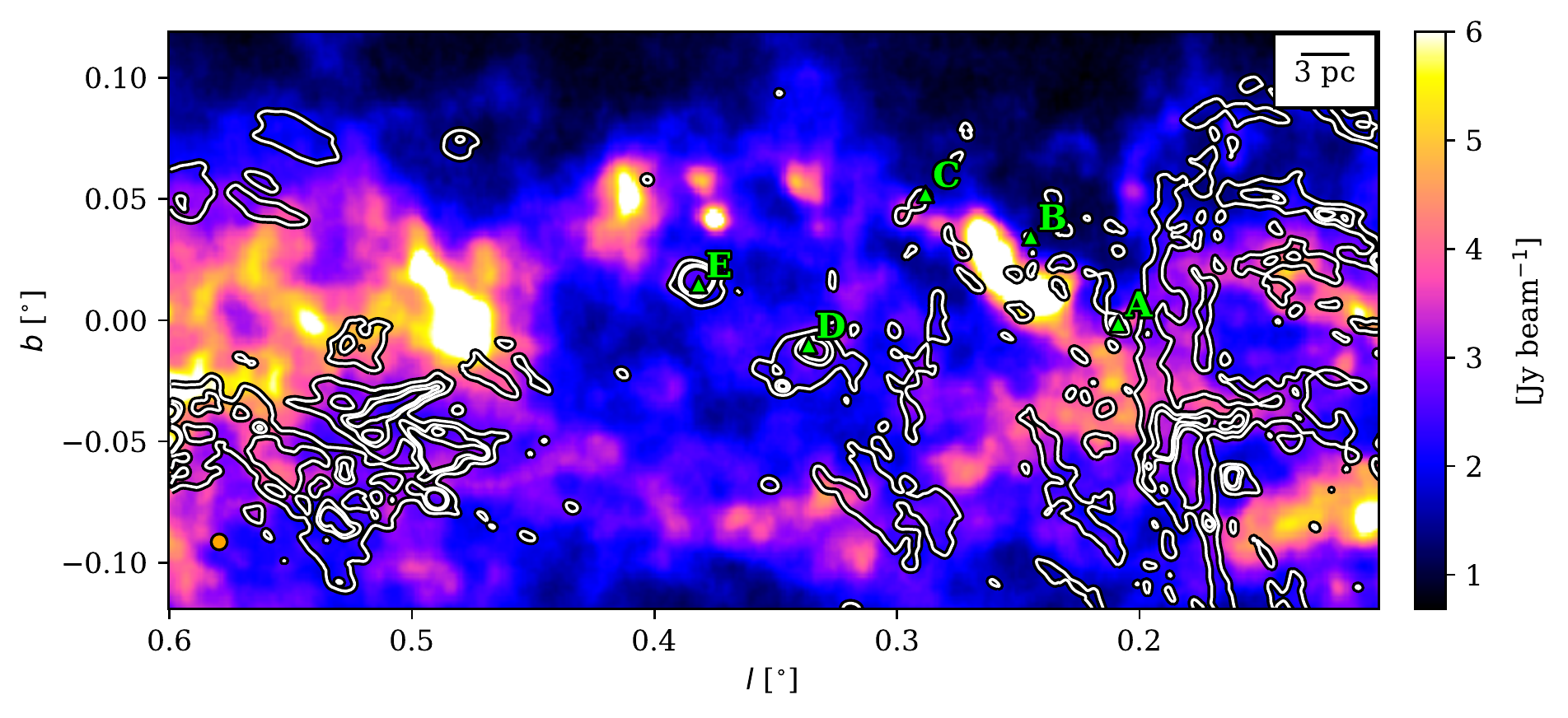}\\
\includegraphics[width=0.33\textwidth]{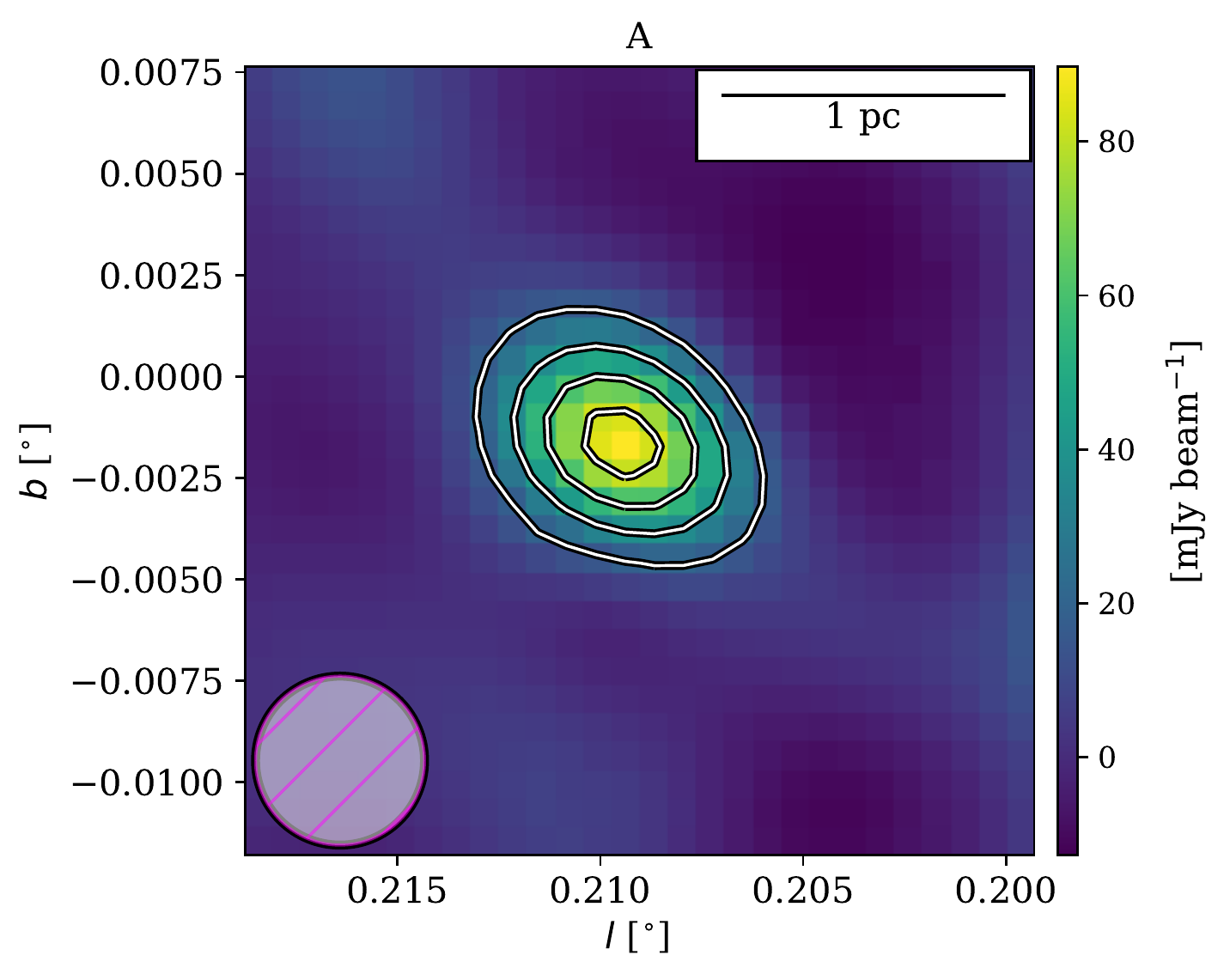}
\includegraphics[width=0.33\textwidth]{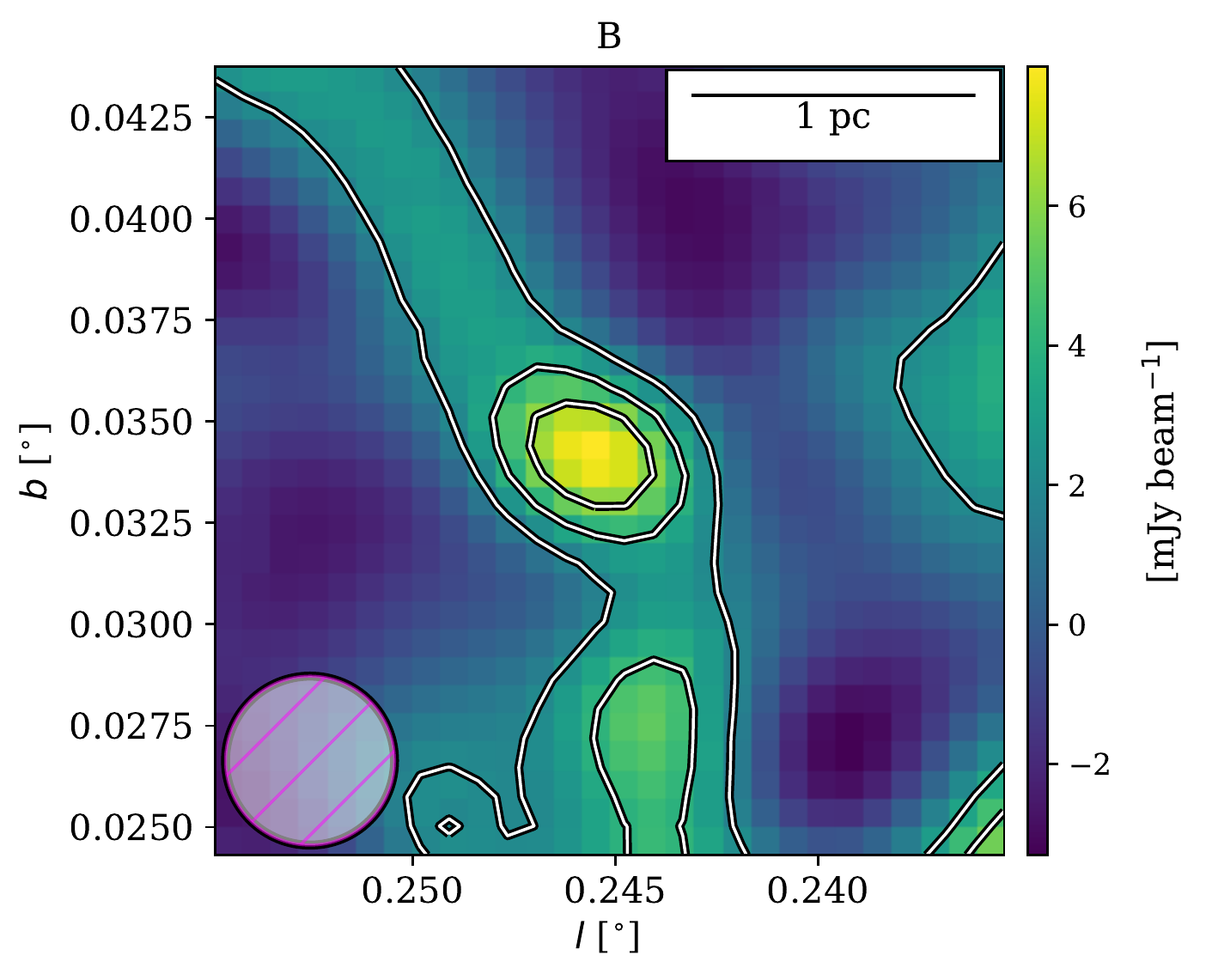}
\includegraphics[width=0.33\textwidth]{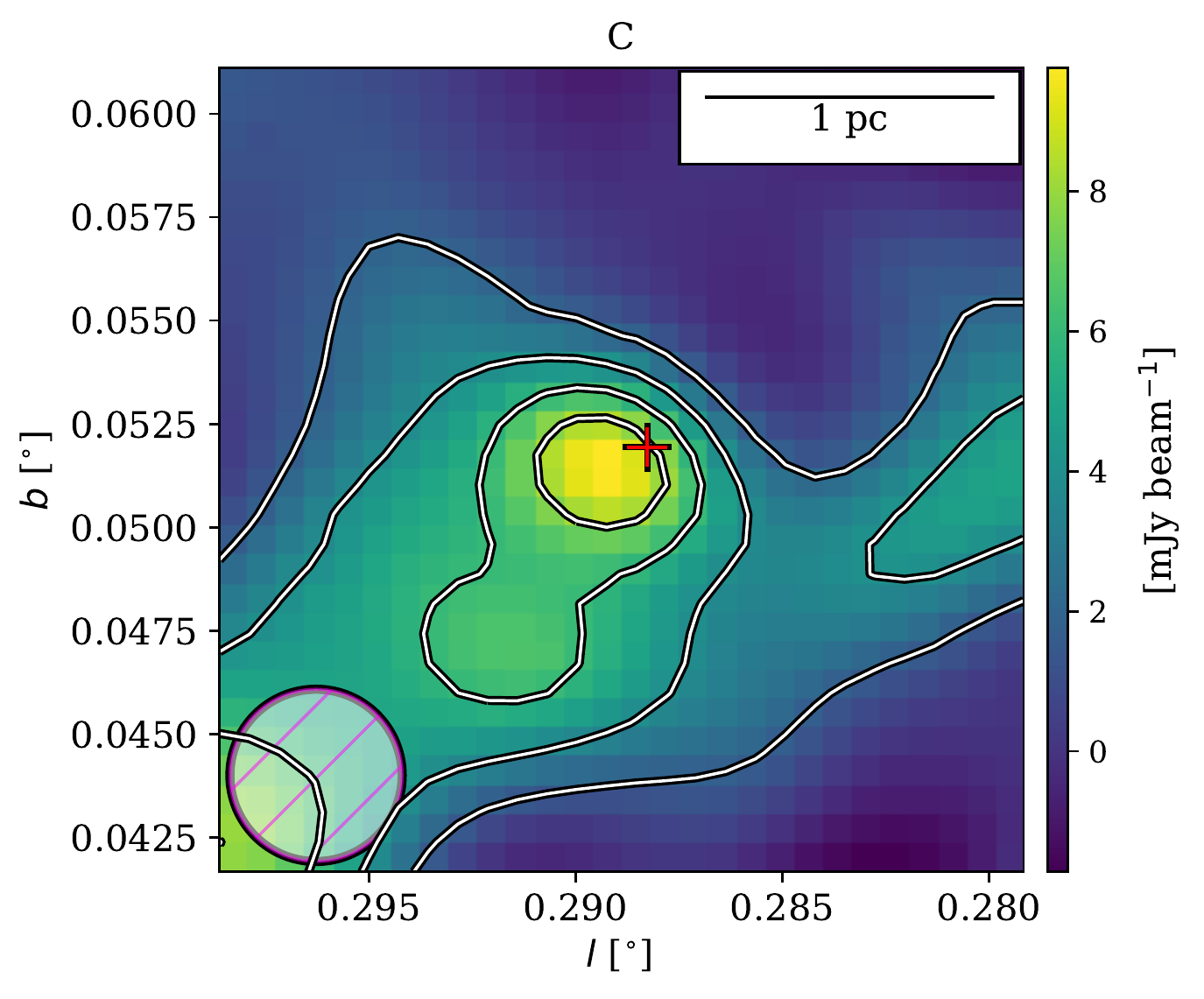}\\
\includegraphics[width=0.33\textwidth]{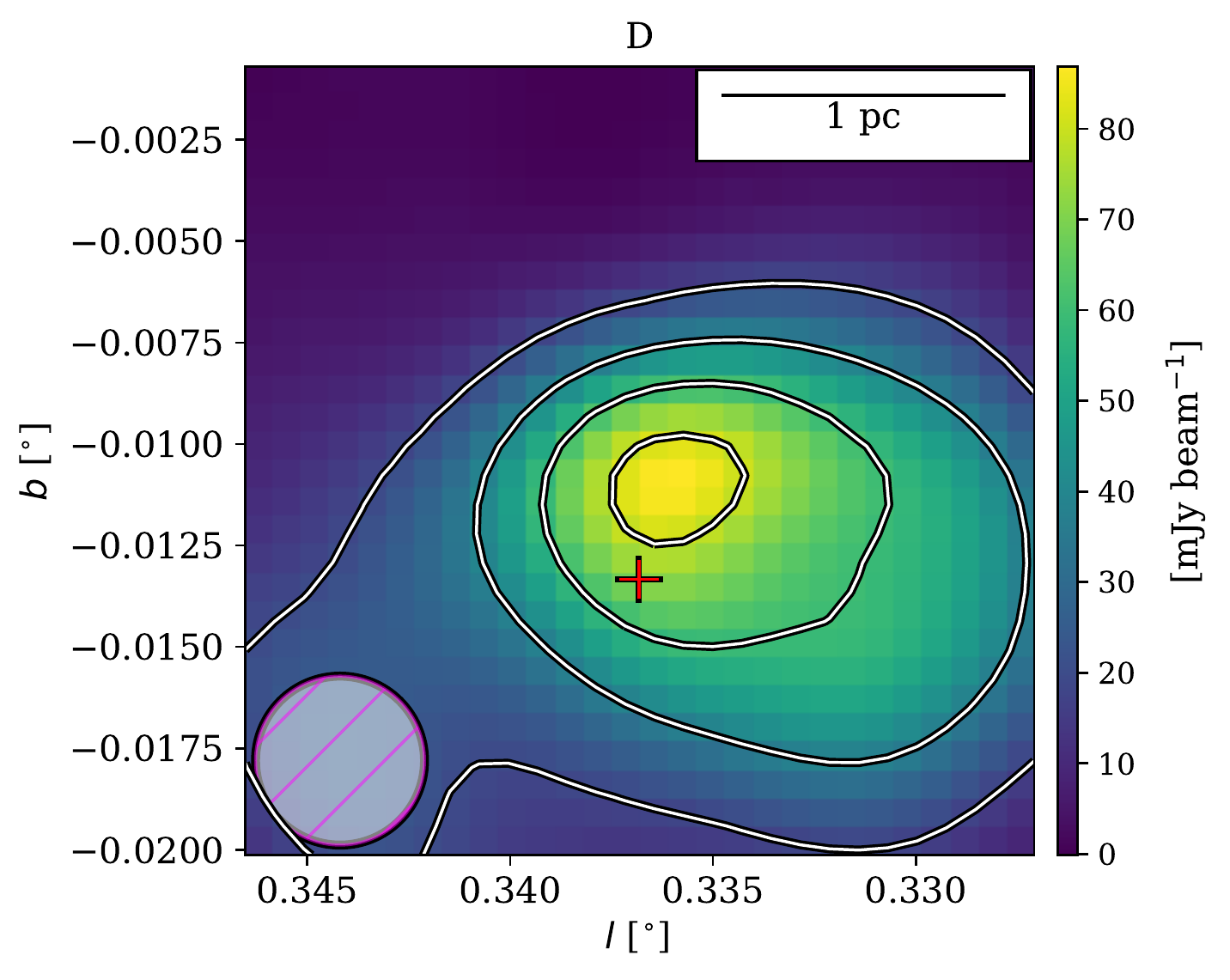}
\includegraphics[width=0.33\textwidth]{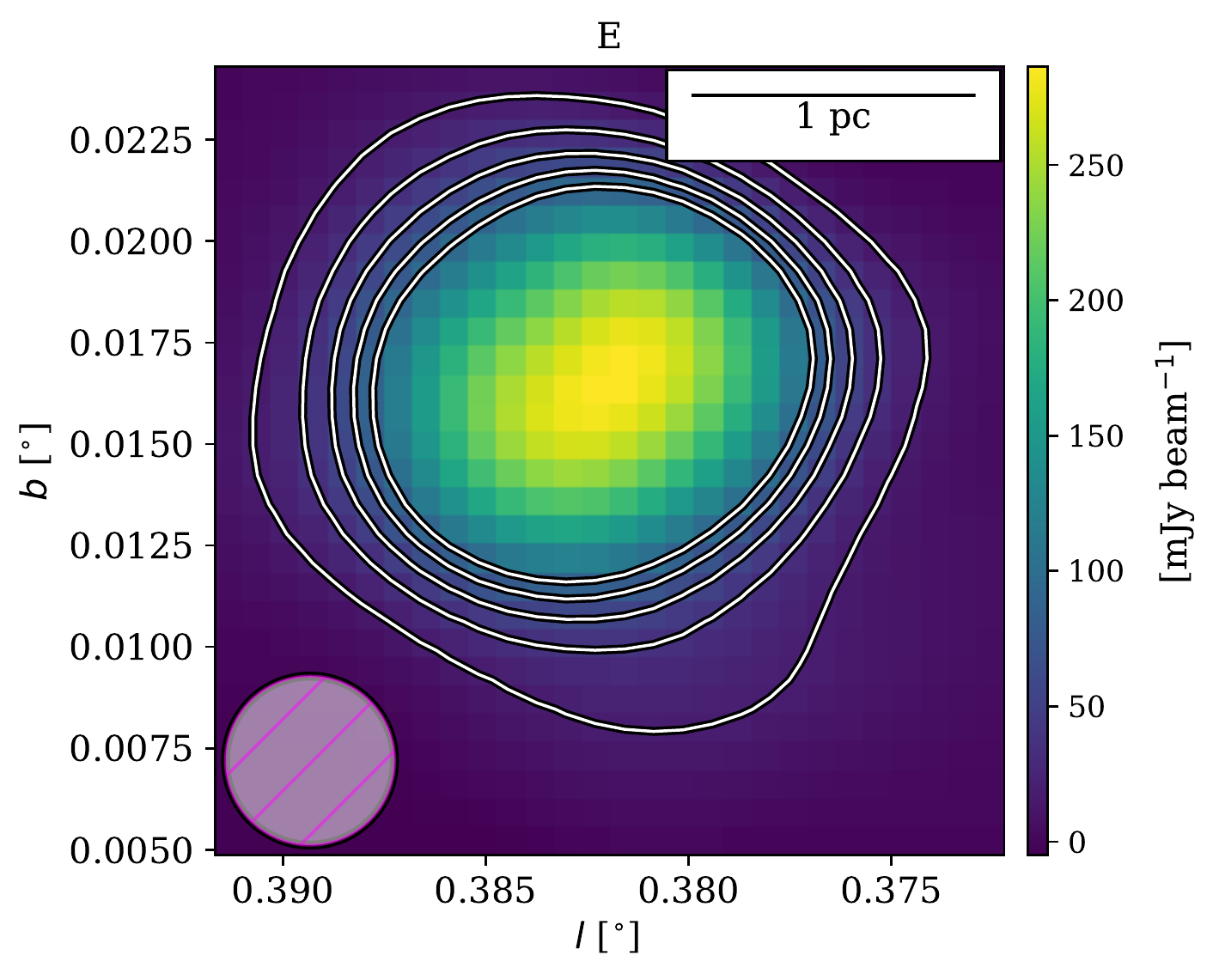}\\
\end{tabular}
\caption{Maps of the dust ridge presented similar to Fig.~5 in \cite{immer2012}. \textit{Top}: ATLASGAL 870\,$\mu$m dust emission towards the dust ridge with GLOSTAR 5.8\,GHz radio continuum contours overlaid with contour levels of 2\%, 8\%, and 13\% of the maximum. \textit{Bottom}: GLOSTAR VLA cutouts of sources A-E cut to a larger angular size compared to \cite{immer2012}. For sources A, D, and E, the contours are from 10$\sigma$ to 50$\sigma$ in steps of 10$\sigma$. For sources B and C, the contour levels are 4$\sigma$,
12$\sigma$, and 20$\sigma$. Red crosses denote the positions of YSOs.}
\label{fig:immer_cf}
\end{figure*}

\subsubsection{The dust ridge}
\label{sect:immer_cf}
A known feature in the CMZ, the so-called dust ridge is a narrow string of massive clumps that connect the radio continuum sources
G0.18--0.04 and Sgr~B1 \citep{dustridge}. \cite{immer2012} detected five radio continuum sources in X-band (labelled A-E) on the periphery of the dust ridge, likely hosting HMSF. This region is also covered in our GLOSTAR-VLA data in C-band. We make a comparison of this region between the X- and C-band studies and display the region and cutouts in
Fig.~\ref{fig:immer_cf} with the same angular size as in \cite{immer2012}. 
The \cite{immer2012} observations were centred at 8.4\,GHz using the VLA in CnB-hybrid configuration and have a restored elliptical Gaussian beam with FWHM of roughly 3\fasec6\,$\times$\,2\fasec5. Compared to their VLA data, our D-configuration VLA data are much coarser
and trace the more extended features that are likely resolved out in the \cite{immer2012} observations. We calculated effective radii and integrated flux densities
and list them in Table~\ref{tab:immer_flux} using CASA's \texttt{imfit} task which does Gaussian fitting of the sources. This was done to compare them with the Gaussian fitted results from \cite{immer2012}. For sources A, B, and C, the source size is larger. Source D shows the most striking difference
in terms of the flux and radius, which can be attributed to the fact that in our VLA D-configuration
data, we have a much lower angular resolution and trace more extended features while the source has been resolved into multiple components in the work done by \cite{immer2012}. It shows significant extended emission in a morphology that is offset from the position of the compact source (potentially cometary). Source E is of a comparable size in both this study and in \cite{immer2012}. Upcoming comparison with GLOSTAR-VLA B\,configuration data can further our investigation of these sources.

Furthermore, to put our GLOSTAR-VLA data into perspective, we are interested in what the \hii~region associated with the ONC would look like at radio wavelengths if placed in the CMZ. \cite{immer2012}, who present radio observations at similar frequencies but with higher resolution, show that the radio
size of the Orion Nebula would only be slightly smaller than one of their radio sources (their Source~E; see Fig.~\ref{fig:immer_cf}, lower-right panel). They also show that scaling the integrated flux density of the ONC to the distance of the CMZ would result in a value of 0.98~Jy, which is comparable to that of their Source~E (see our Table~\ref{tab:immer_flux}). 
Inspecting our list of YSOs with radio counterparts (Table \ref{tab:res}), we find that YSO 234, with O7.5--O8, has a spectral type that is closest, if slightly later, than that of $\theta_1$~C~Ori (O7V), which is the star that provides most of the UV photons that excite the Orion Nebula. The radio flux we determine for this source, 0.44\,Jy, is comparable and slightly lower than the 0.98\,Jy quoted above as one might
expect. As such, it is clear that we are able to detect the radio emission from Orion Nebula-like sources within the GLOSTAR-VLA data.

\begin{table}
\caption{Comparison of the GLOSTAR 5.8\,GHz emission features from the dust ridge with literature, specifically \cite{immer2012}. Integrated flux densities as well as their angular radii, $\Theta_{R}$, are shown.}
\label{tab:immer_flux}
\centering

\begin{threeparttable}

\begin{tabular}{r r r r | r r}  
\hline\hline             
Source& $S_{i, 8.4\,\text{GHz}}$\,$^{a}$ & $S_{i, 5\,\text{GHz}}$\,$^{b}$ & $\Theta_{R}$\,$^{a}$ & $S_{i, 5.8\,\text{GHz}}$ &$\Theta_{R}$\\
 & [mJy] & [mJy] &[\asec] & [mJy] & [\asec] \\
\hline
A & 180$\pm$2 & 154 & 5.1 & 120$\pm$10 & 15\\ 
B & 9$\pm$1 & <9 & 1.8 & 16$\pm$3 & 12\\
C & 10$\pm$1 & <9 & 1.8 & 15$\pm$3 & 15\\
D & 145$\pm$20 & 134& 14.1$^{b}$& 436$\pm$42 & 30\\
E & 886$\pm$57 & 1417& 32.0& 1040$\pm$10 & 33\\
\hline
\end{tabular}

\begin{tablenotes}
\item[]Notes. $^{a}$from \cite{immer2012},  $^{b}$ from \cite{becker1994}
\end{tablenotes}

\end{threeparttable}

\end{table}

\subsubsection{The Arches cluster}\label{sect:arches}

\begin{figure}[!h]
    \centering
    \includegraphics[width=0.5\textwidth]{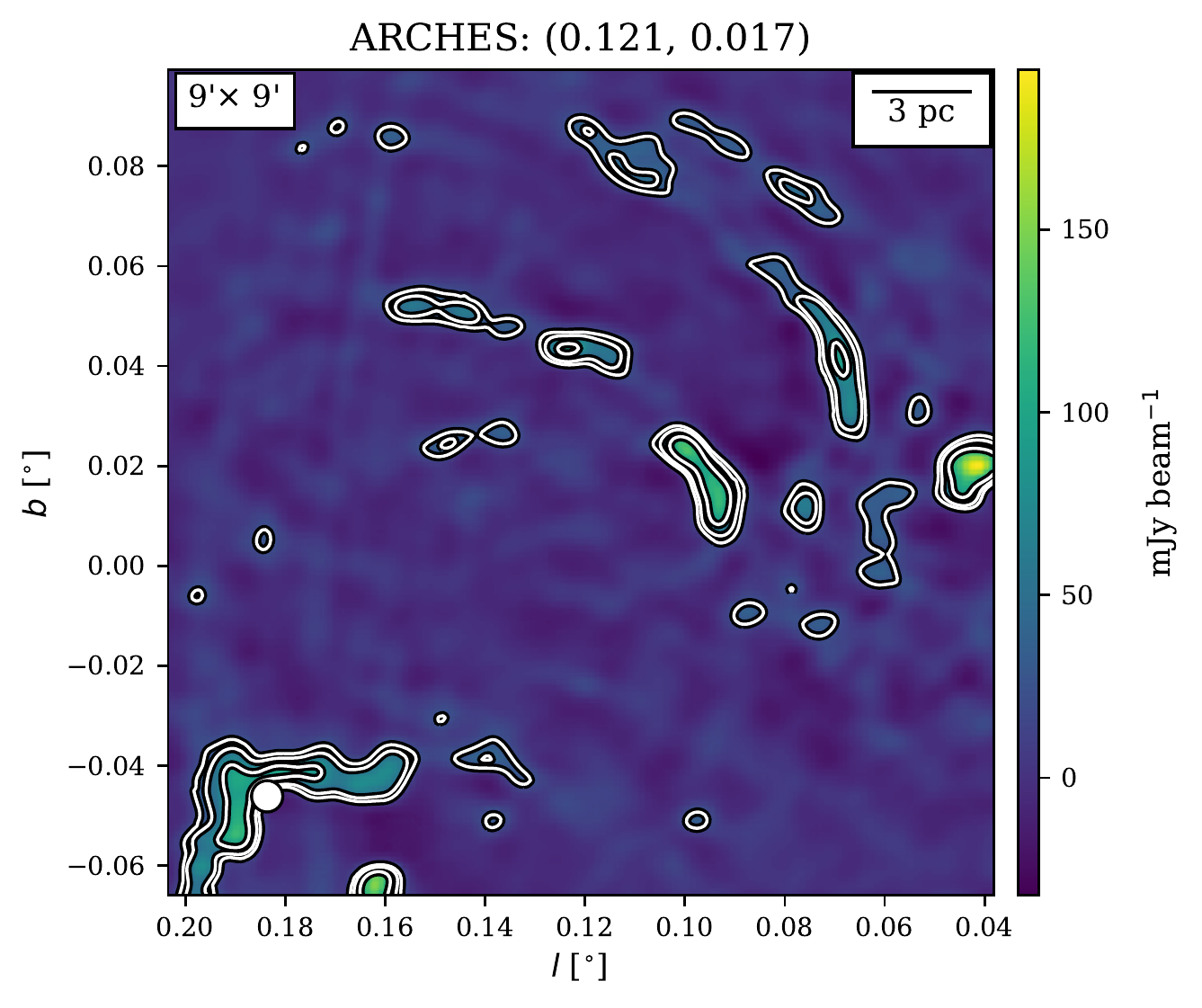}

    \caption{ GLOSTAR cutout of a 9\amin$\times$\,9\amin area centred on the Arches cluster. Plotted in white contours are 3, 5, and 10 times the local rms (8\,mJy~beam$^{-1}$).}
    \label{fig:arches}
\end{figure}

The Arches cluster, otherwise known as G0.121+0.017, is a massive ($7\times10^4$ \msol) cluster of massive young (1--2\,Myr) stars situated near the Galactic Centre that was identified by NIR imaging
(e.g. \citealt{cotera1996}, \citealt{figer2002}).
It has also been extensively studied at radio wavelengths (e.g. \citealt{Lang_2001}, \citealt{yusef2002a}, \citealt{arches_multiFreq}, \citealt{gallego2021} and references therein) in which it shows clear filamentary `arches' surrounding it. These are thought to be ionised by hot stars in the star cluster. High-resolution VLA observations \citep{arches_multiFreq} reveal ten radio sources on 
scales of <0.5\asec that are believed to be due to stellar winds.
With the D-configuration VLA data from our GLOSTAR-VLA survey, we do not currently see any convincing detection of 
the Arches cluster itself, but we do see the filamentary namesake arcs nearby as shown in Fig.~\ref{fig:arches}. The NIR sources of the cluster are spread over $\lesssim30$\asec,
which is close to our beam size of $\sim$18\asec. 
It is likely that the extended emission from these nearby \hii~regions or filaments confuse the point-like emission of the stellar cluster. Filtering out the extended emission during the imaging process may result in a higher sensitivity to compact radio sources at the location of the Arches cluster, but it is left for future analysis.

\subsubsection{The Brick}\label{sect:theBrick}
\begin{figure}[!h]
    \centering
    \includegraphics[width=0.5\textwidth]{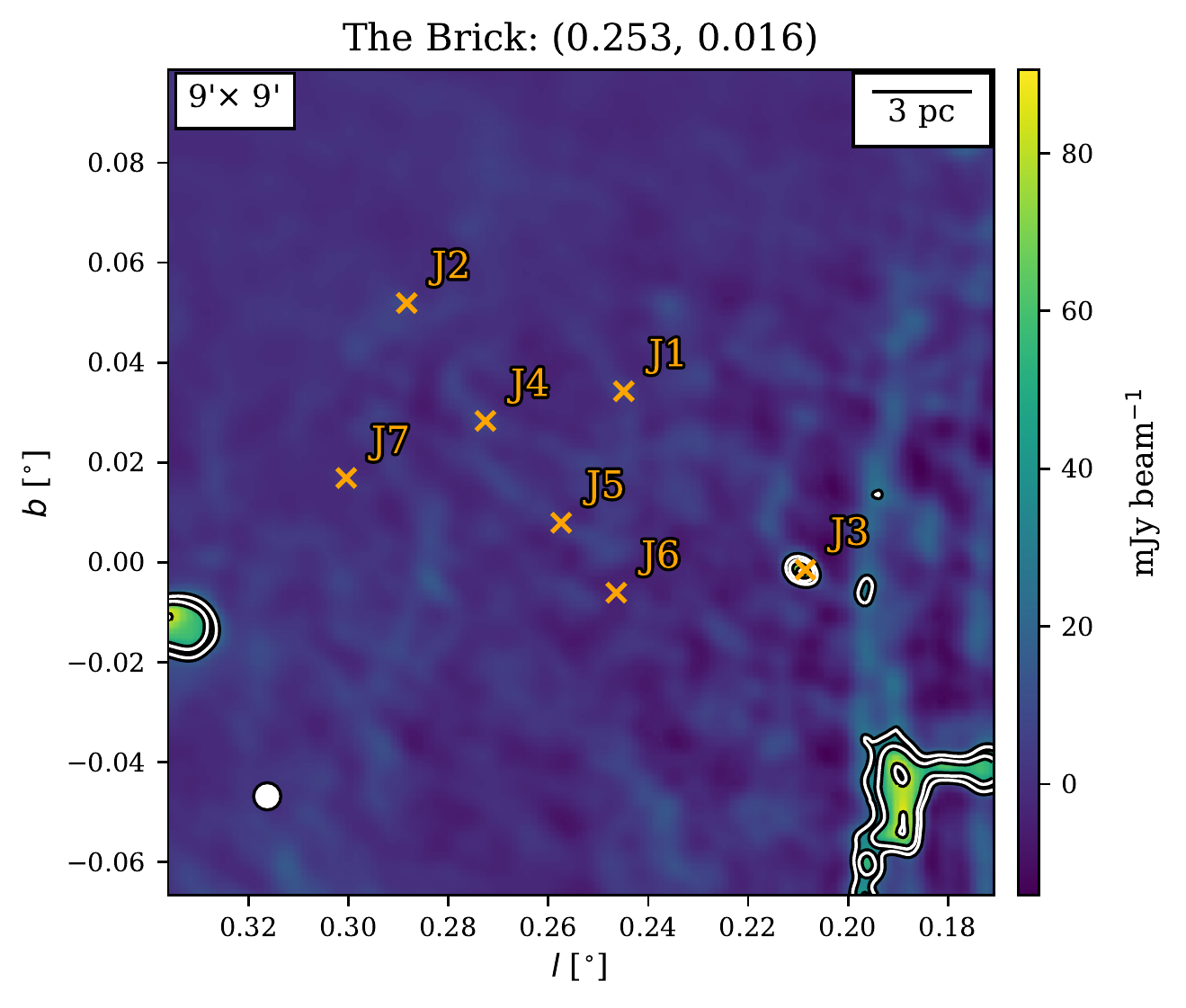}

    \caption{GLOSTAR cutout of a 9\amin$\times$\,9\amin area encompassing the region known as `the Brick'. Plotted in white contours are 3, 5, and 10 times the local rms (8\,mJy~beam$^{-1}$). Labelled in orange are the positions and names of the seven radio sources found by \cite{brickJVLA} using the VLA B- and C-configuration. }
    \label{fig:the_brick}
\end{figure}

G0.253+0.016, otherwise known as `the Brick', is considered to be the prototypical infrared dark cloud \citep{LisMenten1998}. It is one of
the densest and most massive molecular clouds within the Galaxy and the only one above $10^{5}$\,\msol that
does not show significant star formation (e.g. \citealt{Henshaw_2019} and references therein).
We only find a radio continuum counterpart for one of the seven radio
sources detected by \cite{brickJVLA} at 5.307\,GHz and 20.943\,GHz using the VLA in B- and C-configurations,
respectively, where both have an approximate beamsize of $\sim$1\asec. The sources were determined to be \hii~regions or non-thermal sources of an unknown
nature. These sources are displayed in Fig.~\ref{fig:the_brick} using their names from \cite{brickJVLA}. Except for source J3, which is a known \hii~region also detected in other surveys (e.g. \citealt{Anderson2014}, \citealt{becker1994}), the rest are not easily identifiable in our data. While it does not seem to be a result of sidelobe
noise, it is unclear how to determine if these sources are detected above
the local noise. However, with our own B-configuration data, we may be able to resolve these sources in future works.


\subsection{Star formation in the CMZ}\label{sect:SFR_CMZ}

\subsubsection{Mass estimation}\label{sect:mass_calc}

Star formation activity is integral in the evolution, both chemically and structurally,
of galaxies, and by extension, the large-scale structures of the universe. The rate at which the ISM
is converted into stars (SFR) is thus an important quantity in studying star formation.
Here we estimate a lower limit of the current SFR in the CMZ, given that the \hii~regions that we
characterise are associated with YSOs that trace the early stages of star formation.
In order to calculate the SFR in the CMZ from these
\hii~regions, we need to estimate the masses of the individual zero-age-main-sequence (ZAMS) 
stars that are ionising the \hii~regions.
Knowing the flux and the size of these radio sources, we can already calculate further properties as outlined in \cite{immer2012}
where our main interest is the number of Lyman continuum photons, $N_{\text{Lyc}}$, 
associated with each source as it can be used to determine the spectral
type of a new-born star if we assume each \hii~region has only one star. This approximation
generally holds as the most massive star dominates the contribution of Lyman continuum photons.
We relate $N_{\text{Lyc}}$ with our observables as follows:
\begin{equation}
    \left[\frac{N_{\text{Lyc}}}{\text{photons\,s}^{-1}}\right] = 2.35\times10^{35} \left[\frac{S}{\text{Jy}}\right]
    \left[\frac{T}{10^{4}\,\text{K}}\right]^{2}
    \left[\frac{Dist}{\text{kpc}}\right]^{2}
    b(\nu,T)^{5},
\end{equation}

\noindent with

\begin{equation}
    b(\nu,T) =1+0.3195\log\left(\frac{T}{10^{4}\,\text{K}}\right)
    -0.213\log\left(\frac{\nu}{1\,\text{GHz}} \right),
\end{equation}

\noindent where $S$ is the flux density, $T$ is the electron temperature, $Dist$ is the distance to the 
source, $\nu$ is the frequency of the observation, and $b(\nu,T)$ is taken from \cite{panagia1978}. To obtain this relation, we followed \cite{tielens2005} where they have related $N_{\text{Lyc}}$ to the emission measure, $EM$, of an \hii~region assuming that it is an idealised ionised source with spherical geometry, assuming a constant electron density, $n_e$, of the \hii~region:

\begin{equation}
    EM=4.3\times10^{-11}
    \left[\frac{n_e}{10^{3}\,\text{cm}^{-3}}\right]^{\frac{4}{3}}
    \left[\frac{N_{\text{Lyc}}}{\text{photons\,s}^{-1}}\right]^{\frac{1}{3}}
    \text{cm}^{-6}\,\text{pc},
\end{equation}

\noindent and solved for $N_{\text{Lyc}}$ by using expressions for $EM$ and $n_e$ from \cite{panagia1978}:

\begin{equation}
    EM=5.638\times10^4
    \left[\frac{S}{\text{Jy}}\right]
    \left[\frac{T}{10^{4}\,\text{K}}\right]
    b(\nu,T)\theta_{R}^{-2}\text{cm}^{-6}\,\text{pc},
\end{equation}

\begin{equation}
    n_{e} = 311.3\times\left[\frac{S}{\text{Jy}}\right]^{0.5}
    \left[\frac{T}{10^{4}\,\text{K}}\right]^{0.25}
    \left[\frac{Dist}{\text{kpc}}\right]^{-0.5}
    b(\nu,T)^{-0.5}\theta_{R}^{-1.5}\text{cm}^{-3}.
\end{equation}

While $S$ and $\theta_{R}$ (the angular radius of the source in arcminutes) were derived from the data, we assumed a temperature
of $10^{4}$\,K and a distance of 8.2\,kpc for all sources as they
reside in the same general area of the CMZ.

Finally, following again \cite{tielens2005}, we can estimate the mass, $M_{\text{\hii}}$, of the \hii~region within which the ionising star resides as

\begin{equation}
    M_{\text{\hii}}\approx1.6\times10^{-48}
    \left[\frac{n_{e}}{10^{3}\,\text{cm}^{-3}}\right]^{-1}
    \left[\frac{N_{\text{Lyc}}}{\text{photons\,s}^{-1}}\right]
    M_{\odot}.
\end{equation}

\begin{figure}
    \centering
    \includegraphics[width=0.5\textwidth]{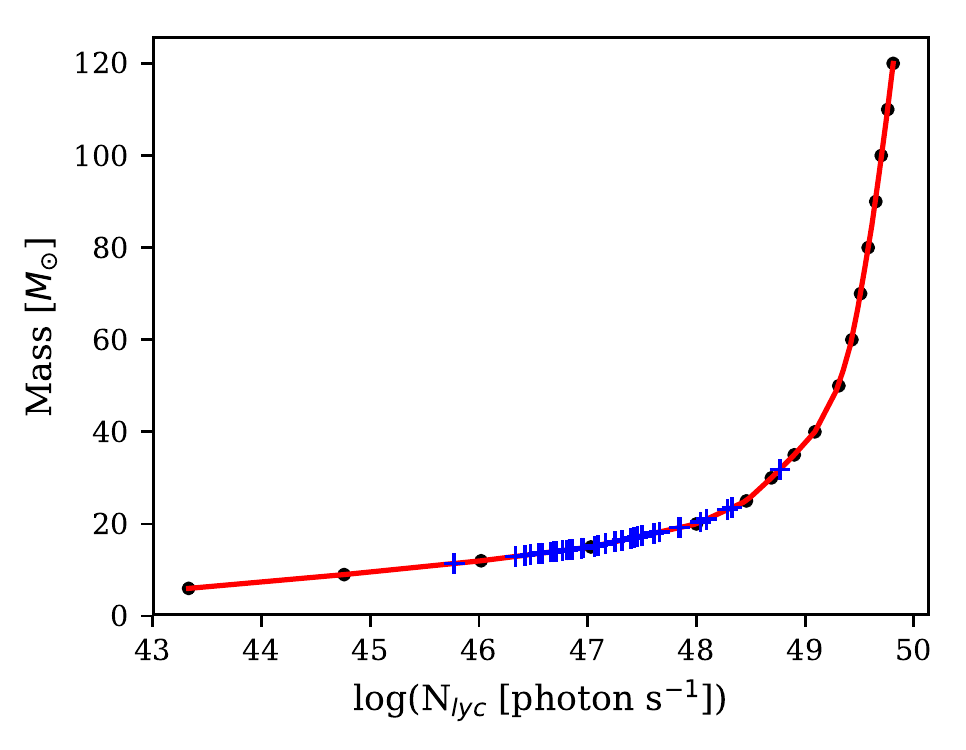}
    \caption{ZAMS masses plotted against the Lyman continuum photon flux. The red line is interpolated from the data in table~1 from \cite{davies2011}. The derived masses for the N$_{Lyc}$ values determined in this work are highlighted as blue crosses.
    }
    \label{fig:interpolate}
\end{figure}

In order to determine the SFR, we determined the mass of the ZAMS that are ionising the \hii~regions by interpolating the stellar masses given as a function of the Lyman continuum flux in \cite{davies2011}, as shown in Fig.~\ref{fig:interpolate}.
The derived masses of the stars range from 12 to 49\,\msol with a mean and median of 16.5
and 15\,\msol, respectively. This corresponds to a spectral type range of B1 to O6 with a mean of B0-B0.5. Derived stellar properties are summarised in Table~\ref{tab:res}.

\begin{figure}
    \centering
    \includegraphics[width=0.5\textwidth]{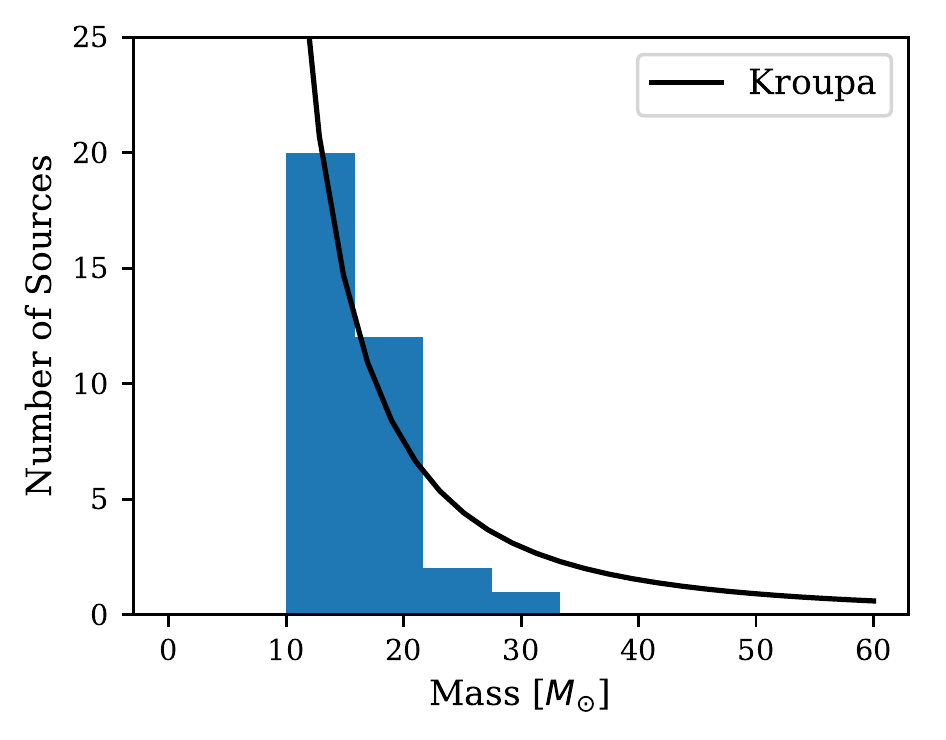}
    \caption{Mass distribution of calculated ZAMS masses for the \hii~regions with associated YSOs. The black line represents the fitted Kroupa IMF \citep{kroupa2001}.}
    \label{fig:imf}
\end{figure}

We show the distribution of the masses of these stars in Fig.~\ref{fig:imf} ,
where masses $M_{*}<10$\,\msol and $M_{*}>40$\,\msol are not represented by our sample.
The majority of the luminosity in a given star cluster comes from the massive stars, the majority of the mass \TBF{of the star cluster}, however, is distributed among the low mass stars. Our distribution clearly shows that we do not cover low mass stars and therefore need a way to infer the total mass of stars by interpolating the distribution of low mass stars. From this distribution of stellar masses, one can calculate the 
SFR. To do this we used an initial mass function (IMF; \citealt{salpeter1955}, \citealt{kroupa2001}) to obtain an estimate of the total mass.
As our sample clearly does not represent all masses,
especially the lower masses, we used the IMF ($\xi(M)$) from \cite{kroupa2001} that better estimates 
the contribution of lower mass stars to the total stellar population and is given by

\begin{align}
    \label{eqn:kroupaA}
    \xi(M)&=\xi_{0,1}M^{-2.3}\,\text{for\,}\,0.5\,M_{\odot}\leq M\leq 120\,M_{\odot}\\
    \xi(M)&=\xi_{0,2}M^{-1.3}\,\text{for\,}\,0.08\,M_{\odot}\leq M\leq 0.5\,M_{\odot}\\
    \xi(M)&=\xi_{0,3}M^{-0.3}\,\text{for\,}\,0.01\,M_{\odot}\leq M\leq 0.08\,M_{\odot}
\end{align}

\noindent where $\xi_{0,1}$, $\xi_{0,2}$, and $\xi_{0,3}$ are scaling factors. 
Following the method detailed in \cite{immerCMZ}, 
we determined the scaling factor $\xi_{0,1}$ by a non-linear least square fit to our data over the mass range 
$10\msol<M_{*}<40\msol$ (Fig.~\ref{fig:imf}).   By requiring $\xi(M)$ to be continuous, we scaled 
$\xi_{0,2}$ and $\xi_{0,3}$ accordingly.  For the scaling factors, we obtained $\xi_{0,1}=6152$,
 $\xi_{0,2}=12304$, and $\xi_{0,3}=153807$.

The total mass of the stars was then calculated using 

\begin{equation}
    M_{tot} = \int_{0.01}^{120}M\xi(M)dM,
\end{equation}

\noindent where $\xi(M)dM$ is the number of stars in the mass range of $M$ and $M+dM$ in units of \msol. Therefore we estimate the mass of 
all ZAMSs in the CMZ in the range of $0.01-120$\,\msol to be $\sim$\,$30\,000$\,\msol.
While this is the standard approach, we recognise that our sample is limited. Of all possible YSOs, we only selected those that have an associated \hii~region and, therefore, only the stars that are radio bright. This, however, is not representative of the total number of stars in the CMZ. Thus, we calculated the recent SFR in two methods to account for this bias.

\subsubsection{Average SFR}\label{sect:avg}
First, we derived the SFR based on the estimated total mass of young stars.
In order to calculate the SFR, we needed to determine the time over which the YSOs were formed. We estimated this using the average age of a YSO.
YSOs need to be embedded in a surrounding envelope of dust in order to
be visible in the mid-infrared \citep{woodChurchwell1989}. This phase is only 
$\sim$10$\%$ of the full lifetime for a typical O or early B star and for an average B0 type star, this is $\sim$1\,Myr \citep{woodChurchwell1989}. 
As such, YSO candidates that are observed are at most $\sim$1\,Myr in age. 
In following the more conservative estimation of the YSO timescale from \cite{nandakumar2018} of $0.75\pm0.25$\,Myr, we calculated the average ongoing SFR as $\dot{M}_{\rm{SF}}=M_{tot}/\tau_{\rm{YSO}}$, where $\tau_{\rm{YSO}}$ is the considered timescale and $M_{tot}$ is calculated in Section~\ref{sect:mass_calc}. We obtain a 
SFR of $0.04\pm0.02$\,\msol\,yr$^{-1}$, which is consistent with the 
results obtained with the YSO counting method by \cite{nandakumar2018}, which is interesting as we have a much smaller sample size. If we instead use the Salpeter IMF, 
we find a SFR of the order of $\sim$0.1\,\msol\,yr$^{-1}$. Given that we do not have a
representative sample that covers low-mass stars, the Salpeter IMF especially may overestimate the SFR from our small sample and, in our case, the SFR is double what was estimated by a Kroupa IMF.

Secondly, we considered the total ionising flux from the \hii~regions and followed the statistical approach from \cite{kauffmann2017a} that relates the SFR to the number of \hii~regions. In their work, they also adopted a \cite{kroupa2001} IMF, where the power law covering the largest masses is $\alpha=2.7$. The distribution has a mean stellar mass of $\langle m_{*}\rangle = 0.29\,\msol$. They derived a relationship between the number of cluster members of an \hii~region, which includes masses of 0.01\,\msol and greater, to the mass of the largest member as,

\begin{equation}
    N_{\rm{cl}} = 20.5\times(M_{\rm{max}}/\msol)^{1.7}.
\end{equation}

The total mass in the given
\hii~region is then $\langle m_{*}\rangle \times N_{\rm{cl}}$. To calculate the SFR
contribution from a given \hii~region, we again need to consider an appropriate timescale over which this
mass is produced. Following \cite{kauffmann2017a}, we consider a timescale, $\tau_{\hii}$, of 1.1\,Myr.
This value is estimated based on the comparison of the ratio between the number of \hii~regions and the statistical estimate of the  number of high-mass stars (here, the radio bright YSOs) and the ratio of their respective timescales. A more detailed discussion can be found in their appendix. Using this timescale, we calculated the SFR as $\dot{M}_{\rm{SF}}=\langle m_{*}\rangle \times N_{\rm{cl}}/\tau_{\hii}$. Using our sample of \hii~regions,
we used our calculated masses to find the number of cluster members and then estimate the individual SFR. The sum of these rates provides a total
ongoing SFR of 0.023\,\msol\,yr$^{-1}$. Knowing that the sample does not
cover all possible \hii~regions within the CMZ, we used ancillary values compiled in Table~7 of \cite{kauffmann2017a} that considers the SFR in all the major GMCs in the CMZ. We made sure to exclude two sources that overlap with the Sgr~B2 GMC, and one source with the `20\,km\,s$^{-1}$ cloud', and in these cases, we solely used the literature value. In this way, we provide an update on the SFR from \cite{kauffmann2017a} with a final estimate of $\sim 0.068$\,\msol\,yr$^{-1}$.

Other estimates using YSO counting
(\citealt{yusef2009}, \citealt{an2011}, \citealt{immer2012}) 
estimate SFRs of at least a factor of two or more. 
A reference of calculated SFRs using different methods is summarised in Table~4 of
\cite{nandakumar2018}. The numerous different methods for calculating the observed ongoing
SFR is consistent in that they all point to a lower SFR than expected with respect to the available dense molecular gas present (e.g. 0.78\,\msol\,yr$^{-1}$; column density threshold, 0.41\,\msol\,yr$^{-1}$; volumetric star formation relations \citealt{longmore2013a} also show lower SFR). While it is not yet known 
what the definitive reason for this deficiency of star
formation is, \cite{longmore2013a} suggest turbulence 
as a possible counteracting component to gravitational collapse, 
which is supported from observations of the large velocity dispersions
found in the clouds in the CMZ (\citealt{bally1987}, \citealt{Christopher2005}, \citealt{rahul2012}, \citealt{Kauffmann2013}, \citealt{Mills2015}, \citealt{Rathborne2015}). 
\cite{kruijssen2014} discuss additional possible mechanisms that work on different size scales. On larger scales, episodic star formation from the accumulation of dense gas from spiral instabilities and the gas not being self-gravitating may explain the observed SFR and, on smaller scales, high turbulence likely drives up the volume density threshold needed to form stars. Alternatively, simulations by \cite{sormani2020} suggest that the SFR might indeed be variable and that such variability is a reflection of changes in the mass of the CMZ instead of changes in the star formation efficiency.


\clearpage
\begin{table*}[!tbhp]
\caption{Physical parameters of the detected radio continuum sources that have candidate
YSO associations (see Table~\ref{tab:sources1}).}             
\label{tab:res}      
\centering          
\begin{threeparttable}
\begin{tabular}{r r r r r r r r r r}     
\hline\hline
Source & $S_{5.8\,\text{GHz}}$  &$D_{\text{eff}}$ & $n_{e}$ & \textit{EM} & $N_{\text{Lyc}}$& $M_{\text{\hii}}$& Spectral type & $M_{*}$ \\
& [mJy]& [pc]& [cm$^{-3}$]& [$10^{5}$cm$^{-6}$ pc]& [log(photon s$^{-1}$)]& [M$_{\odot}$] & (single ZAMS star$^{a}$) & [M$_{\odot}$]\\ 
\hline
3 &2.8  &1.0 &62 &0.029 &46.3 &0.50 &B0-B0.5 & 12.8\\ 
5 &26.9  &1.4 &127 &0.158 &47.3 &2.35 &B0-B0.5 & 16.2\\ 
34 &69.6  &2.0 &115 &0.192 &47.7 &6.66 &O9.5-B0 & 18.4\\ 
44 &10.7  &0.9 &142 &0.136 &46.9 &0.83 &B0-B0.5 & 14.5\\ 
51 &2.1 &0.8 &71 &0.032 &46.2 &0.33 &B0.5-B1& 12.4\\ 
54 &10.1  &1.1&104 &0.088 &46.8 &1.07 &B0-B0.5 & 14.5\\ 
64 &7.7  &0.8 &140 &0.119 &46.7 &0.61 &B0-B0.5& 14.1\\ 
66 &4.5  &1.1 &73 &0.042 &46.5 &0.67 &B0-B0.5 & 13.4\\ 
78 &5.4  &0.9 &106 &0.074 &46.6 &0.56 &B0-B0.5& 13.7\\ 
80 &13.4  &1.1 &115 &0.110 &47.0 &1.29 &B0-B0.5 & 14.8\\ 
82 &6.8  &1.1 &87 &0.061 &46.7 &0.86 &B0-B0.5 & 13.9\\ 
87 &190.1  &2.6 &126 &0.302 &48.1 &16.66 &O8.5-O9 & 21.3\\ 
89 &16.8  &1.3 &108 &0.110 &47.1 &1.71 &B0-B0.5& 15.2\\ 
91 &5.2  &1.1 &76 &0.047 &46.6 &0.76 &B0-B0.5 & 13.6\\ 
93 &5.8  &1.2 &69 &0.042 &46.6 &0.94 &B0-B0.5& 13.7\\ 
115 &13.9  &0.9 &170 &0.189 &47.0 &0.90 &B0-B0.5& 14.9\\ 
135 &9.7  &0.9 &134 &0.122 &46.8 &0.80 &B0-B0.5& 14.4\\ 
140 &60.5  &1.4 &185 &0.345 &47.6 &3.61 &O9.5-B0 & 18.1\\
157 &120.9 &1.2 &326 &0.922 &47.9 &4.10 &O9-O9.5 & 19.6\\
230 &15.9 &1.2 &115 &0.117 &47.0 &1.53 &B0-B0.5& 15.1\\ 
234 &436.9  &2.6 &195 &0.712 &48.5 &24.79 &O7.5-O8& 25.5\\ 
235 &73.6  &1.1 &297 &0.691 &47.7 &2.74 &O9.5-B0 & 18.5\\ 
262 &33.8  &1.2 &162 &0.237 &47.4 &2.31 &O9.5-B0& 16.8\\ 
277 &54.6  &1.2 &222 &0.424 &47.6 &2.72 &O9.5-B0& 17.8\\ 
284 &43.3  &1.4 &147 &0.227 &47.5 &3.25 &O9.5-B0& 17.3\\ 
296 &40.8  &1.9 &92 &0.120 &47.5 &4.87 &O9.5-B0 & 17.2\\ 
299 &240.2  &1.9 &236 &0.754 &48.2 &11.24 &O8-O8.5 & 22.4\\ 
307 &87.0  &1.5 &186 &0.391 &47.8 &5.17 &O9.5-B0 & 18.9\\ 
311 &1010.2  &1.2 &951 &7.775 &48.8 &11.77 &O6-O6.5 & 33.7\\ 
315 &7.3  &0.9 &117 &0.092 &46.7 &0.69 &B0-B0.5& 14.0\\ 
323 &37.8  &1.3 &153 &0.228 &47.4 &2.74 &O9.5-B0 & 17.0\\

\hline      
\end{tabular}

\begin{tablenotes}
\item[] \textbf{Notes.} {\it From left to right:} Source ID, 5.8\,GHz integrated flux ($S_{5.8\,\text{GHz}}$), effective diameter ($D_{\text{eff}}$), electron number density ($n_{e}$), emission measure (\textit{EM}), Lyman continuum photon flux ($N_{\text{Lyc}}$), \hii~region mass ($M_{\text{\hii}}$), spectral type, and interpolated stellar mass from values ($M_{*}$) given by \cite{davies2011}.
We calculated these values using the same caveat as in \cite{immer2012}, where we assume radio sources are spherically symmetric \hii~regions. In our study, however, we adopted this assumption for extended sources as well, which have been marked in Table~\ref{tab:sources1}. \textbf{Footnotes.} $^{(a)}$ Obtained from \cite{panagia1973}. 
\end{tablenotes}

\end{threeparttable}

\end{table*}


\section{Summary and conclusions}\label{sect:summary}

To investigate HMSF in the CMZ, one can use YSOs, that is to say tracers of on-going star formation
to characterise the SFR. YSOs are observed indirectly from the re-emission of
their energy from their surrounding natal dust cloud in the infrared. If these infrared sources are indeed sites of current star formation, we expect them to be currently associated with \hii~regions or for them to be in the future, which can be seen in the radio.
We used a set of 334 YSOs that \cite{nandakumar2018} selected using a new colour-colour-diagram
selection criterion as targets to look for radio sources to see how many of these YSO candidates have already formed an \hii~region. Using the GLOSTAR 5.8\,GHz radio continuum data, we searched for YSO association
candidates and obtain a final sample of 35 YSO sources that
have a potential radio continuum counterpart. We also compared the YSO sample with ATLASGAL and find 94 coincident associations, with 14 having a GLOSTAR counterpart. For those without dust emission and radio emission, \TBF{the lack of emission at these wavelengths suggests that they are potentially much older or are perhaps not high-mass YSOs. A cross-match of the 334 YSOs with the CMZoom survey showed 22 potential counterparts, and of these 22, two have radio counterparts in our data.}

We used these 35 radio sources to estimate the SFR by first characterising their properties. We calculated their flux, size, shape, and their spectral indices. 
\TBF{We also compared these radio sources to the WISE catalogue and found that there are six without WISE counterparts, where five of them are potential new \hii~regions.} We determined the Lyman continuum
photon flux of the ionising ZAMS and determined its mass. For our sub-sample, we found
masses between 10\,\msol$<M_{*}<40$\,\msol. We calculated their contribution to the SFR in the CMZ to be
$0.04\pm0.02$\,\msol\,yr$^{-1}$, which is consistent with the results from \cite{nandakumar2018} and 
other independent investigations that used different methods. However, we note the limitations in our approach of using, in essence, only the radio bright sources in our sample. We therefore adapted the formulation of the SFR from \cite{kauffmann2017a} and used the total ionising flux of \hii~regions to also estimate the SFR in the CMZ to be 0.068\,\msol\,yr$^{-1}$.

\begin{acknowledgements}
    We would like to thank the anonymous referee for their useful comments.
    We would like to thank Denise Riquelme for a careful reading of the manuscript.
    H.N. is a member of the International Max-Planck Research School at the universities of Bonn and Cologne (IMPRS).
    
    This research was partially funded by the ERC Advanced Investigator Grant GLOSTAR (247078).
    
    Contributions from J.K. are in part supported by the National Science Foundation under Grant Number AST–1909097.

    H.B. acknowledges  support from the European Research Council under the European Community's Horizon 2020 framework program (2014-2020) via the ERC Consolidator grant ‘From Cloud to Star Formation (CSF)' (project number 648505). H.B. further acknowledges support from the Deutsche Forschungsgemeinschaft (DFG) via Sonderforschungsbereich (SFB) 881 “The Milky Way System” (sub-project B1).
    
    The National Radio Astronomy Observatory is a facility of the National Science Foundation, operated under a cooperative agreement by Associated Universities, Inc.
    
    It made use of information from the ATLASGAL database at \url{http://atlasgal.mpifr-bonn.mpg.de/cgi-bin/ATLASGAL_DATABASE.cgi} supported by the MPIfR in Bonn.
    
    This publication also makes use of data products from the Wide-field Infrared Survey Explorer (WISE) which is a joint project of the University of California, Los Angeles, and the Jet Propulsion Laboratory/California Institute of Technology, funded by the National Aeronautics and Space Administration. 
    
    This research made use of Astropy,\footnote{http://www.astropy.org} a community-developed core Python package for Astronomy \citep{Astropy-CollaborationRobitaille:2013ab,Astropy-CollaborationPrice-Whelan:2018aa}.
    
    This research has made use of the SIMBAD database, operated at CDS, Strasbourg, France.

\end{acknowledgements}

\bibliographystyle{aa}
\bibliography{cmzbib}


\begin{appendix}


\section{GLOSTAR cutouts}\label{app:cutouts}
Contained in this section are cutouts similar to Fig.~\ref{fig:example_source} for the
remaining sources in the selected sample using GLOSTAR-VLA 5.8\,GHz D-configuration continuum data.
\begin{figure*}[!h]
\begin{tabular}{cc}
\includegraphics[width=0.48\textwidth]{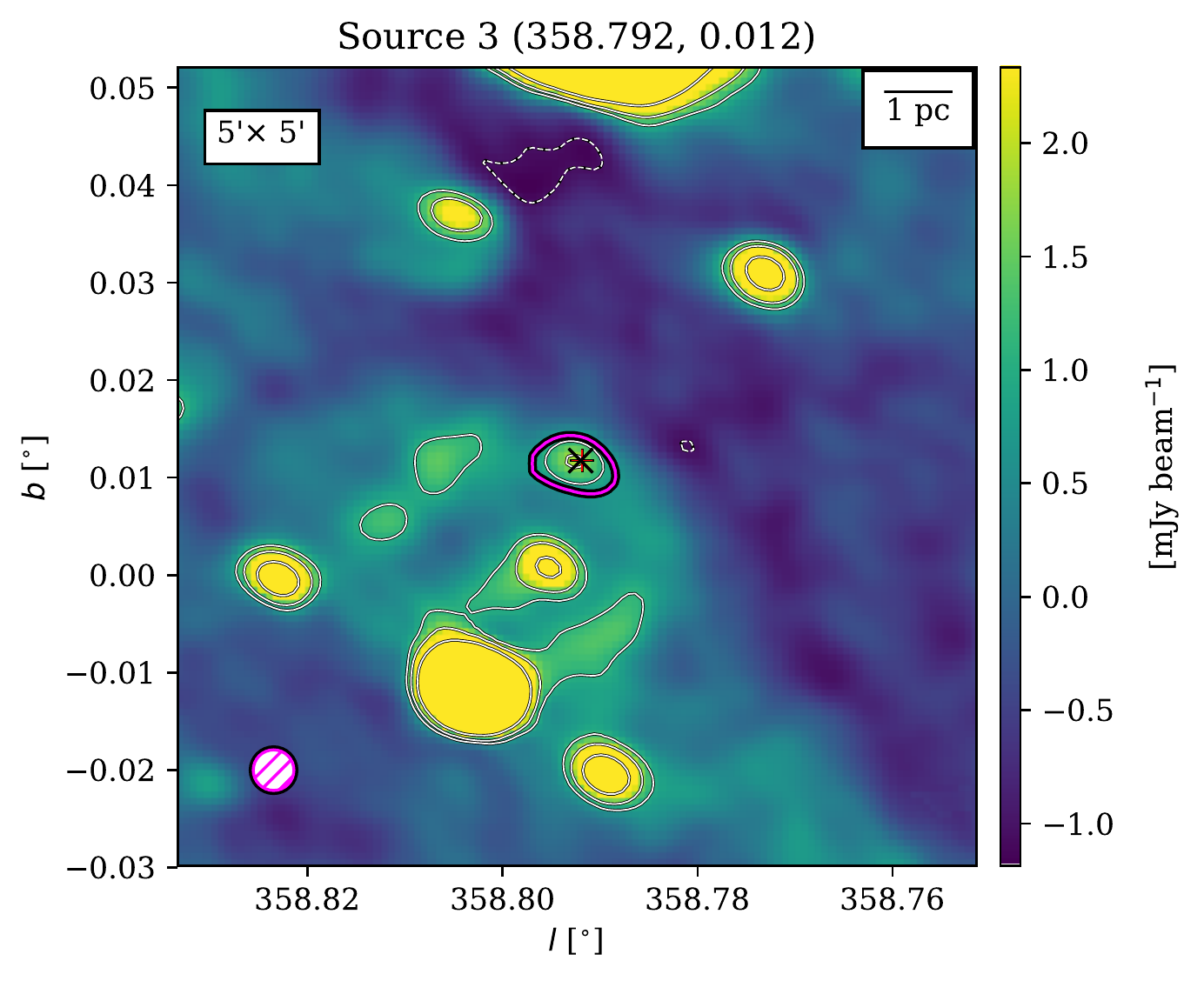}
\includegraphics[width=0.48\textwidth]{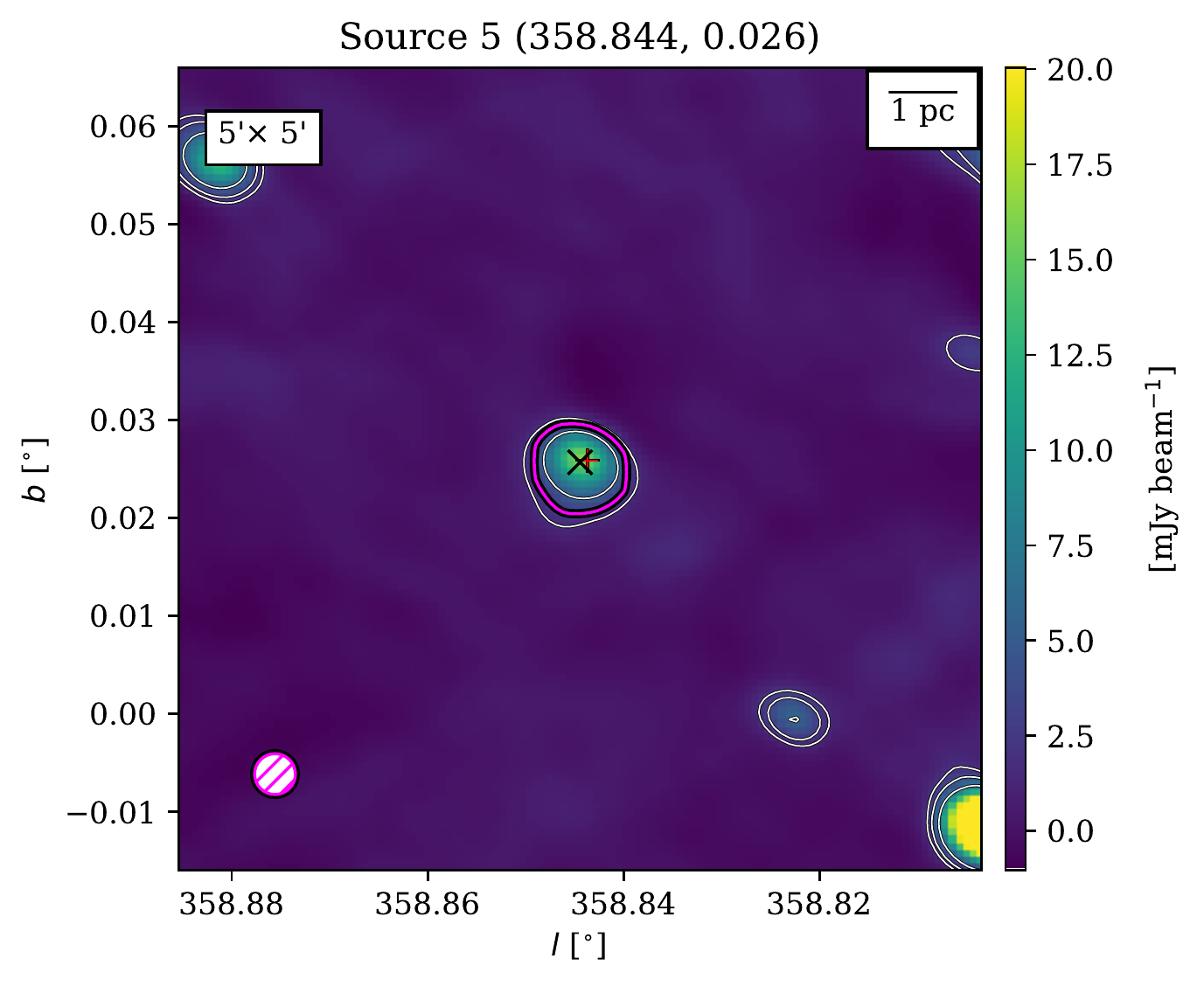}\\
\includegraphics[width=0.48\textwidth]{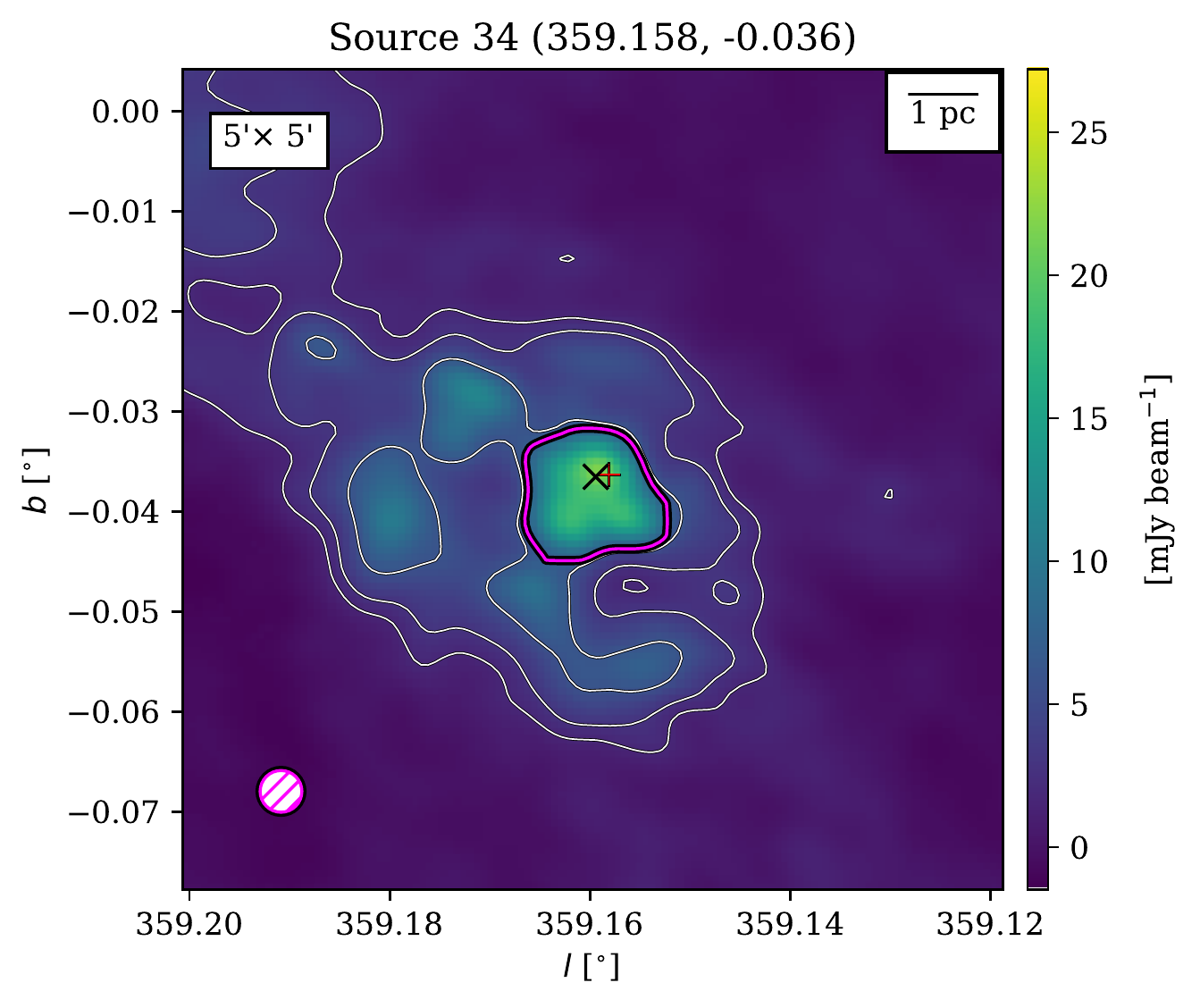}
\includegraphics[width=0.48\textwidth]{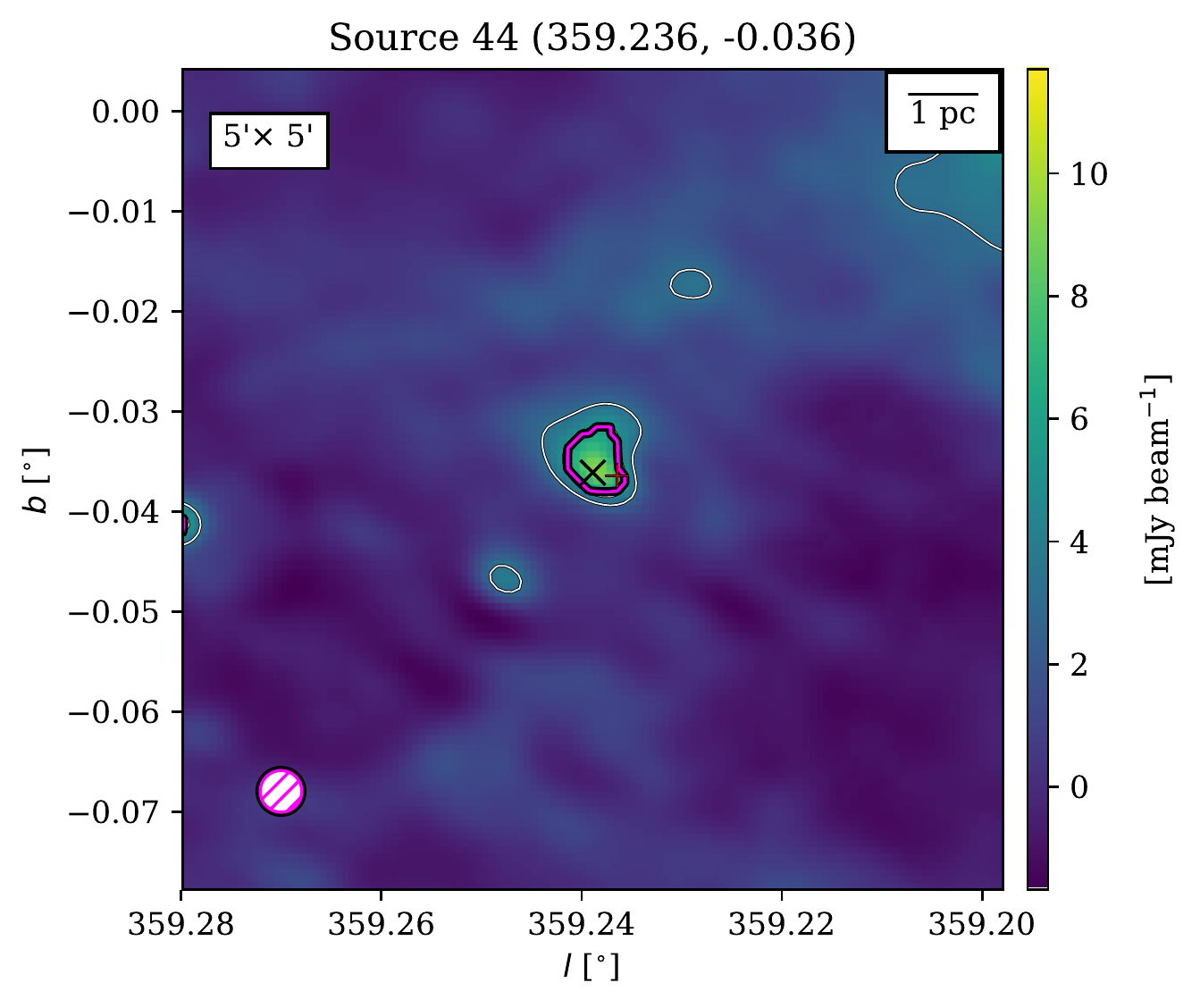}\\
\includegraphics[width=0.48\textwidth]{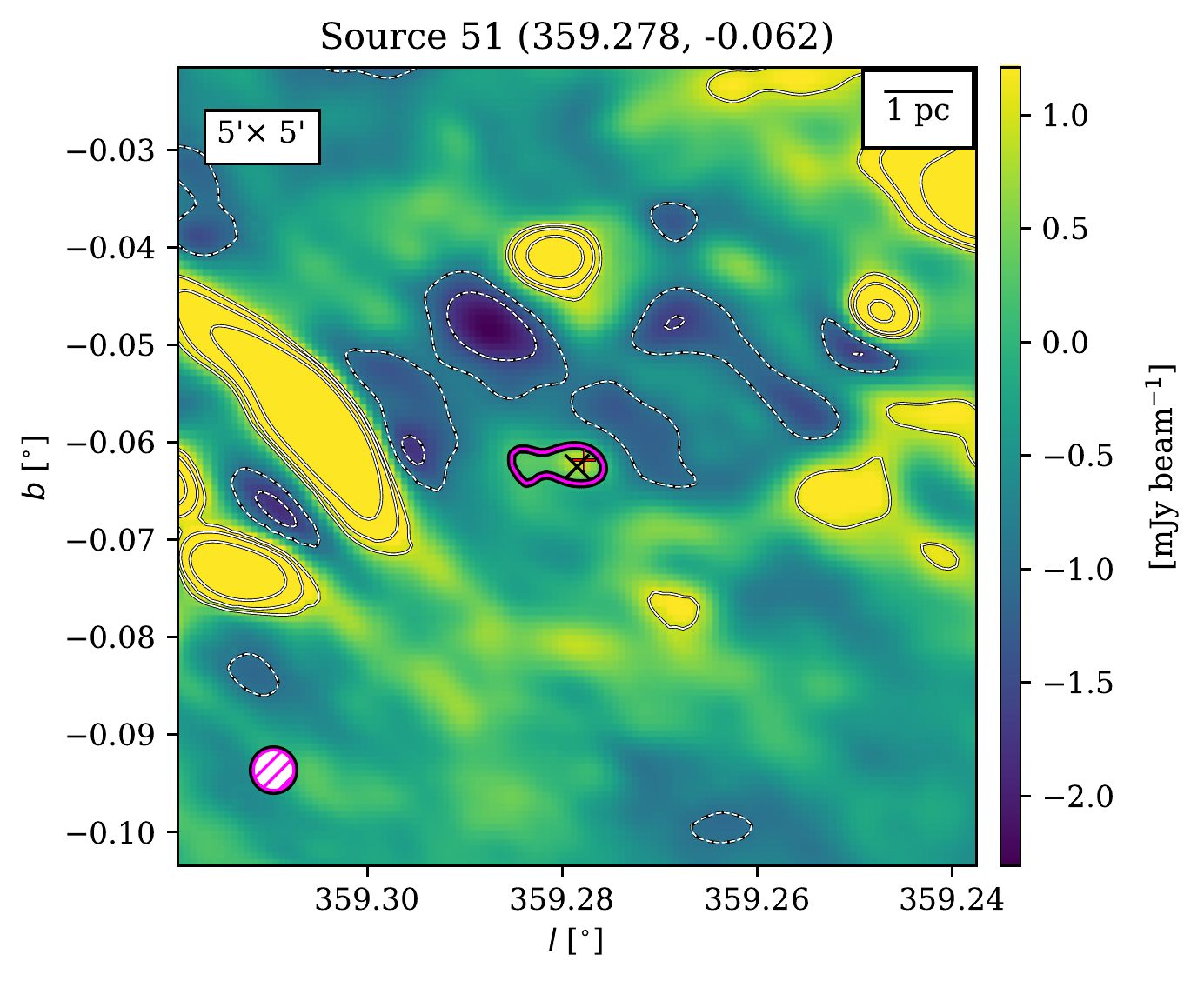}
\includegraphics[width=0.48\textwidth]{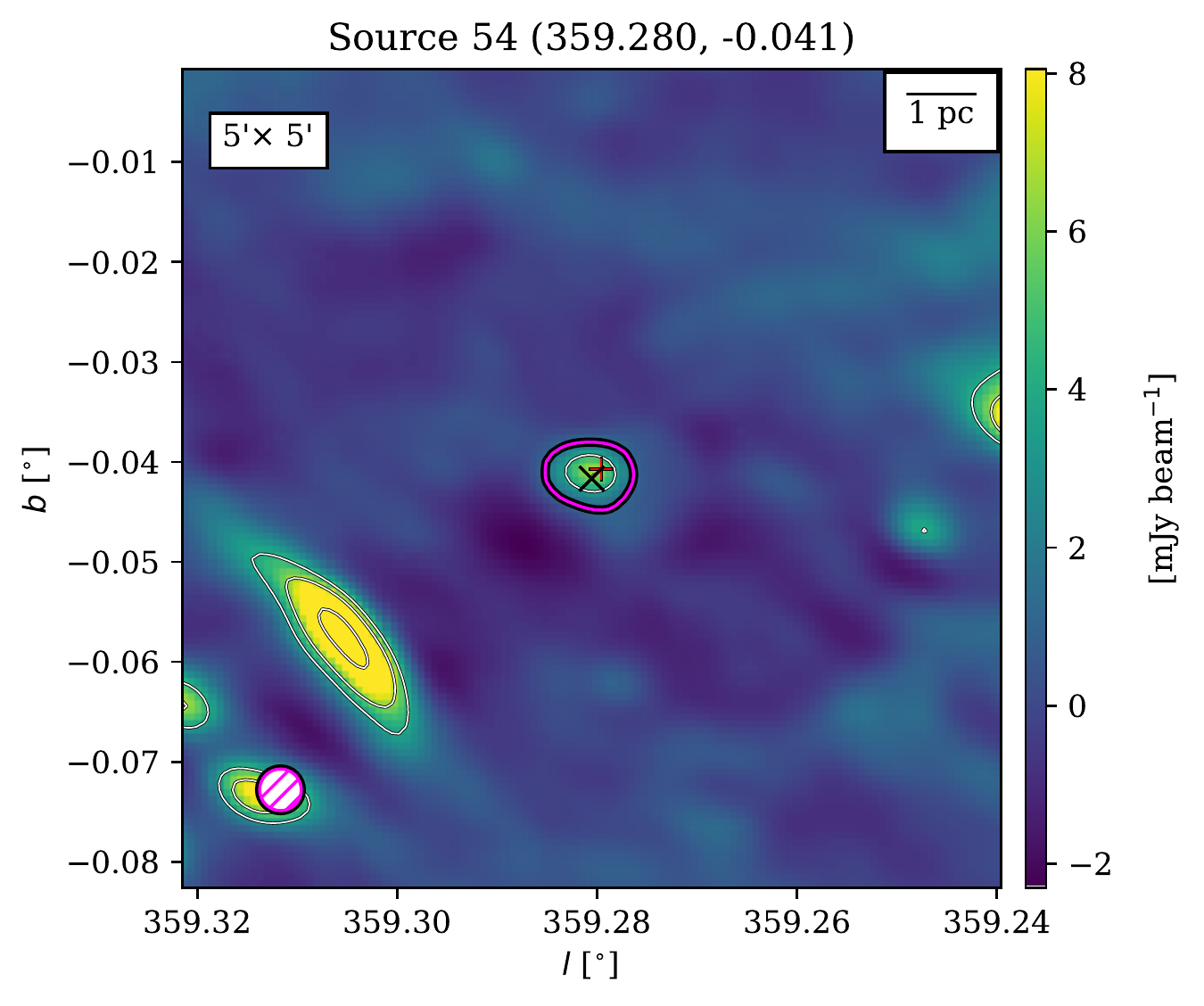}\\
\end{tabular}
\caption{Same as Fig.~\ref{fig:example_source}, but for the remaining sources in Table~\ref{tab:sources1}.}
\label{fig:manySource1}
\end{figure*}

\begin{figure*}[!h]
\begin{tabular}{cc}
\includegraphics[width=0.48\textwidth]{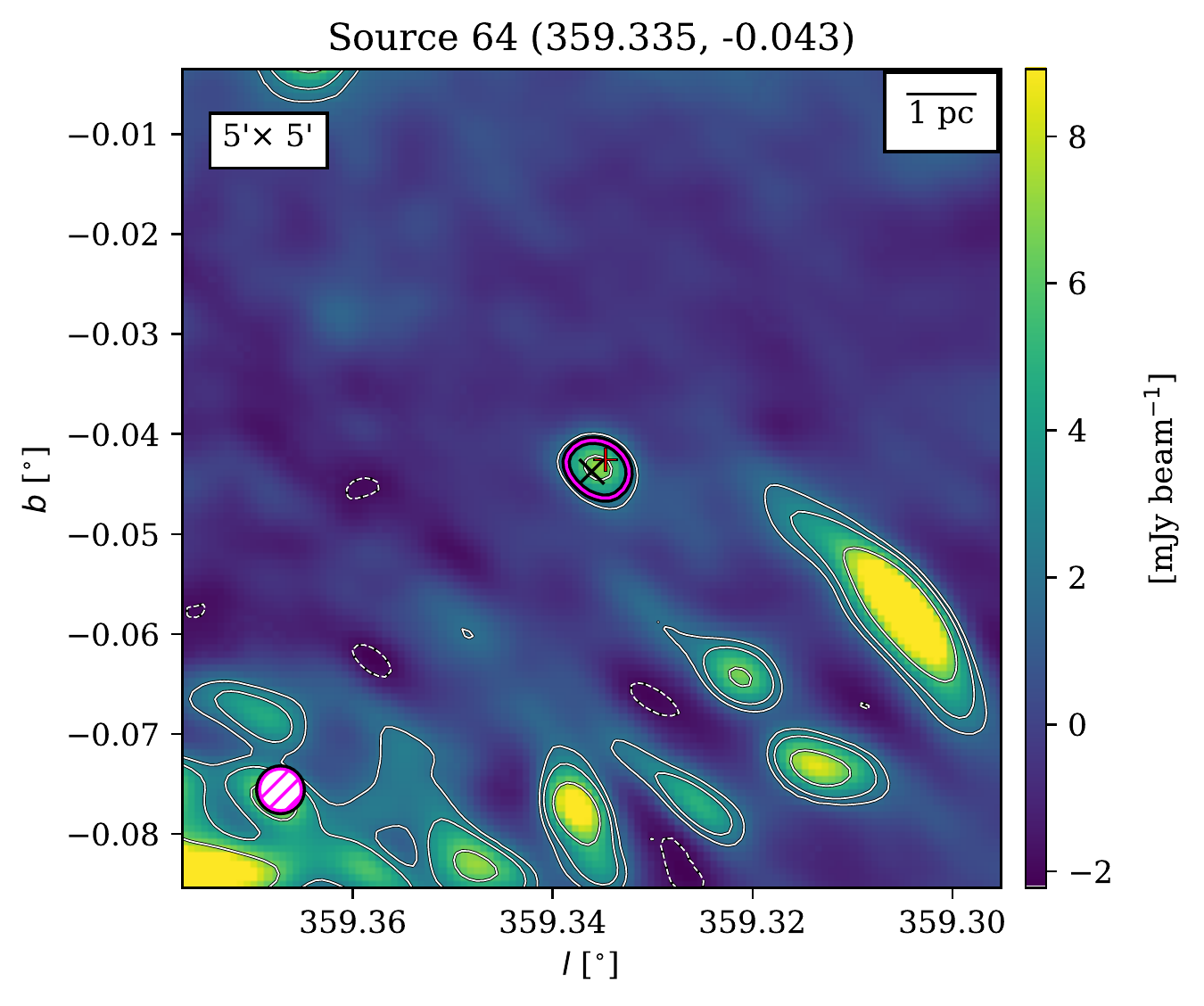}
\includegraphics[width=0.48\textwidth]{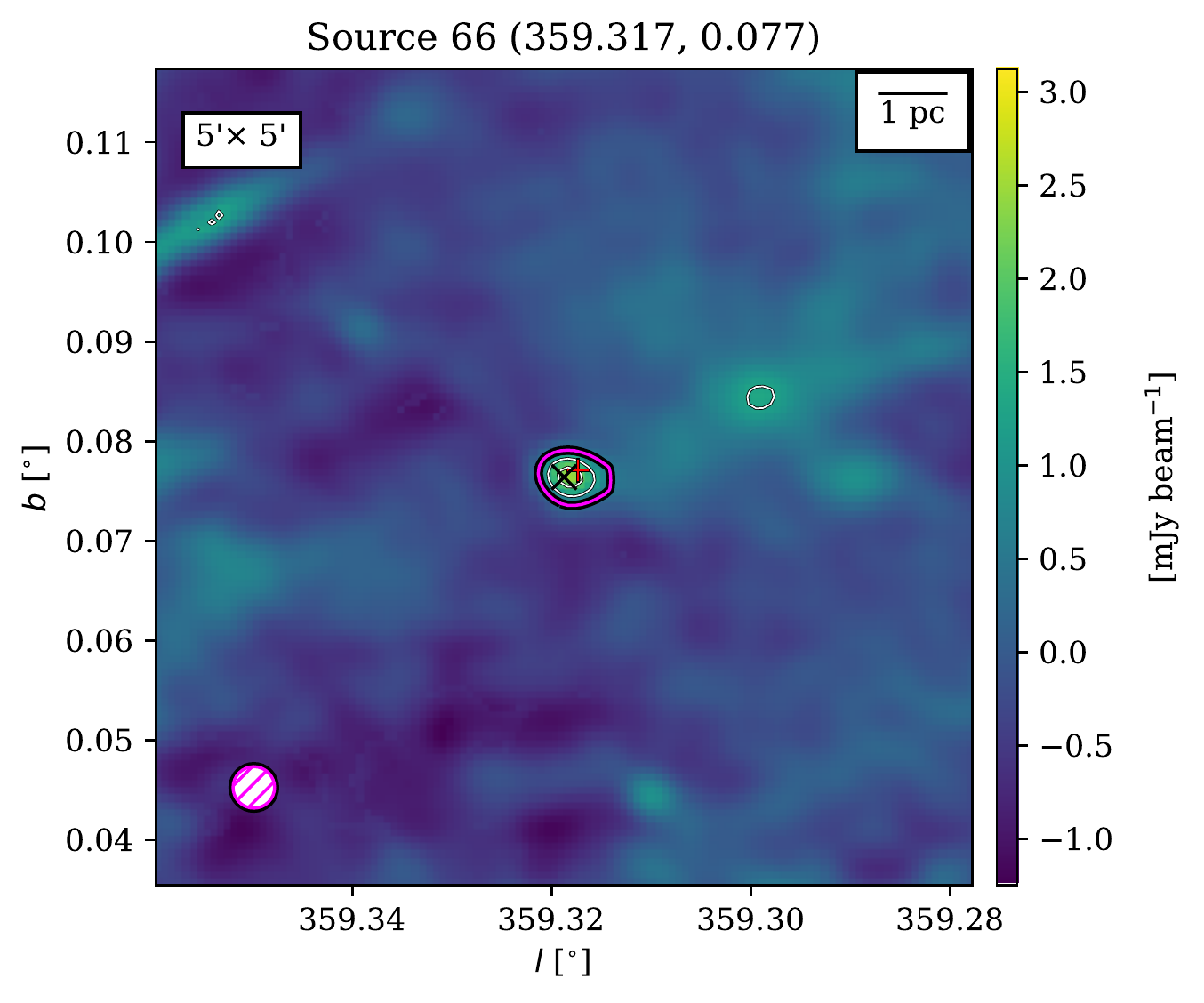}\\
\includegraphics[width=0.48\textwidth]{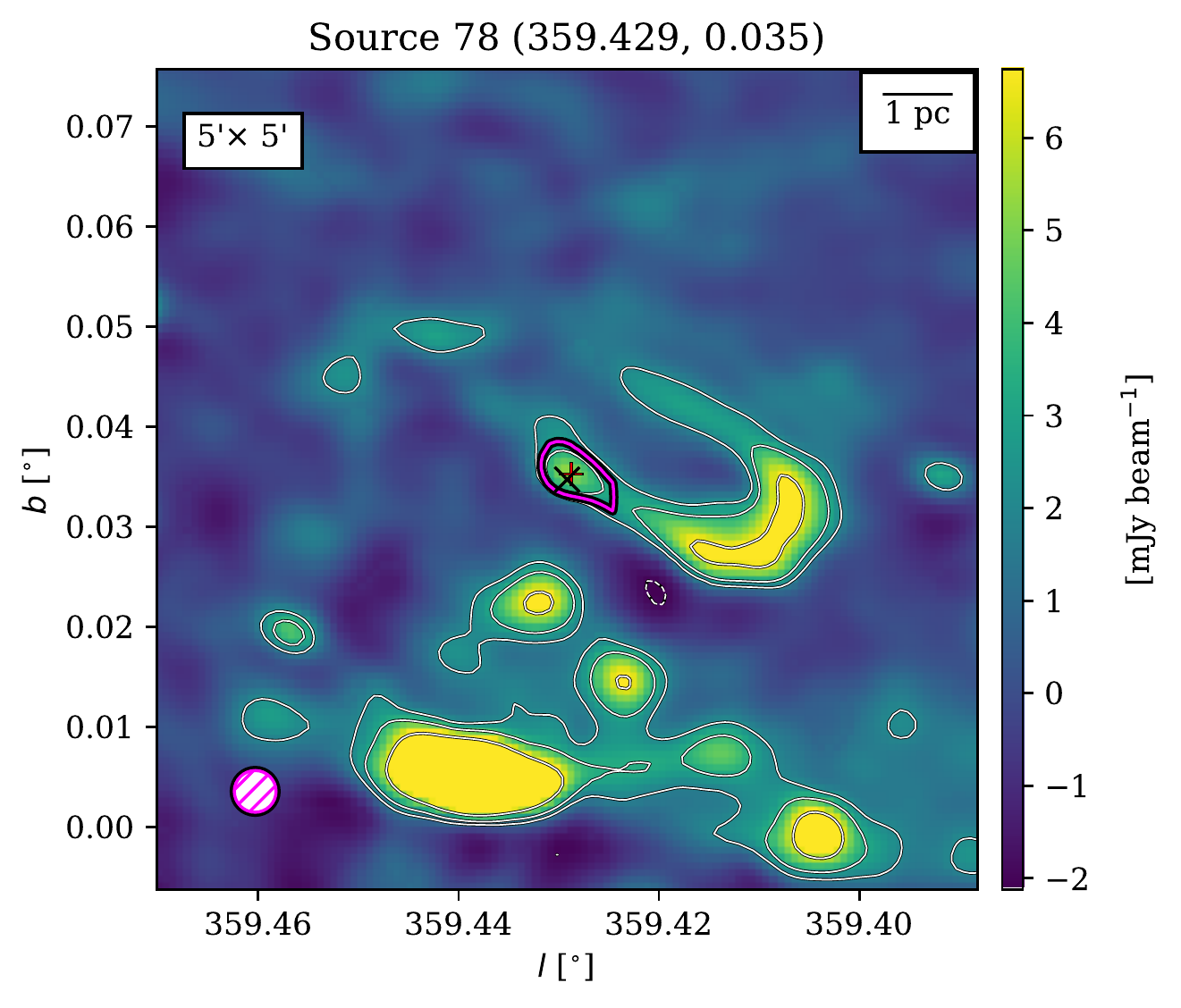}
\includegraphics[width=0.48\textwidth]{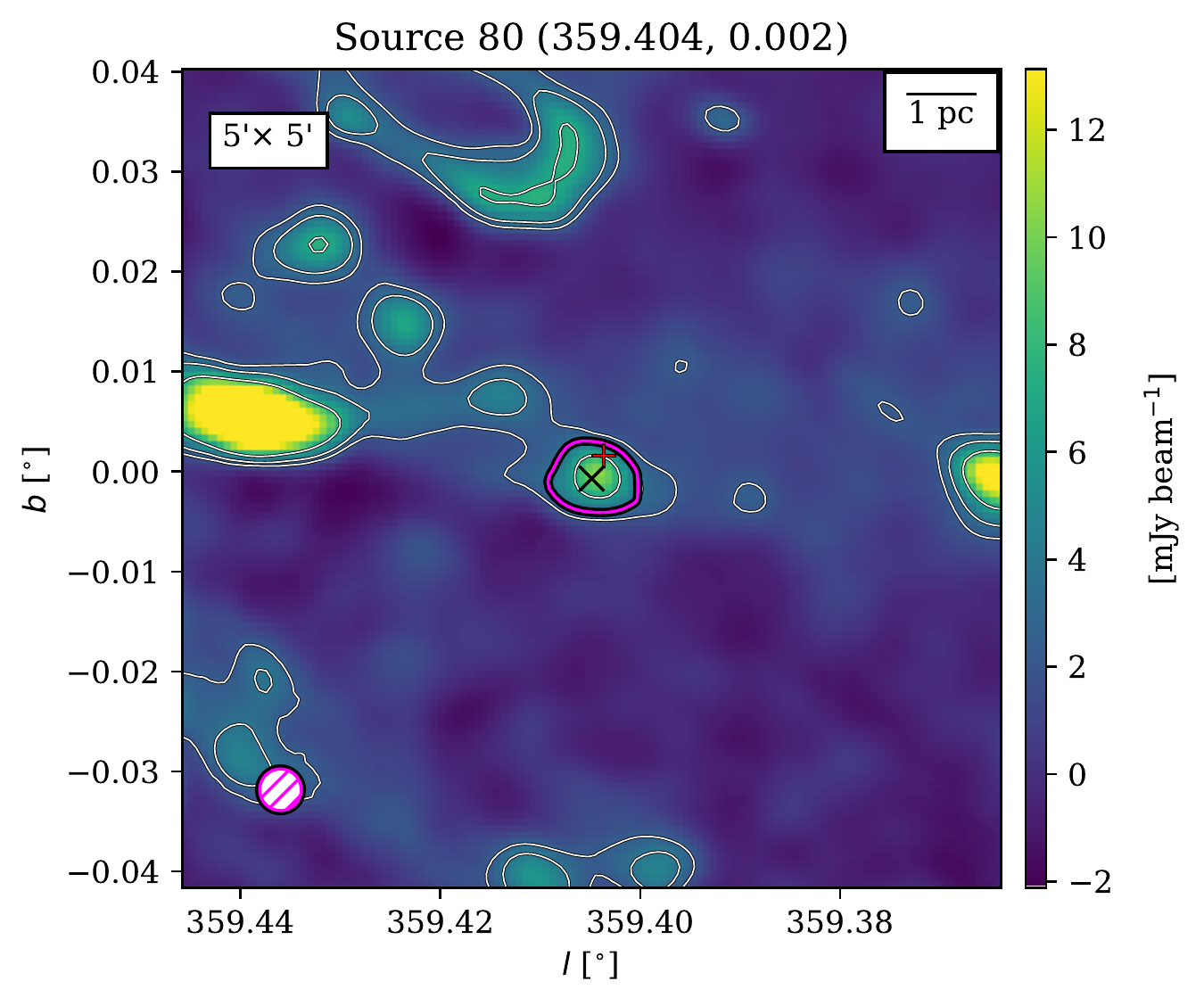}\\
\includegraphics[width=0.48\textwidth]{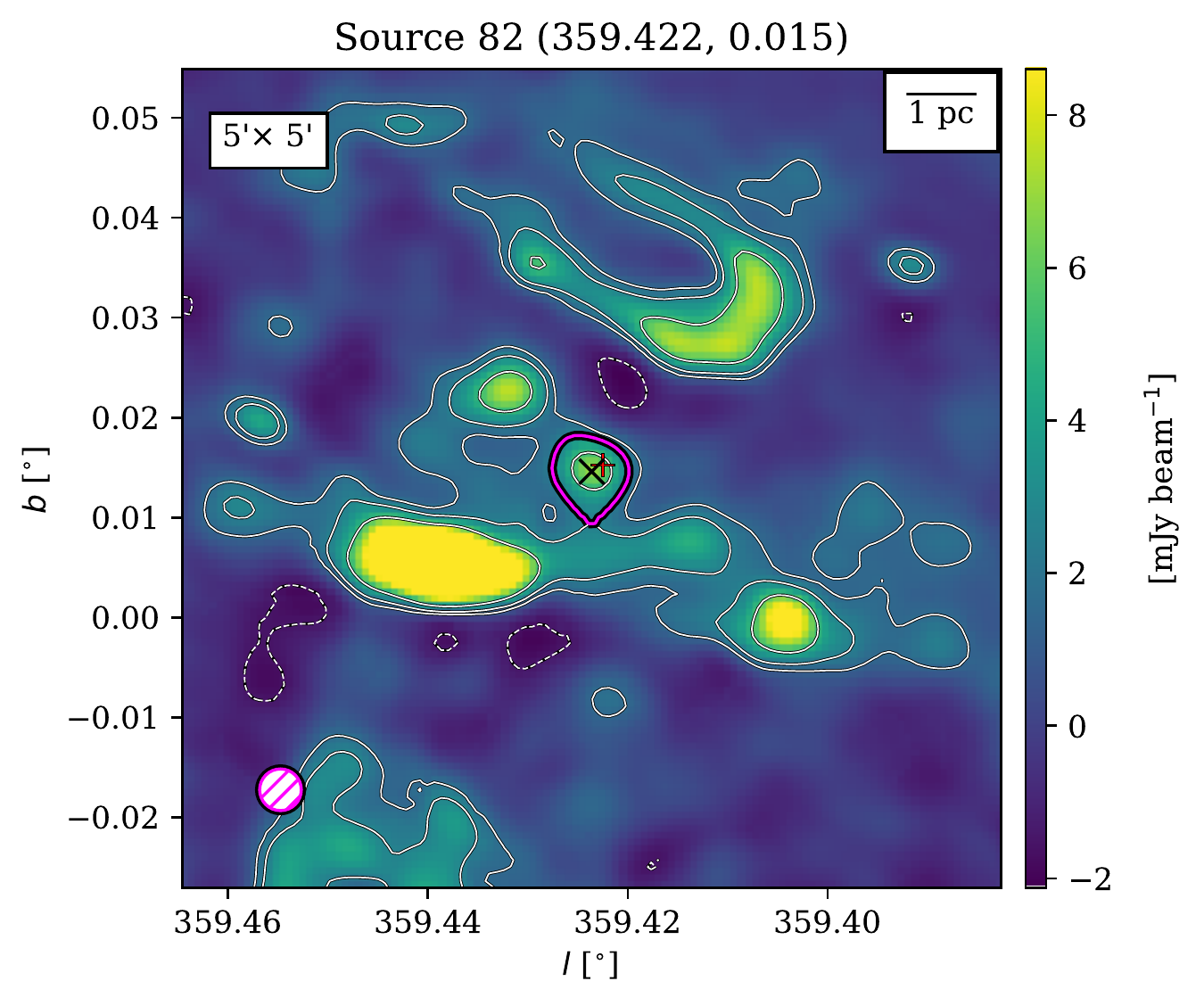}
\includegraphics[width=0.48\textwidth]{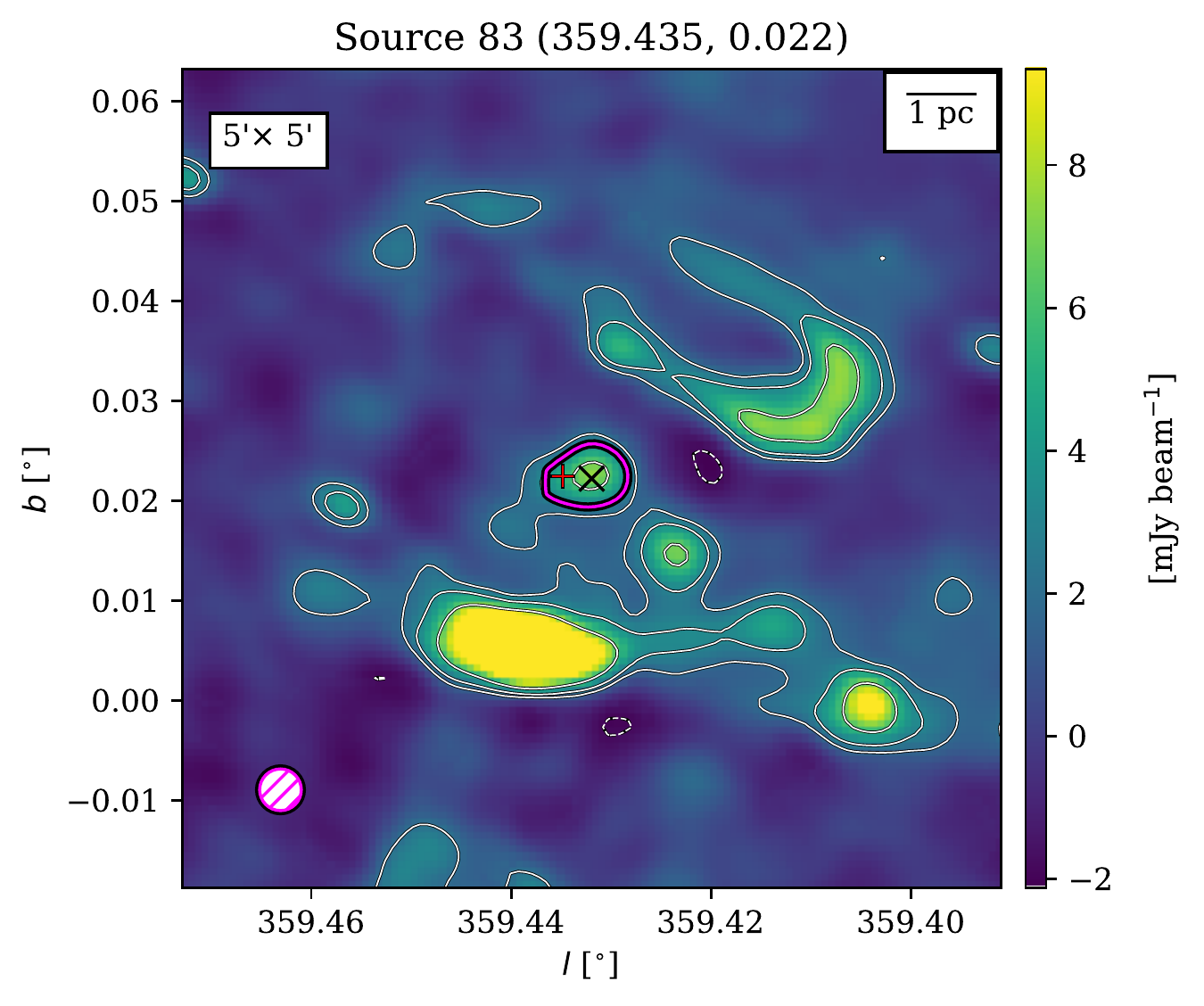}\\
\end{tabular}
\caption{Continued from Fig.~\ref{fig:manySource1}.}
\label{fig:manySource2}
\end{figure*}

\begin{figure*}[!h]
\begin{tabular}{cc}
\includegraphics[width=0.48\textwidth]{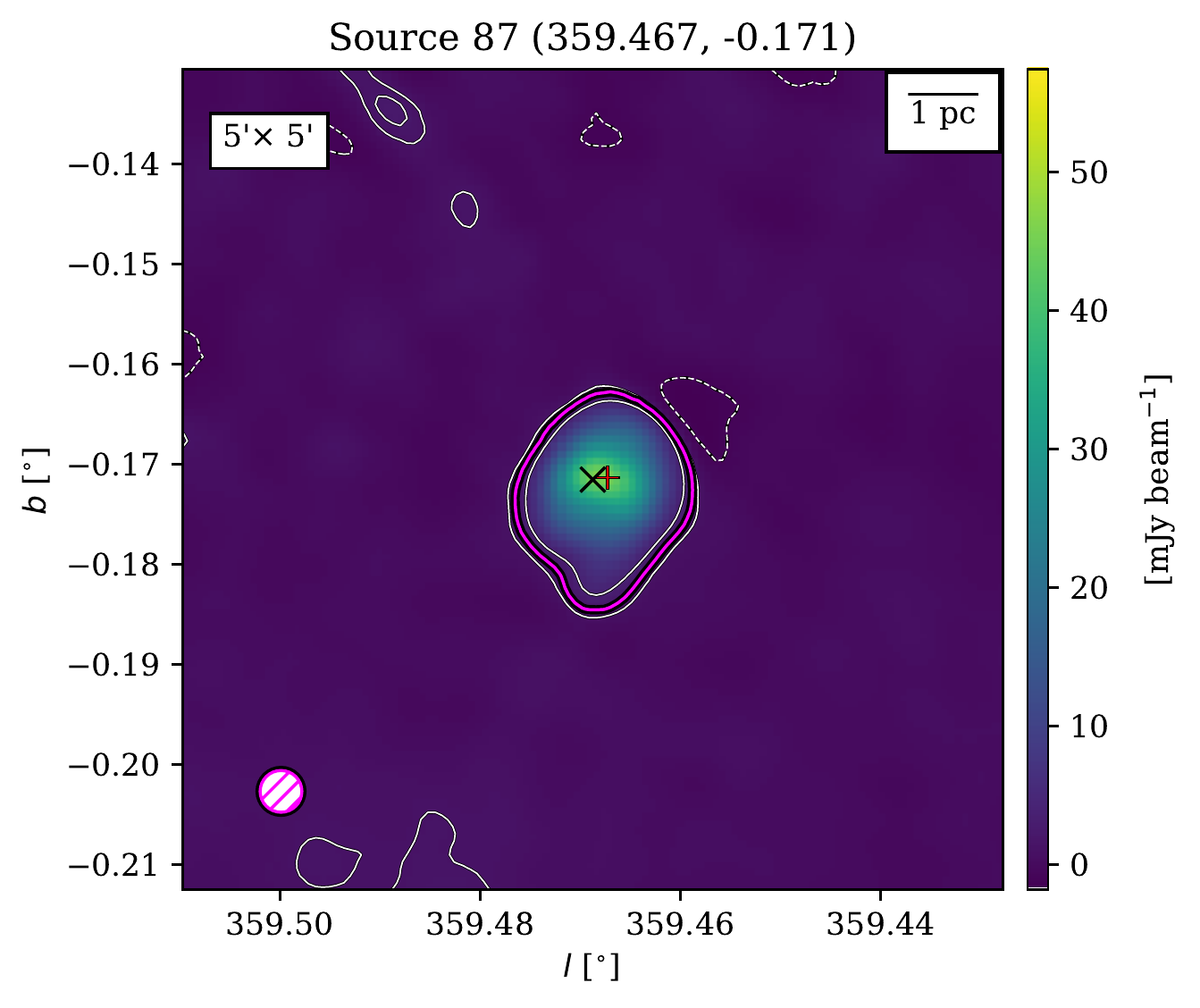}
\includegraphics[width=0.48\textwidth]{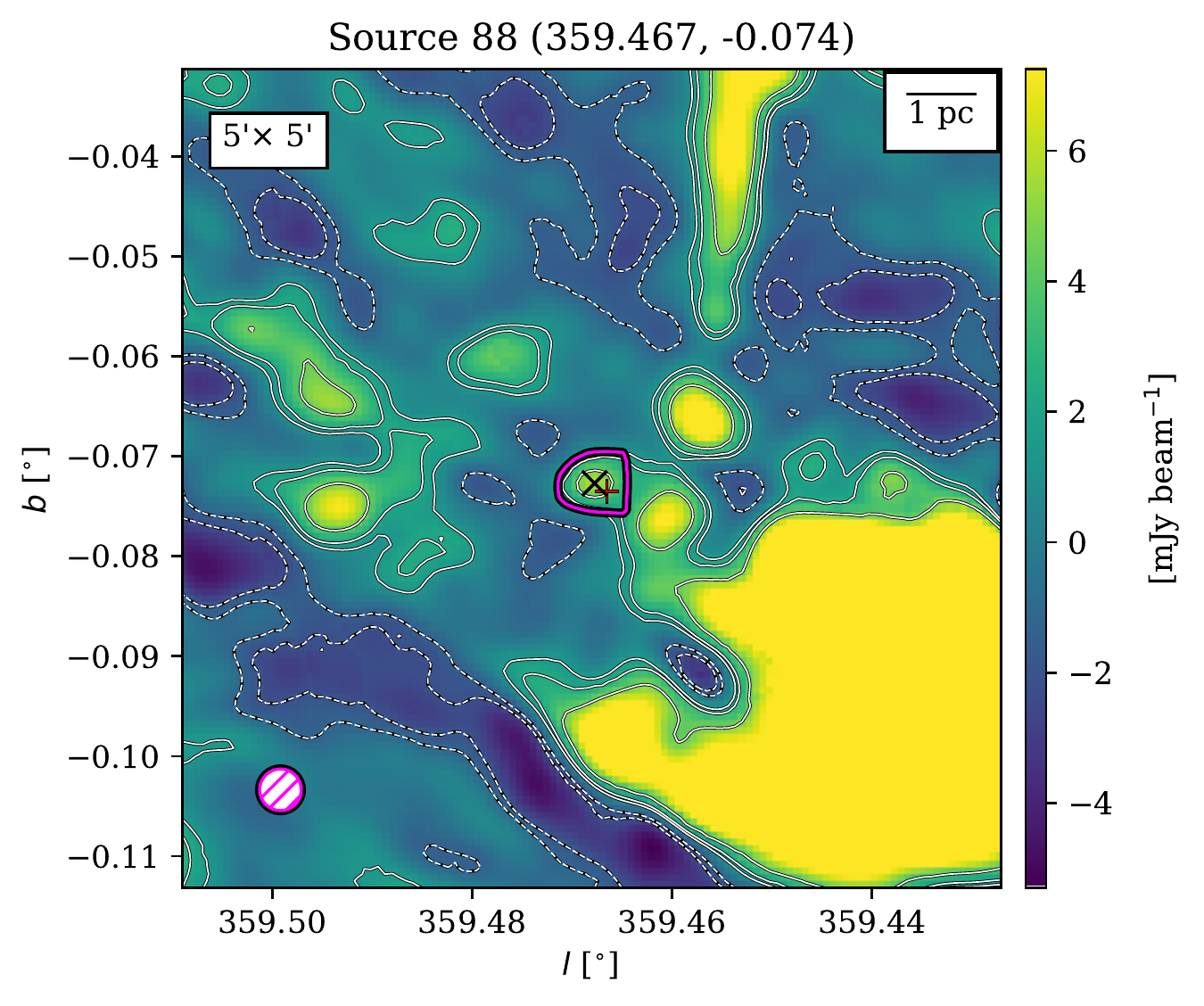}\\
\includegraphics[width=0.48\textwidth]{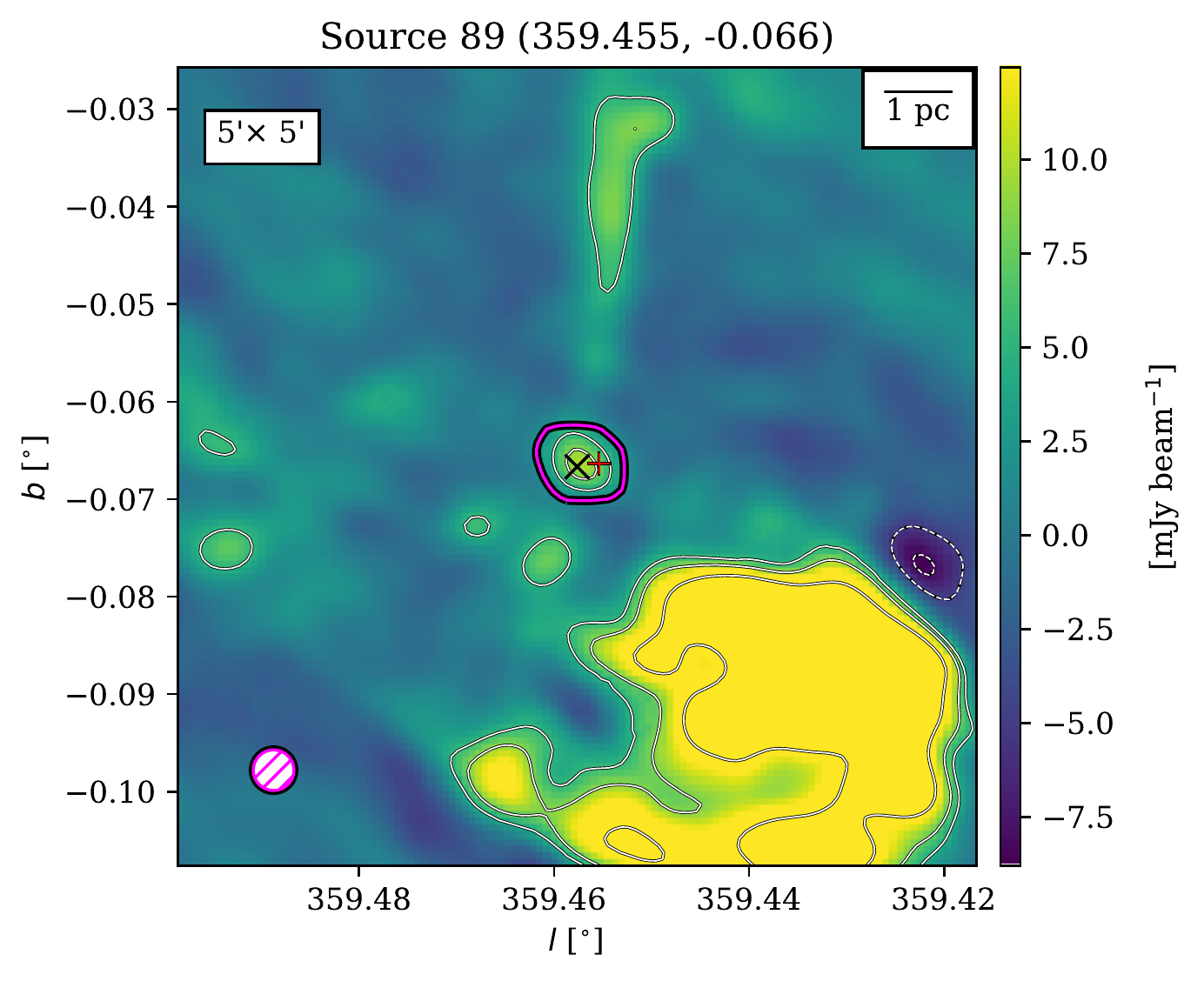}
\includegraphics[width=0.48\textwidth]{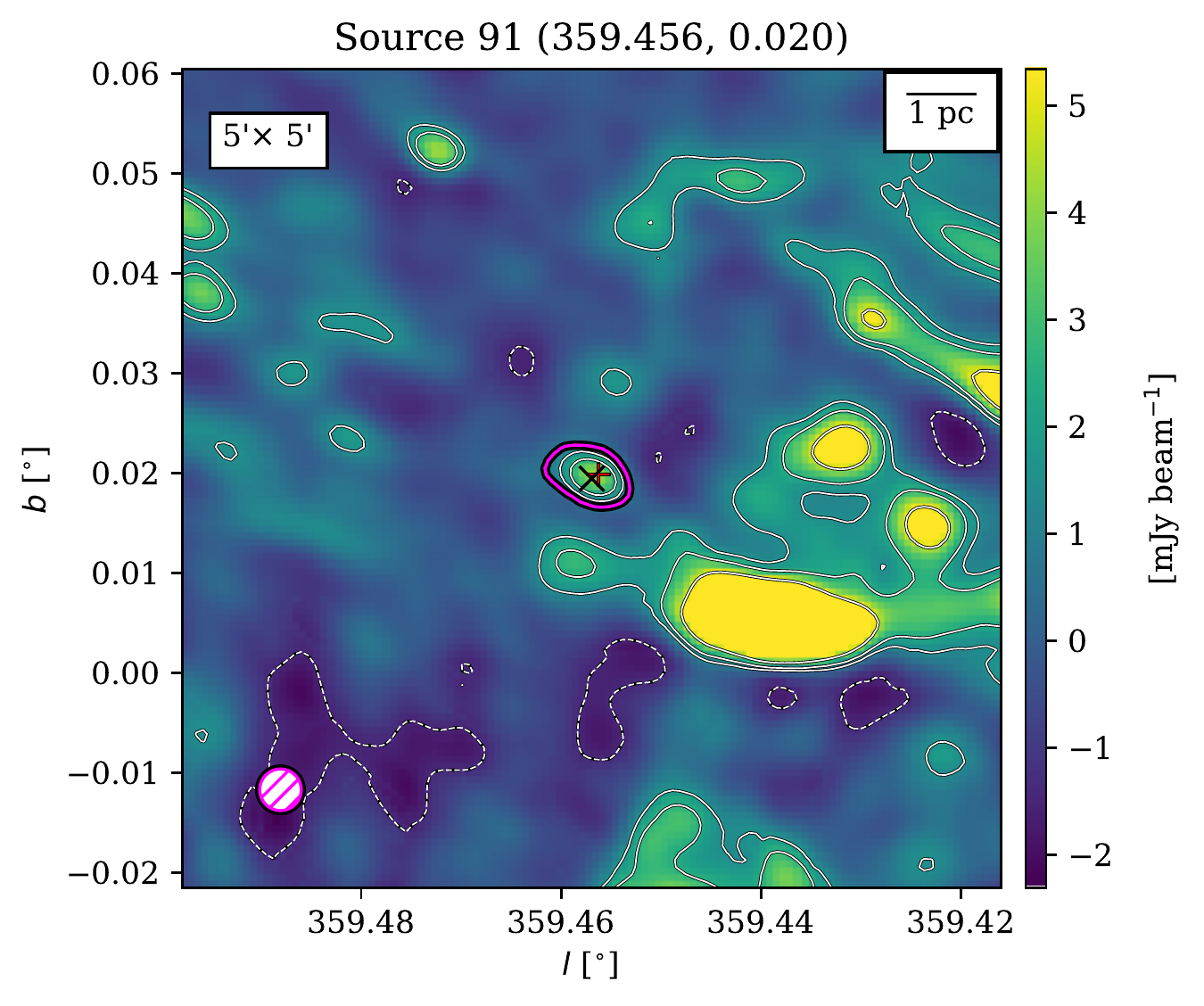}\\
\includegraphics[width=0.48\textwidth]{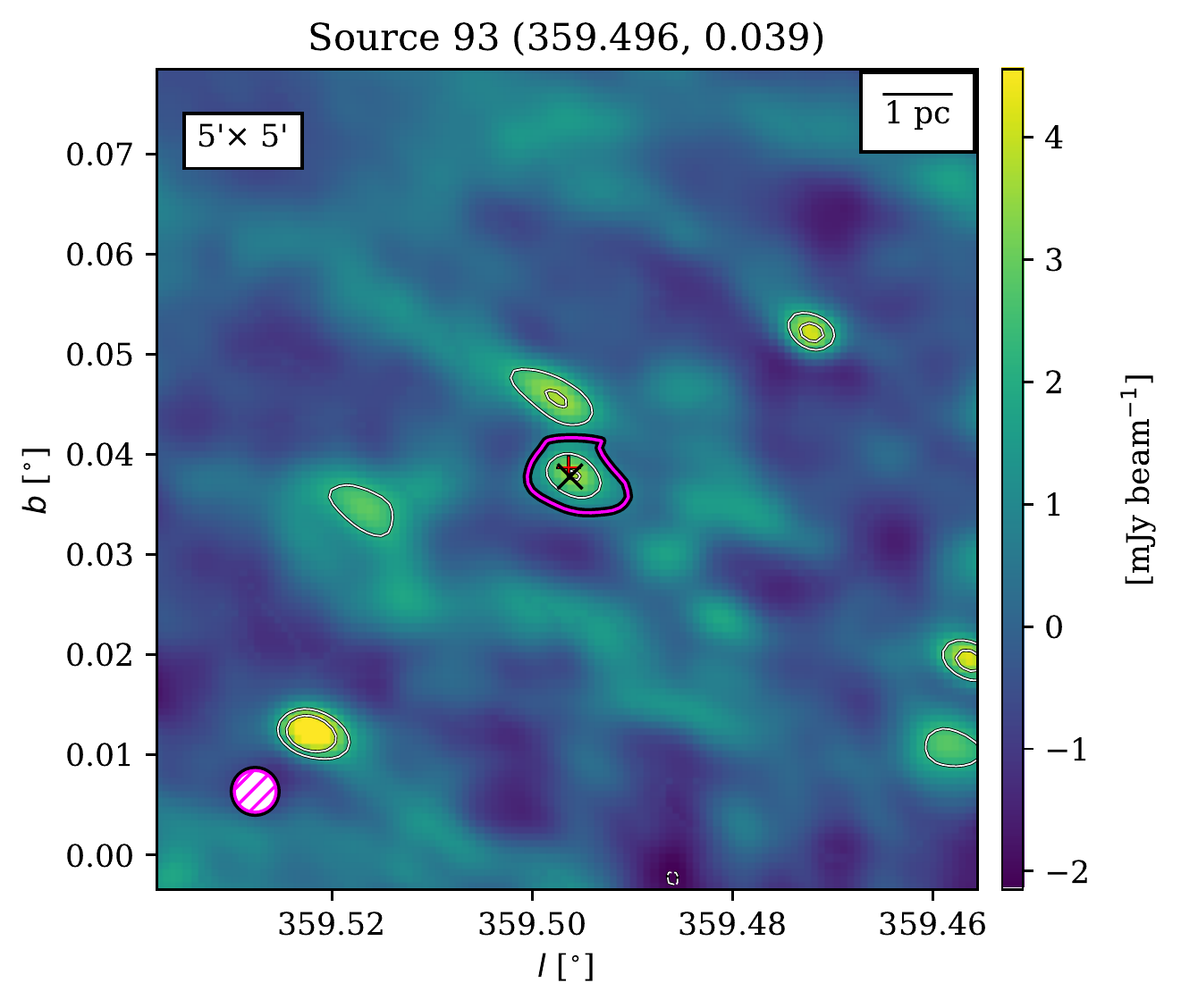}
\includegraphics[width=0.48\textwidth]{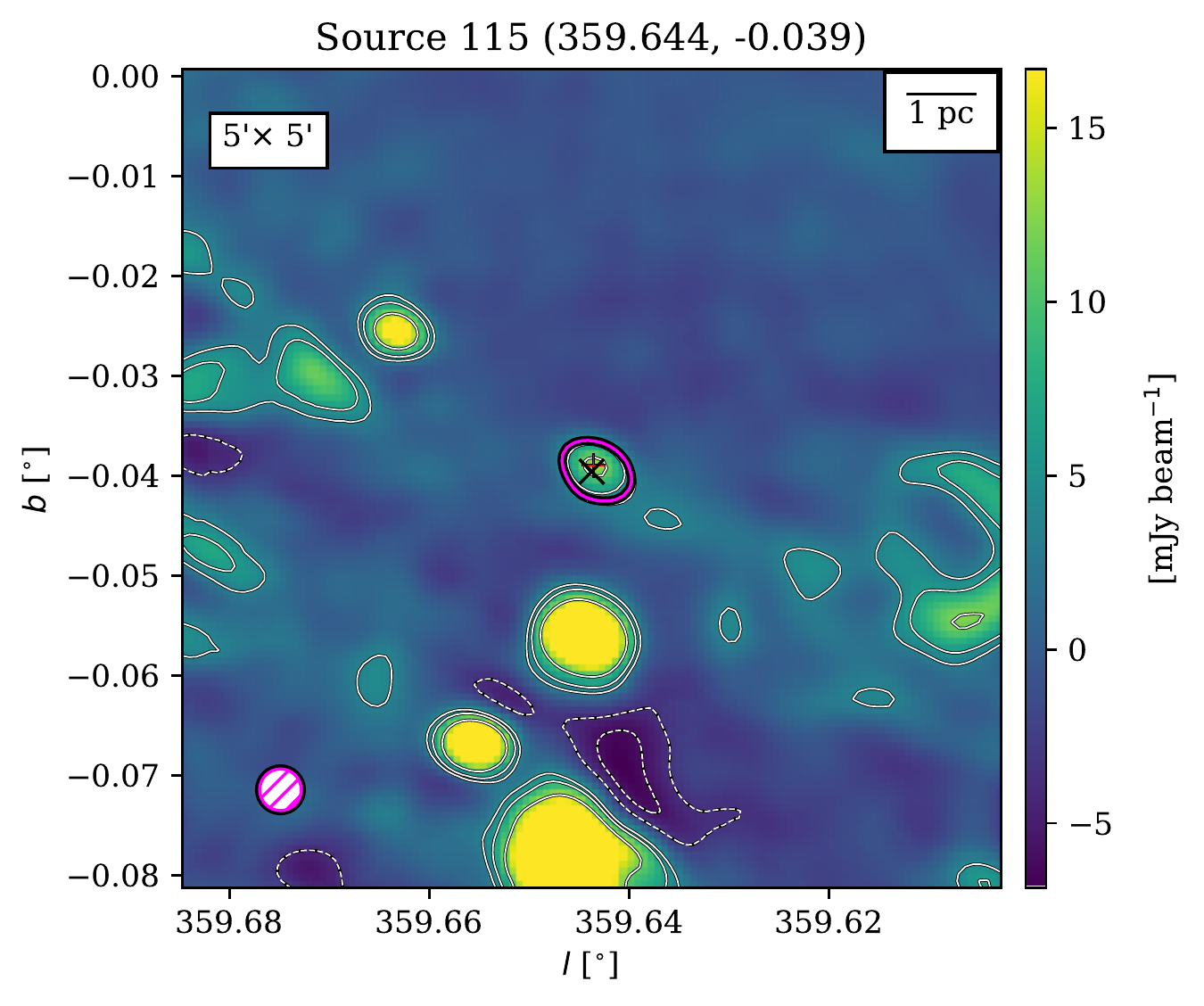}\\
\end{tabular}
\caption{Continued from Fig.~\ref{fig:manySource1}.}
\label{fig:manySource3}
\end{figure*}

\begin{figure*}[!h]
\begin{tabular}{cc}
\includegraphics[width=0.48\textwidth]{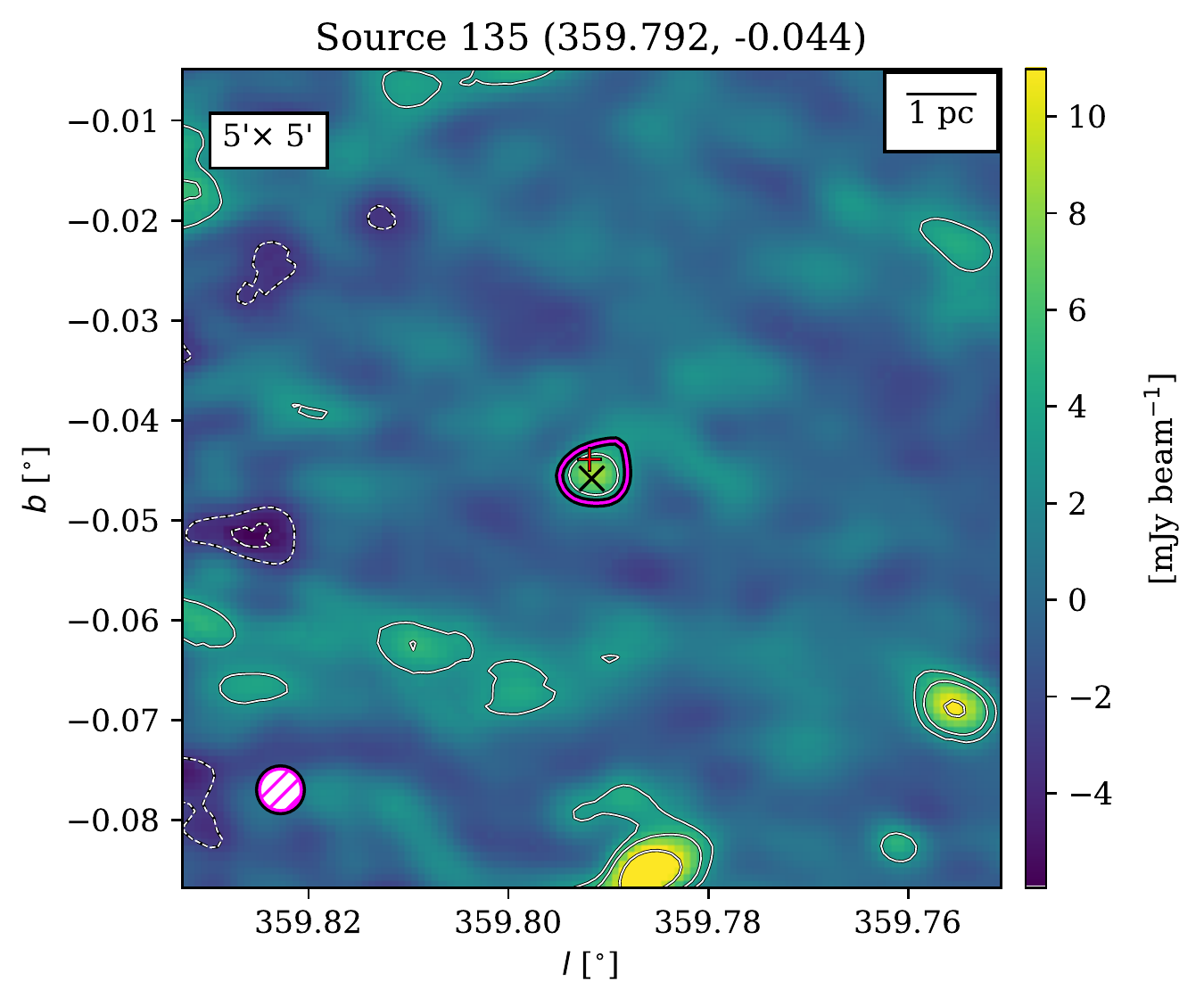}
\includegraphics[width=0.48\textwidth]{cutouts/source_image_s140_5arcmin.pdf}\\
\includegraphics[width=0.48\textwidth]{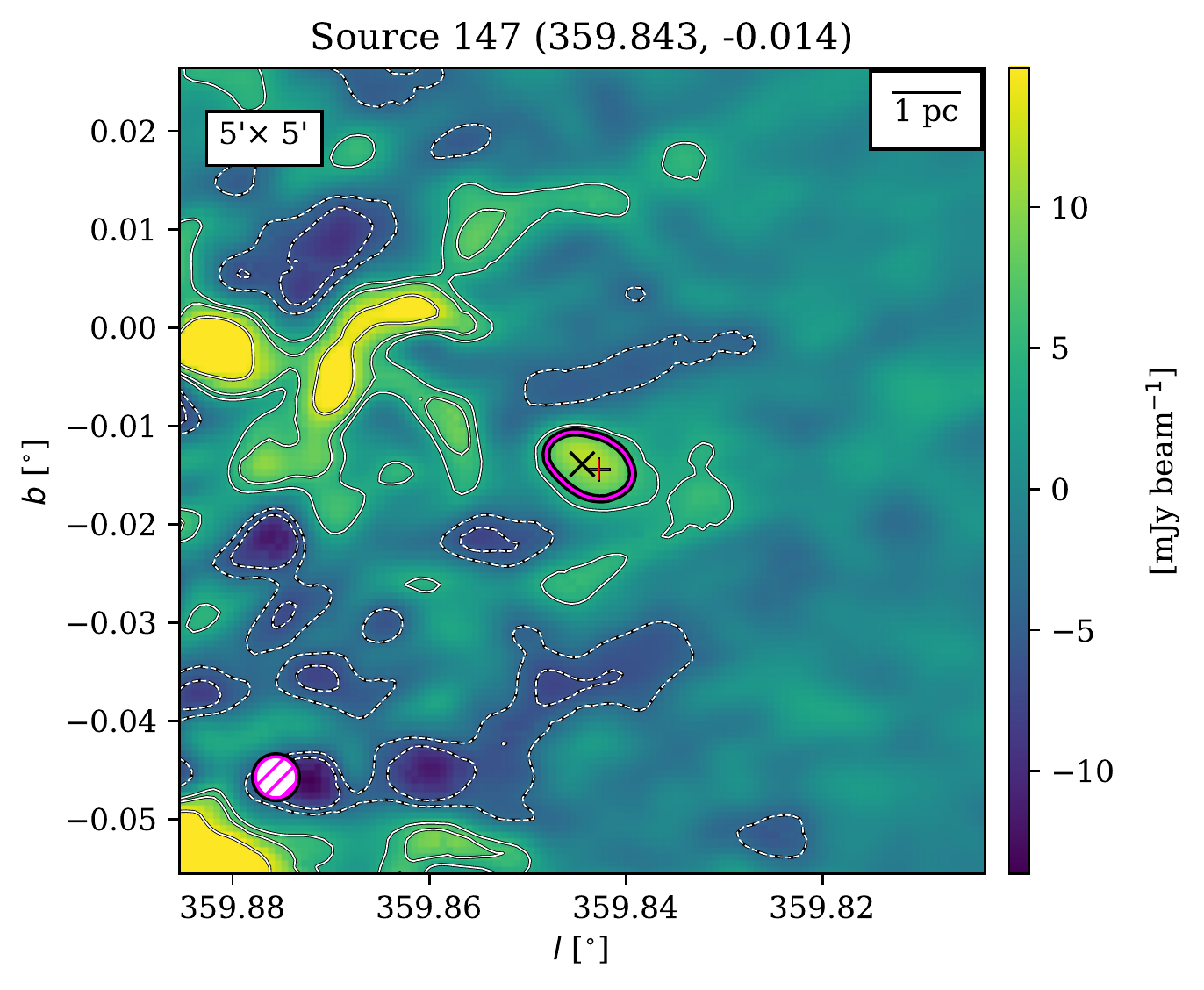}
\includegraphics[width=0.48\textwidth]{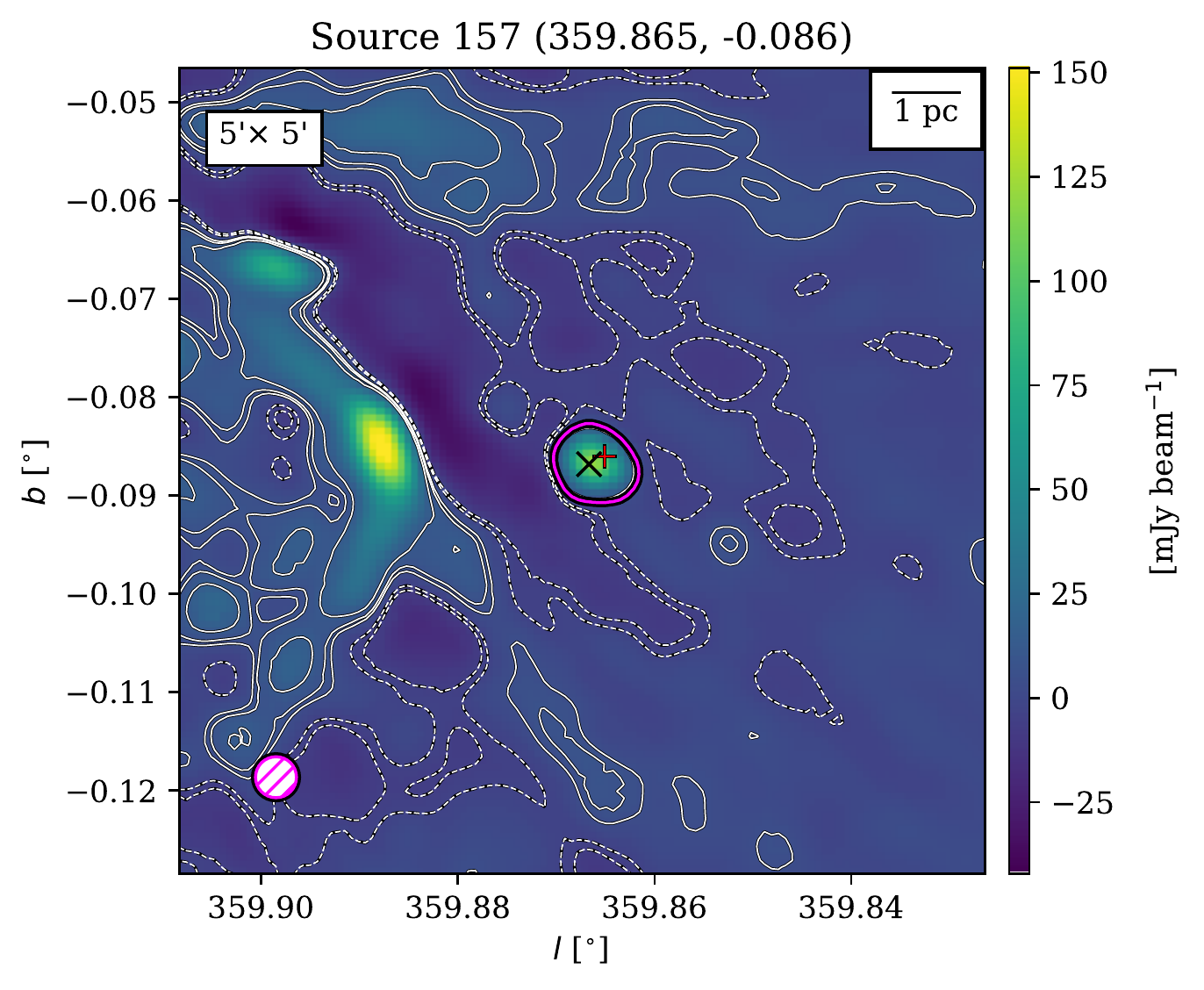}\\
\includegraphics[width=0.48\textwidth]{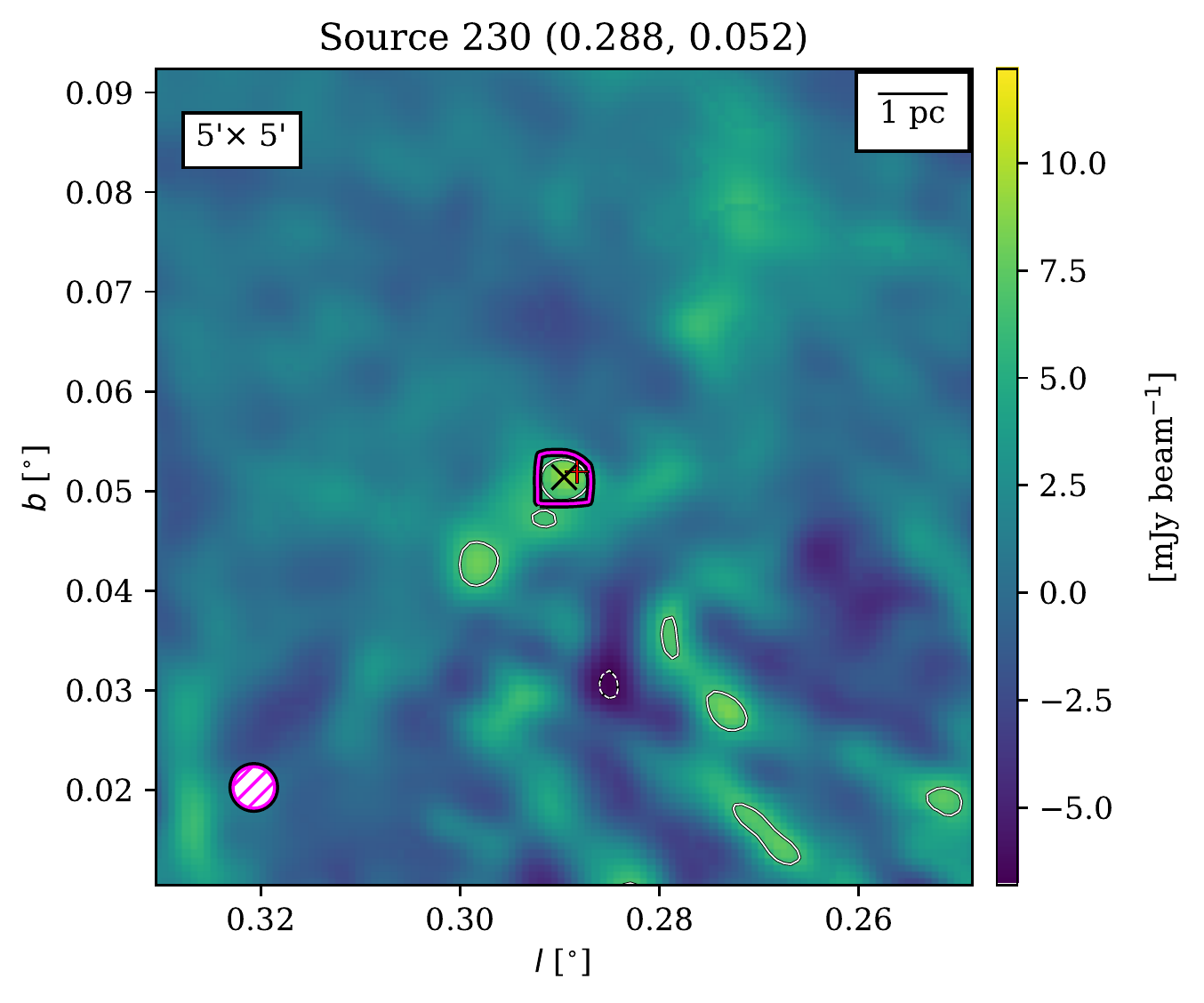}
\includegraphics[width=0.48\textwidth]{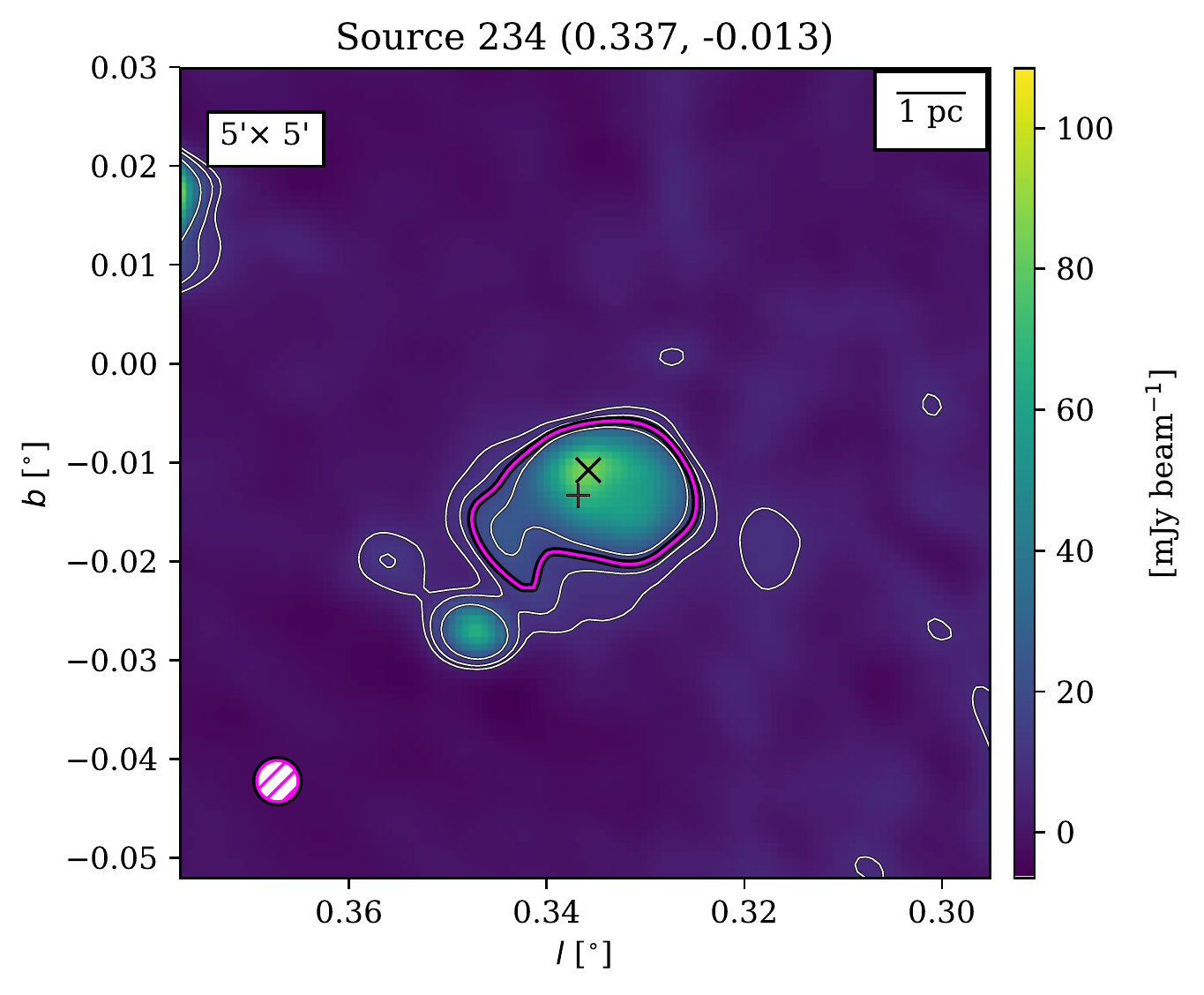}\\
\end{tabular}
\caption{Continued from Fig.~\ref{fig:manySource1}.}
\label{fig:manySource4}
\end{figure*}

\begin{figure*}[!h]
\begin{tabular}{cc}
\includegraphics[width=0.48\textwidth]{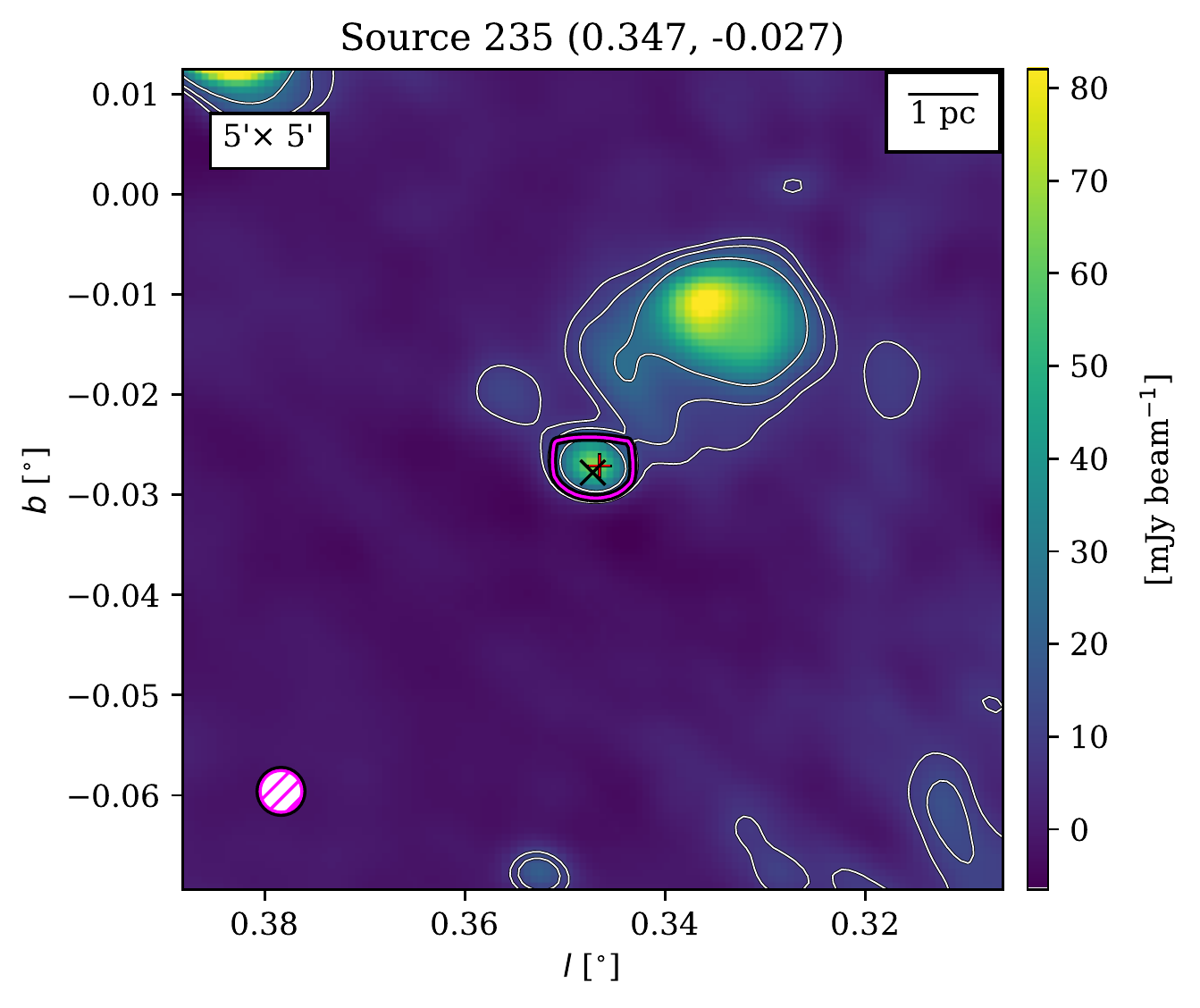}
\includegraphics[width=0.48\textwidth]{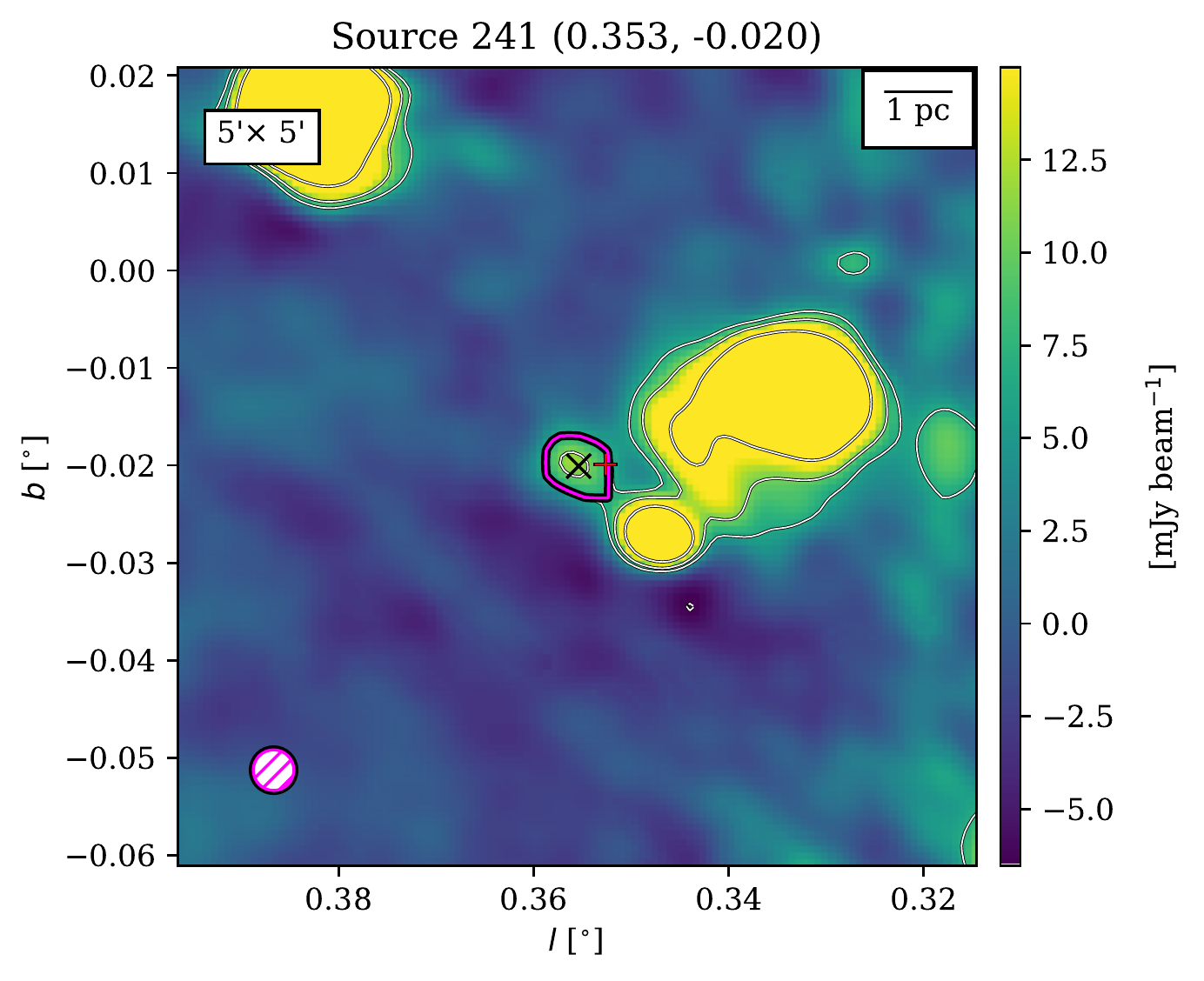}\\
\includegraphics[width=0.48\textwidth]{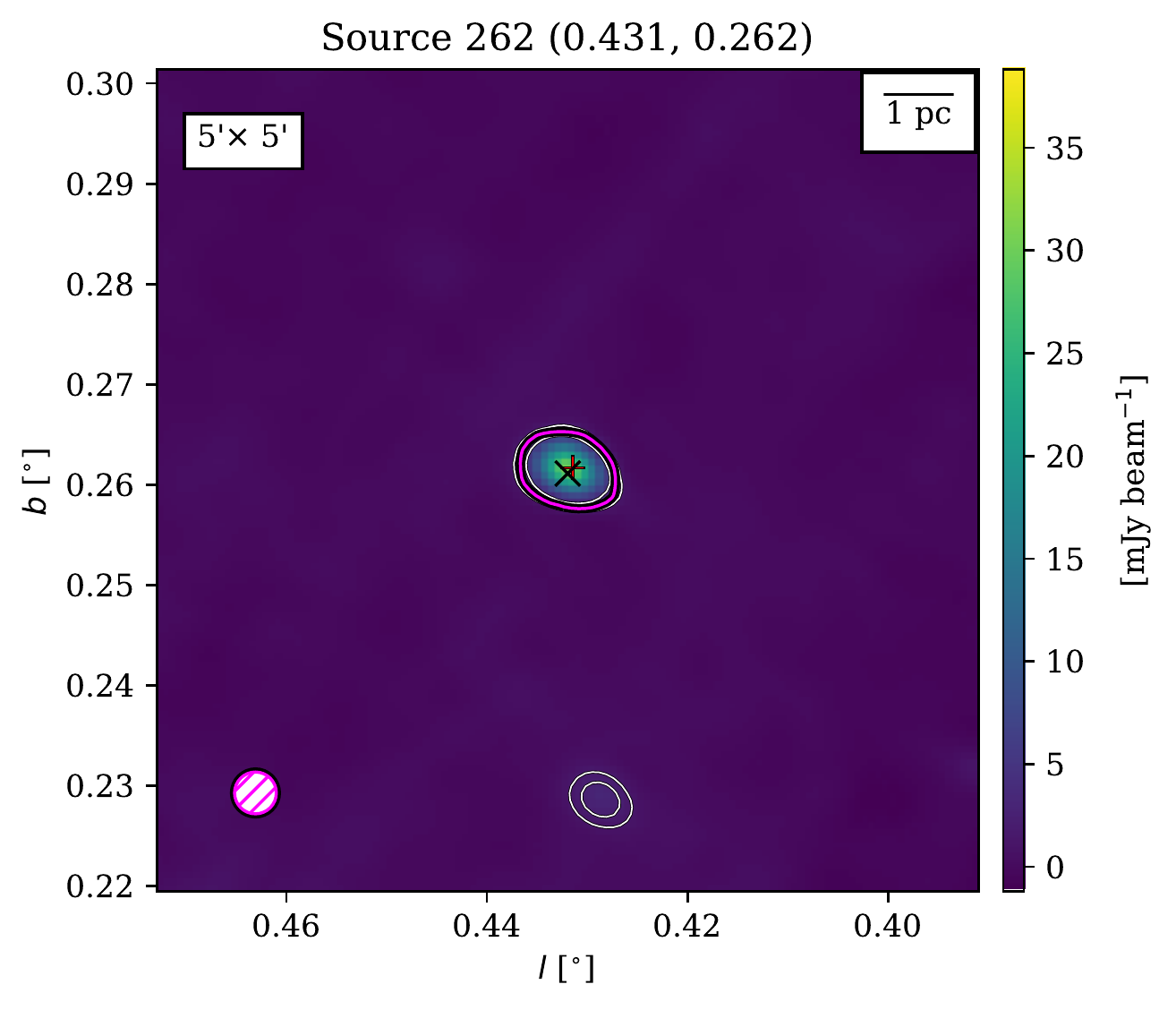}
\includegraphics[width=0.48\textwidth]{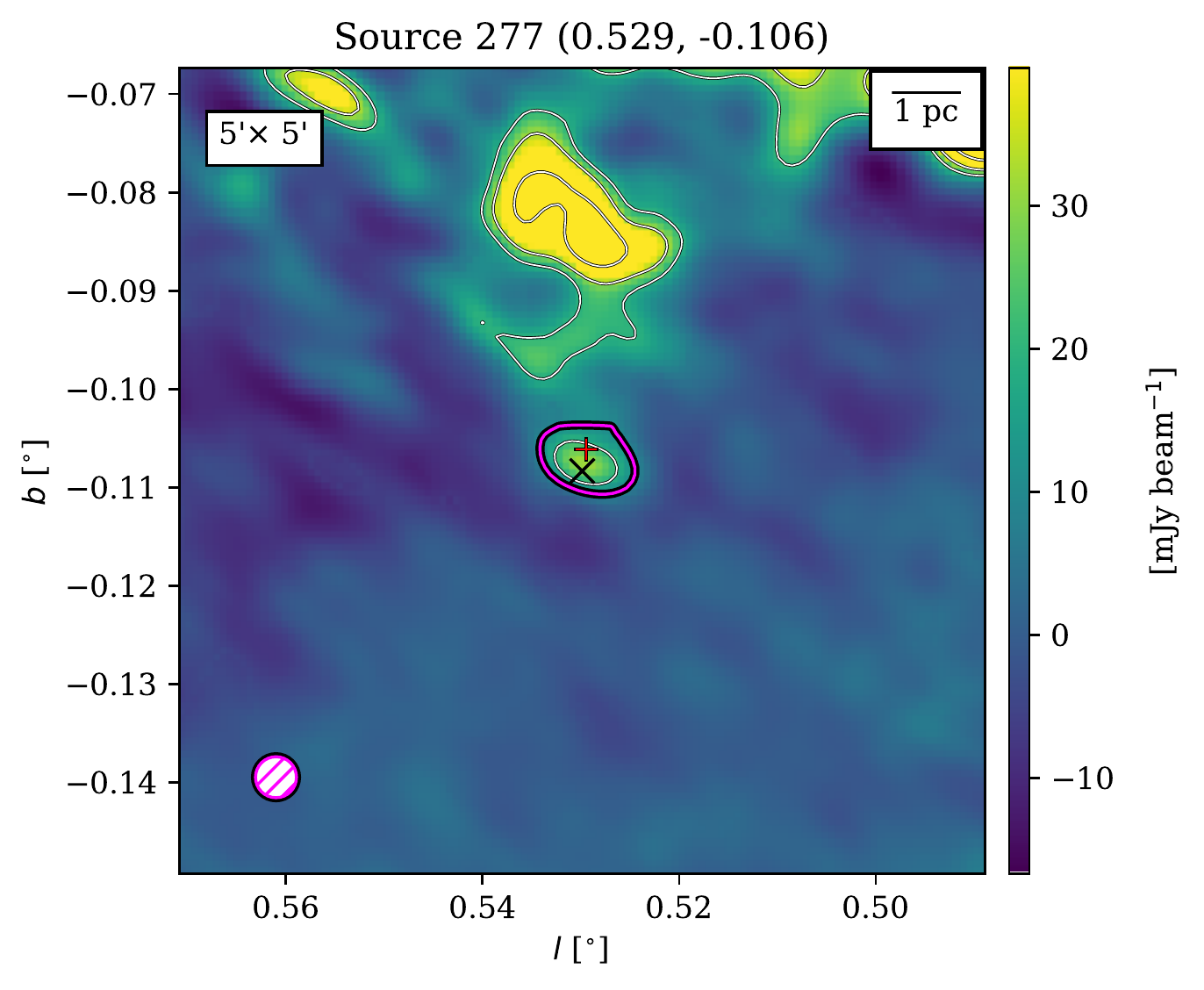}\\
\includegraphics[width=0.48\textwidth]{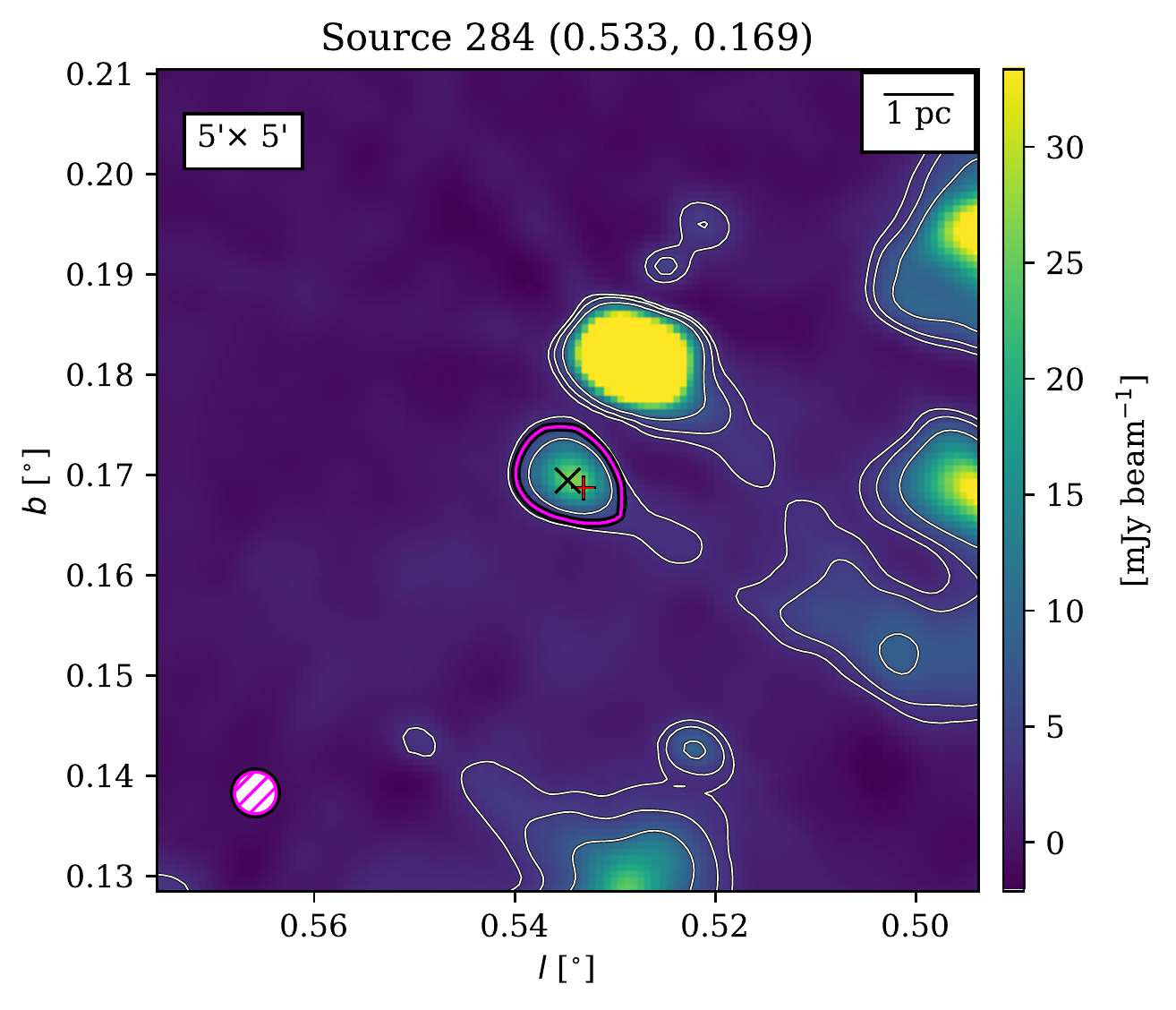}
\includegraphics[width=0.48\textwidth]{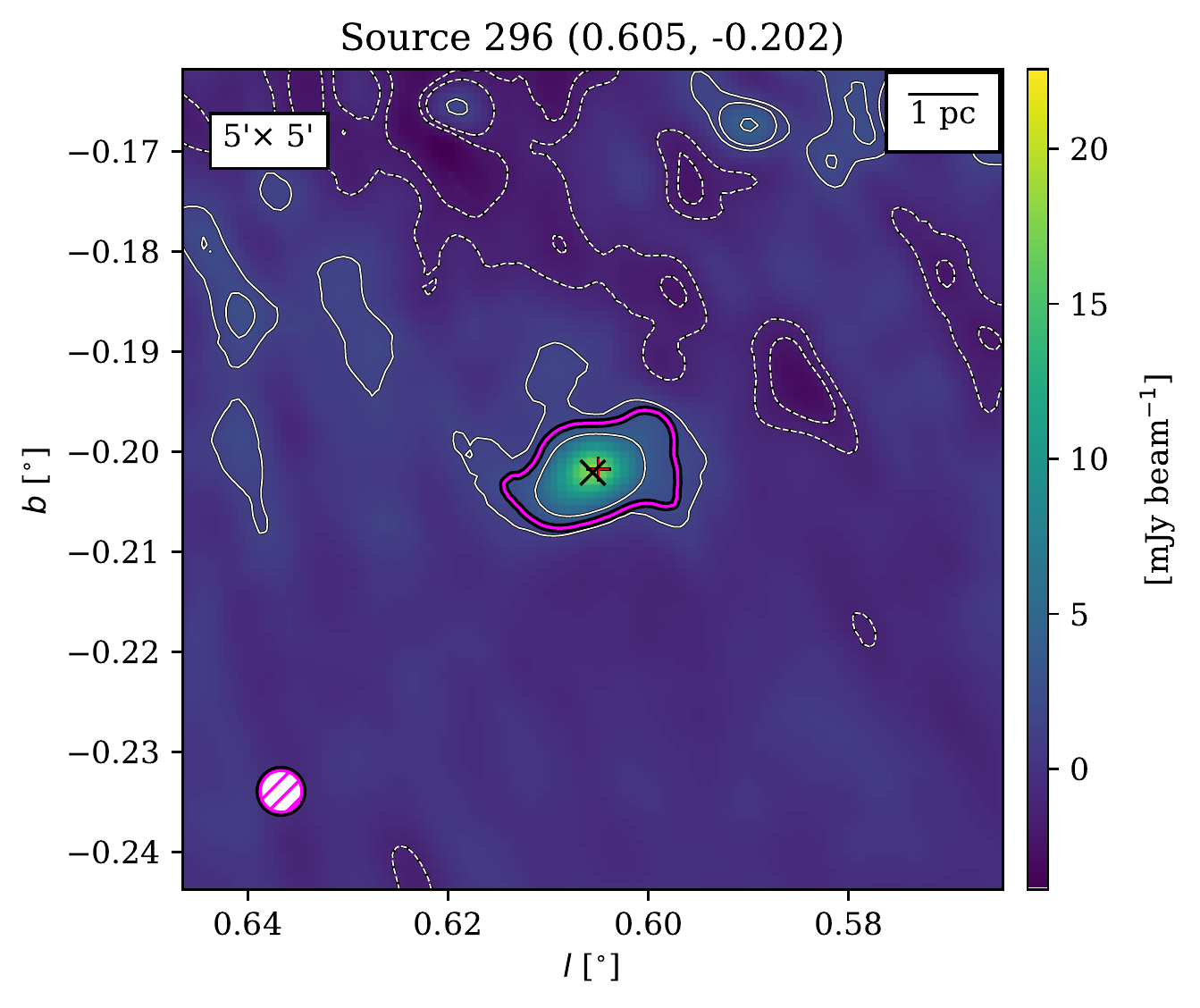}\\
\end{tabular}
\caption{Continued from Fig.~\ref{fig:manySource1}.}
\label{fig:manySource5}
\end{figure*}

\begin{figure*}[!h]
\begin{tabular}{cc}
\includegraphics[width=0.48\textwidth]{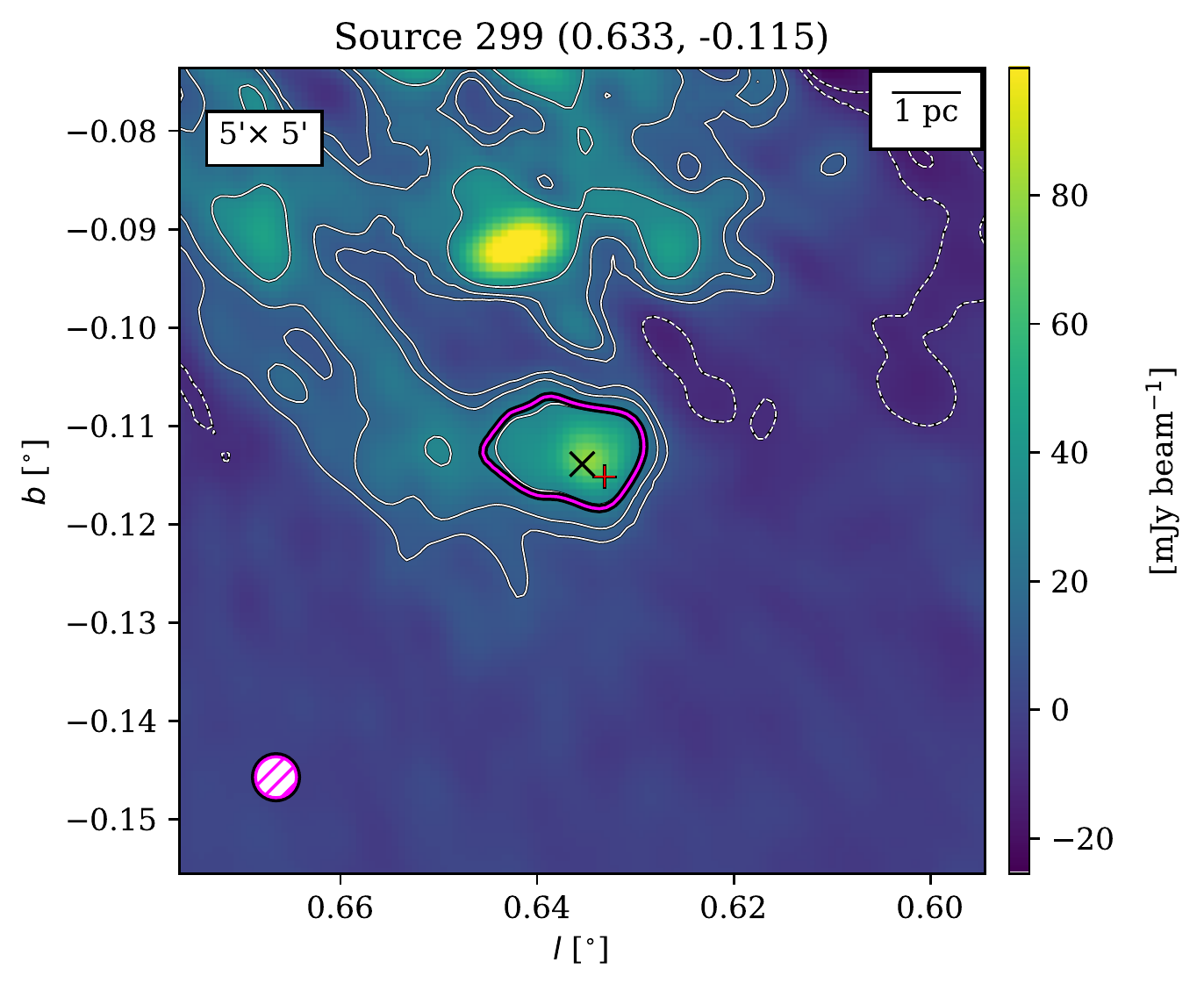}
\includegraphics[width=0.48\textwidth]{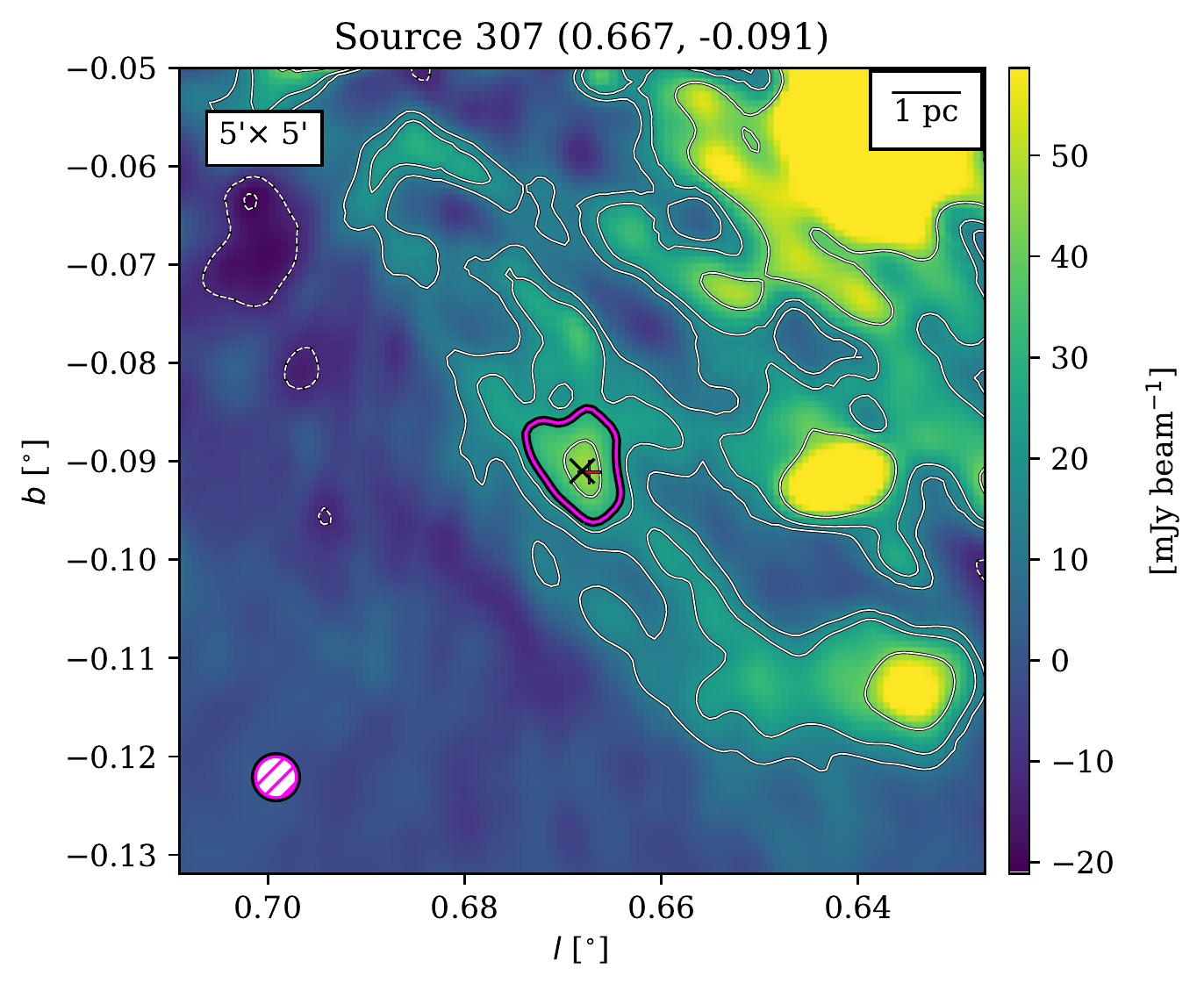}\\
\includegraphics[width=0.48\textwidth]{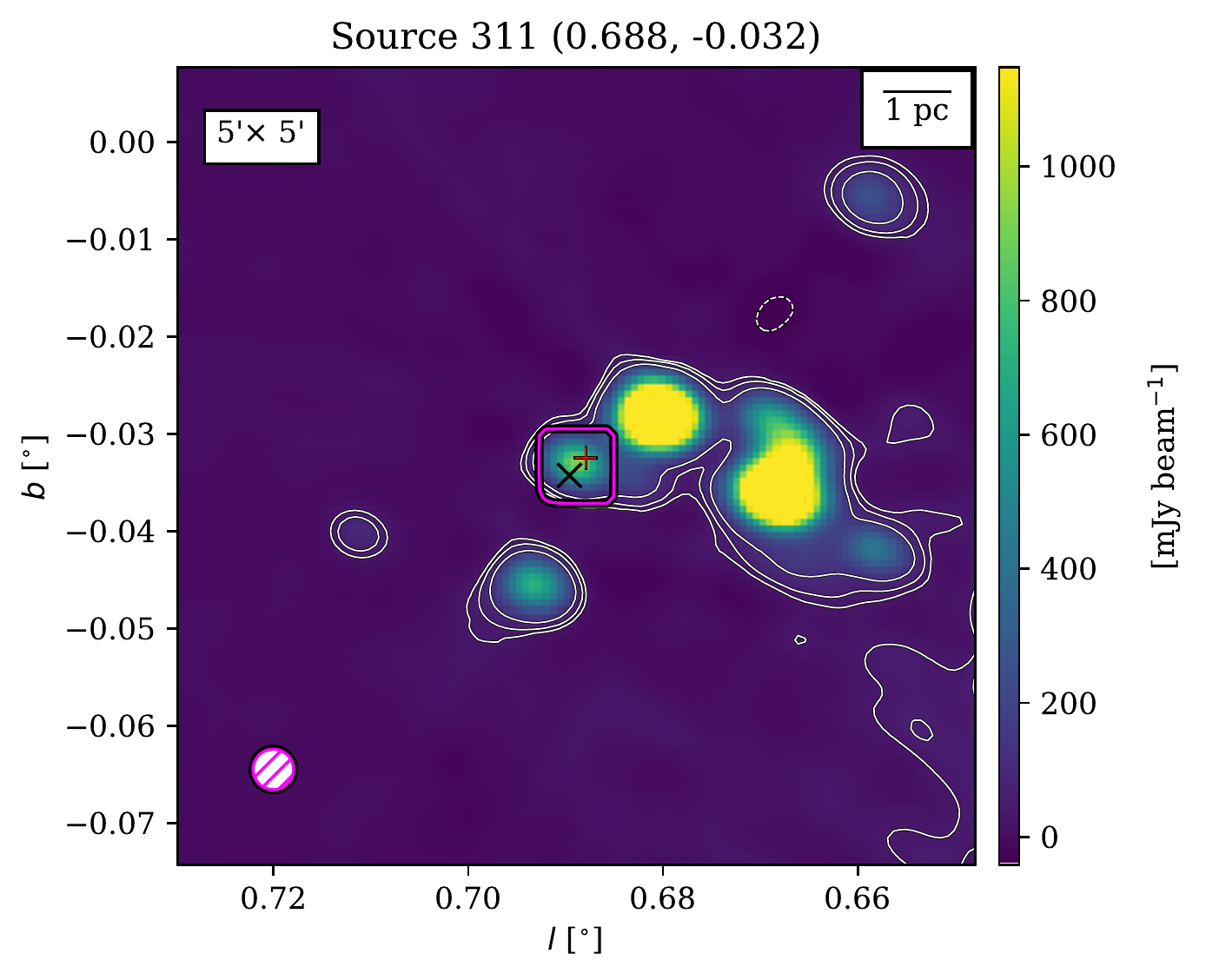}
\includegraphics[width=0.48\textwidth]{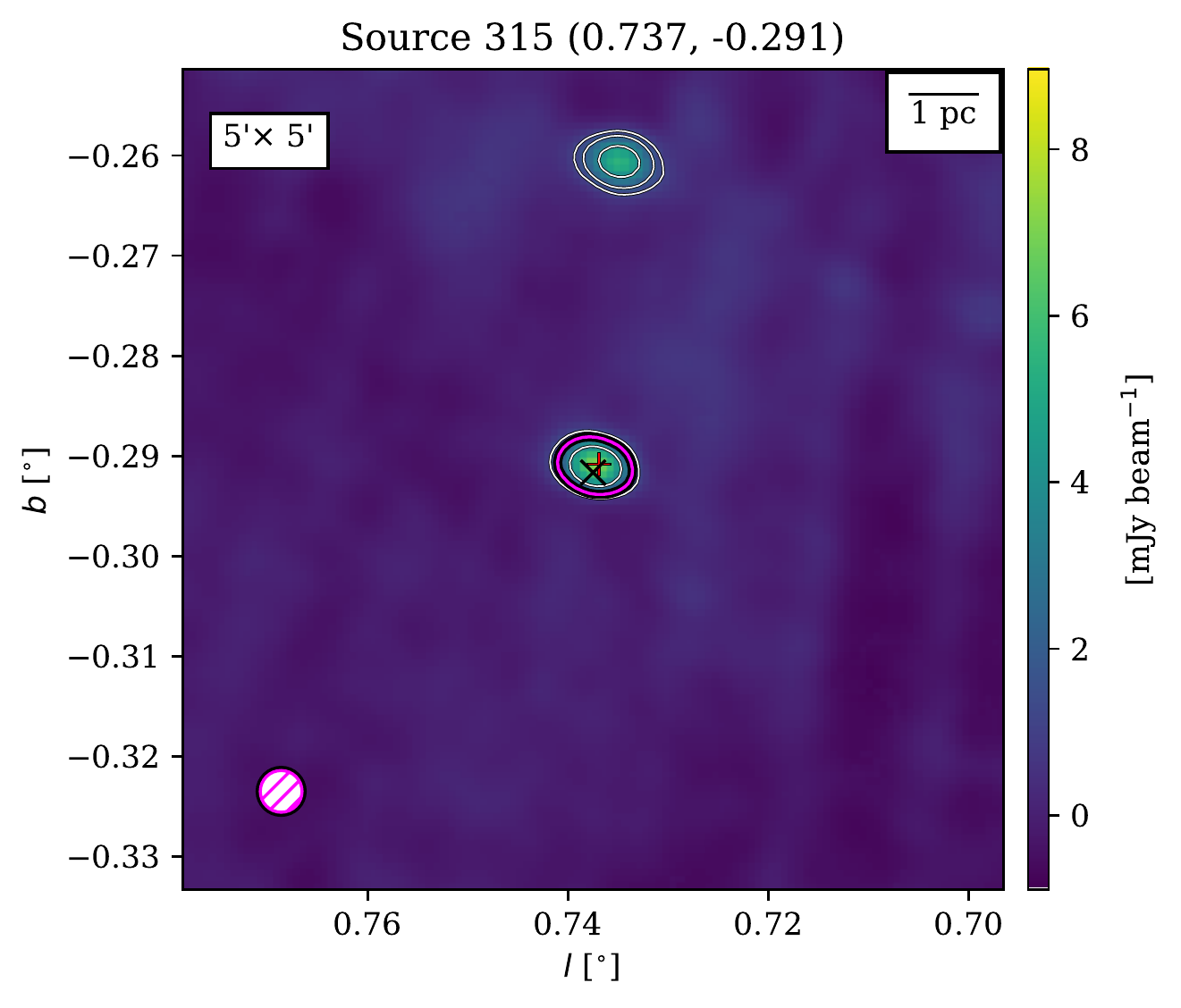}\\
\includegraphics[width=0.48\textwidth]{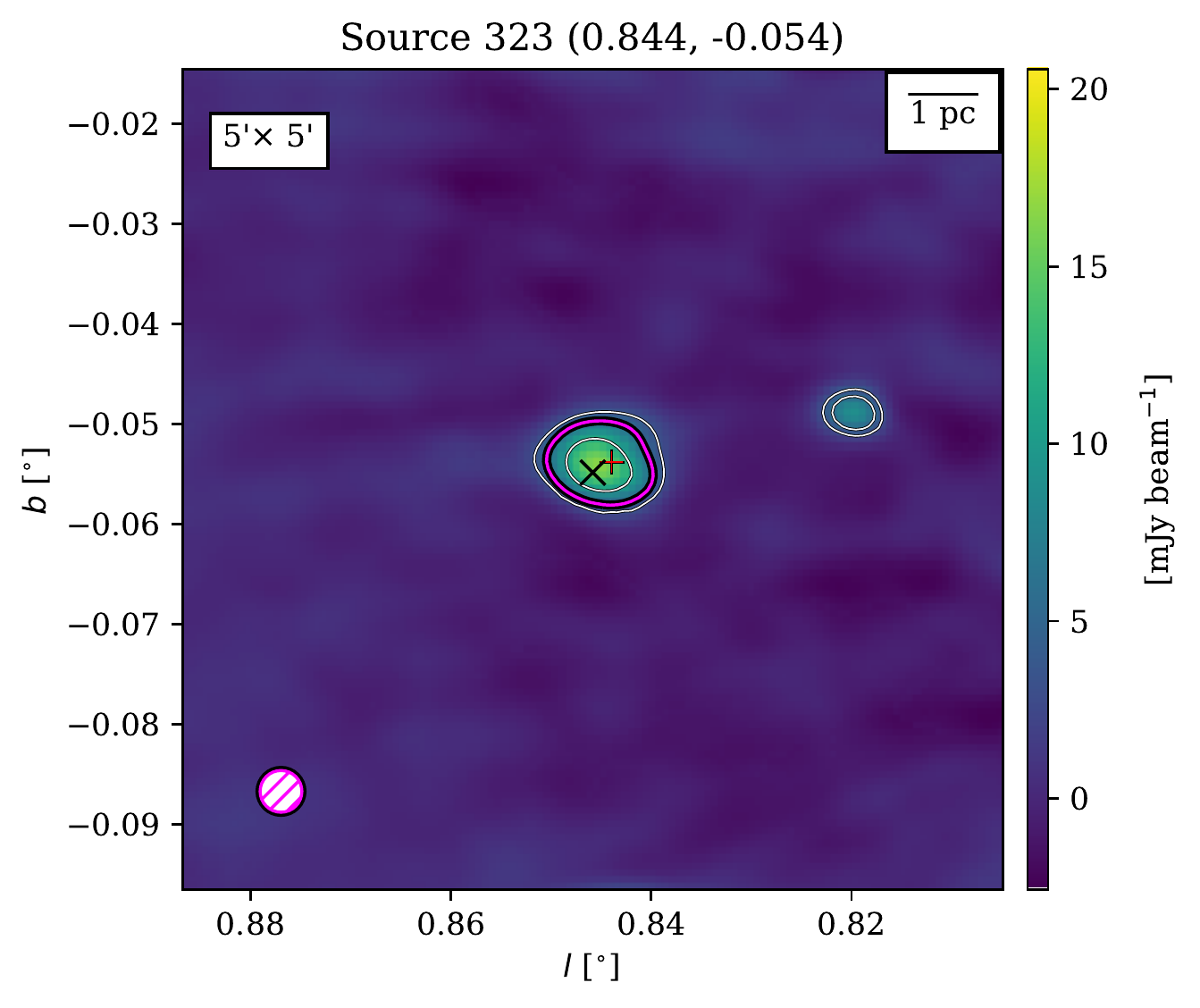}\\
\end{tabular}
\caption{Continued from Fig.~\ref{fig:manySource1}.}
\label{fig:manySource6}
\end{figure*}

\newpage
\section{Other spectral index images}\label{app:specIndex}
Here we provide an example of the different frequency channels used in the spectral index determination for a given source, which shows how the morphology is slightly different in each channel and therefore necessitates using only the peak flux for calculations.

\begin{figure*}[!h]
\begin{tabular}{cc}
\includegraphics[width=0.95\textwidth]{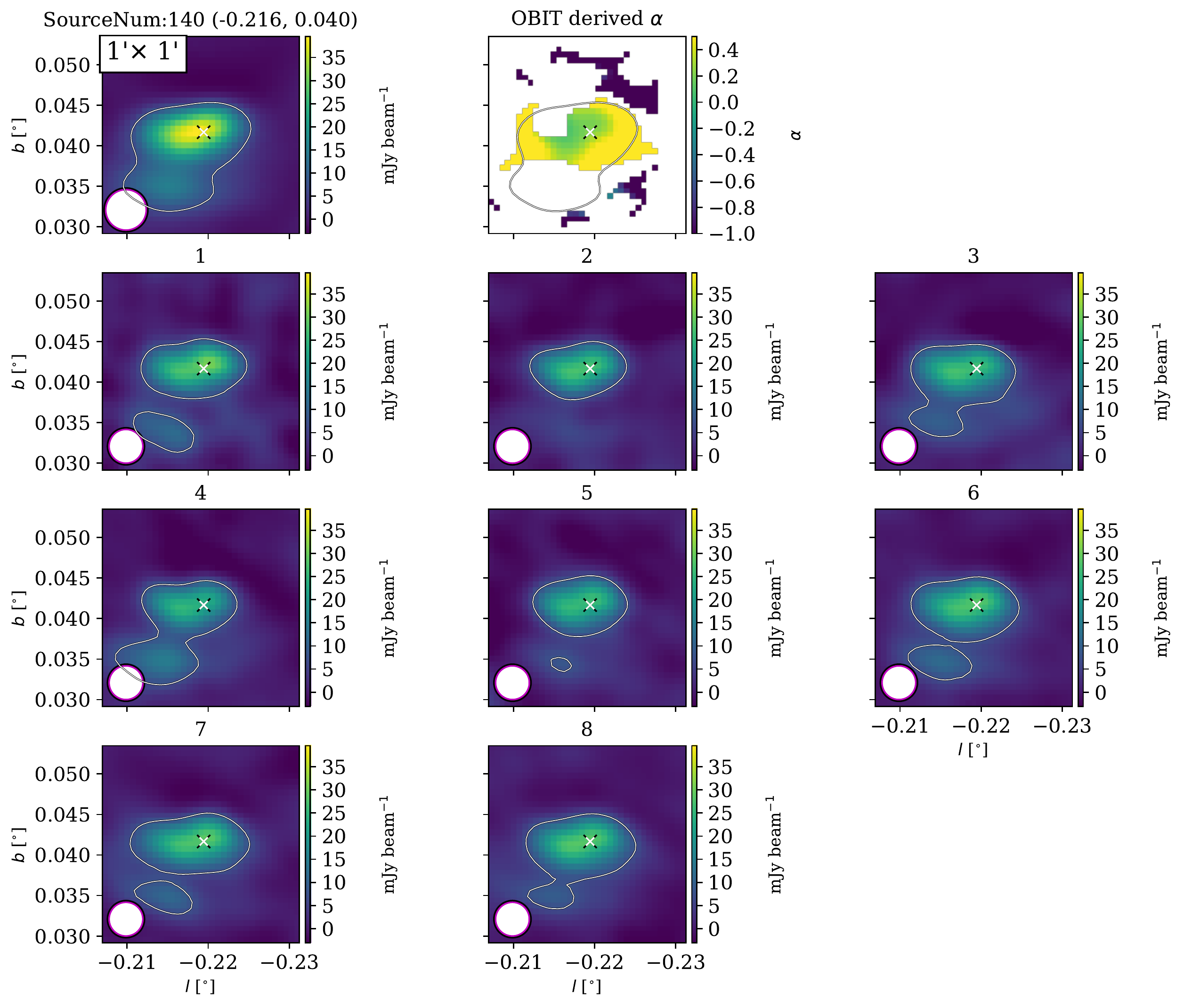}
\end{tabular}
\caption{GLOSTAR continuum images of source 140 used for the determination of the spectral index. The top-left image is the combined image at 5.8\,GHz, while the remaining 1--8 are from the individual frequency bands. Band 9 is omitted due to high noise. Shown also is the spectral index map produced by OBIT. The contour corresponds to the 5$\sigma$ level of the combined image (7.608\,mJy\,beam$^{-1}$) and used for comparison at each frequency. The combined contour is also overplotted atop the spectral index map. The `x' marker denotes the position of the peak pixel from the combined image. The calculated spectral index for this source is $\alpha =0.23\pm0.09$.}
\label{fig:specIndex_example}
\end{figure*}

\begin{figure*}[!tbhp]
\begin{tabular}{cccc}
\includegraphics[width=0.24\textwidth]{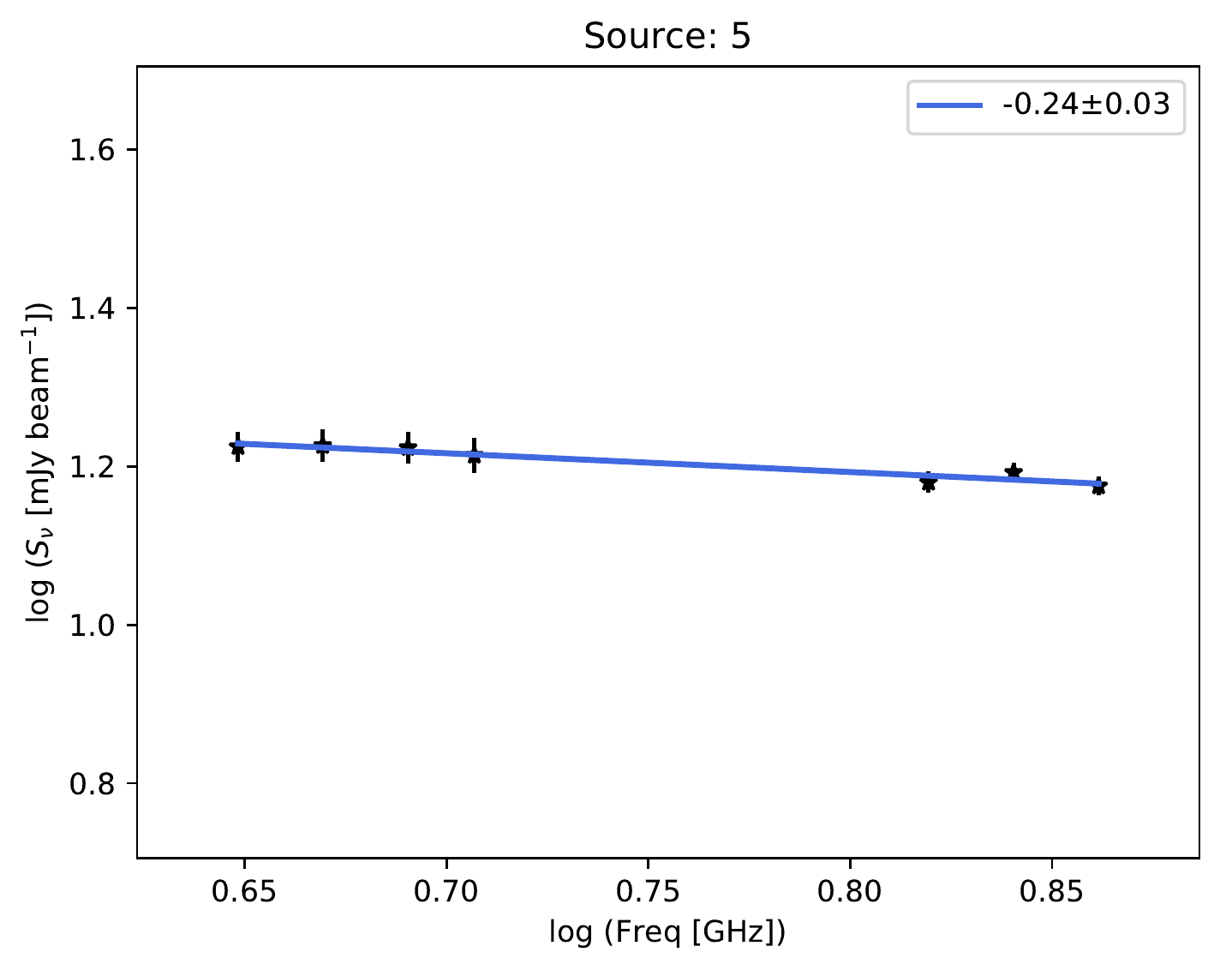}
\includegraphics[width=0.24\textwidth]{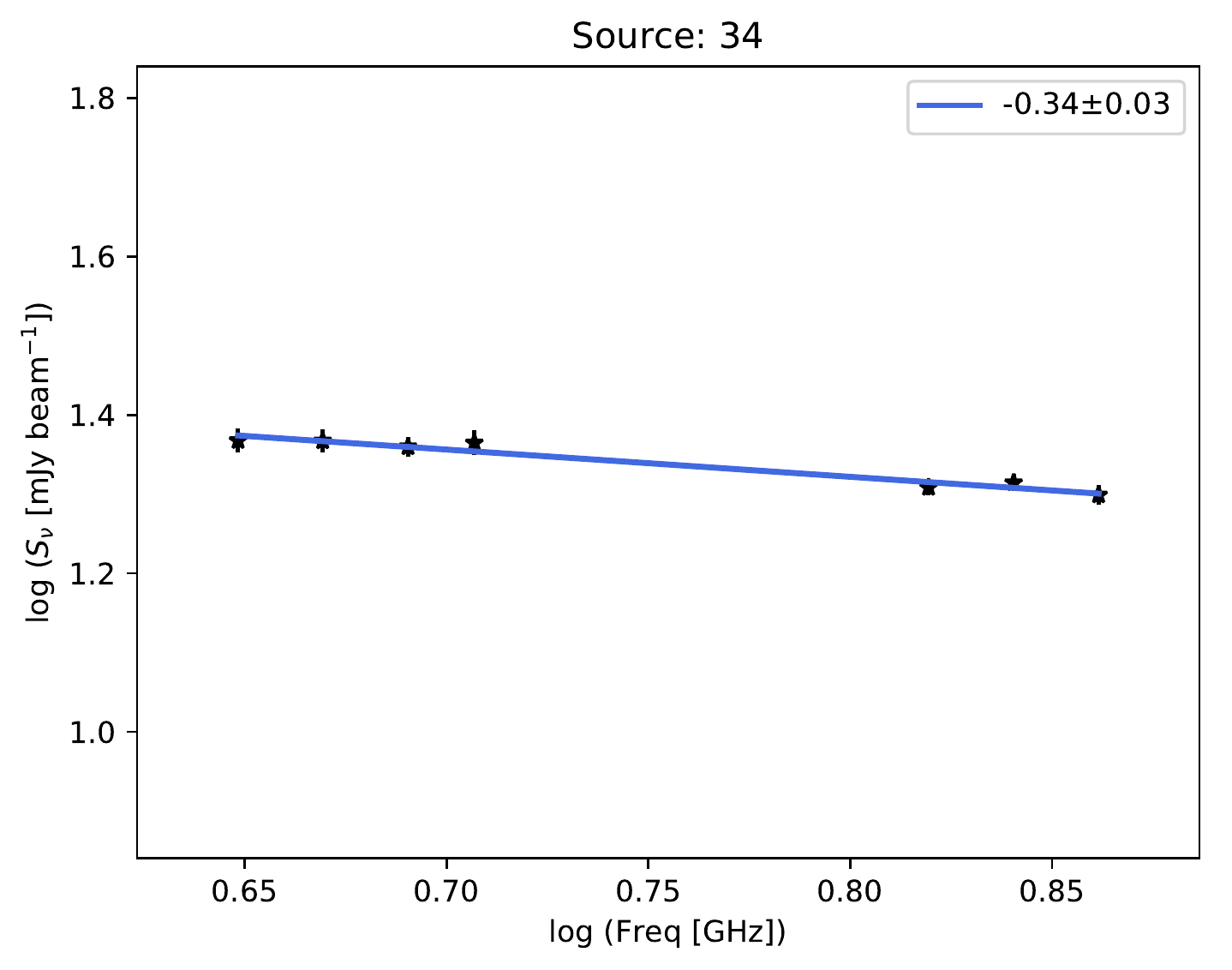}
\includegraphics[width=0.24\textwidth]{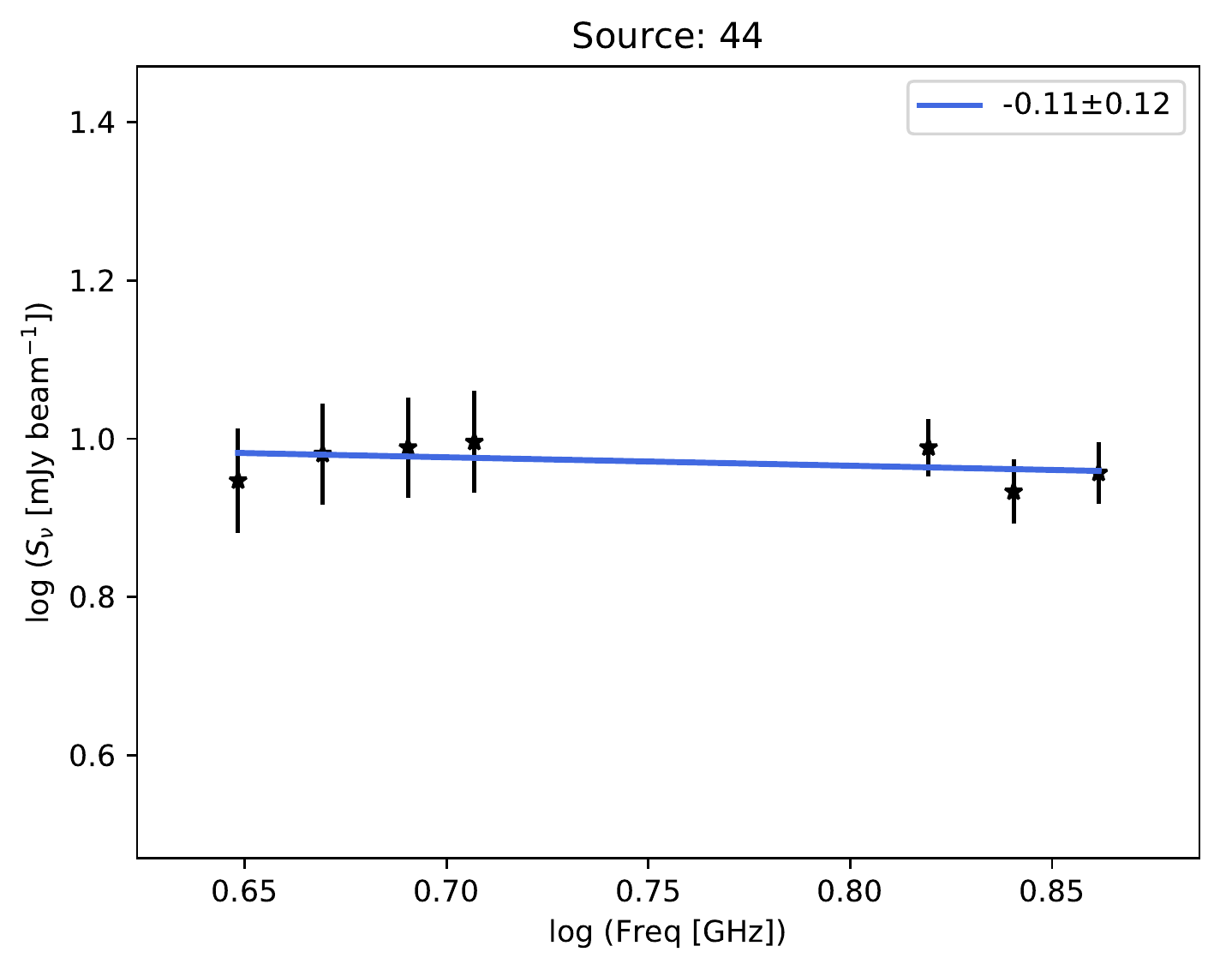}
\includegraphics[width=0.24\textwidth]{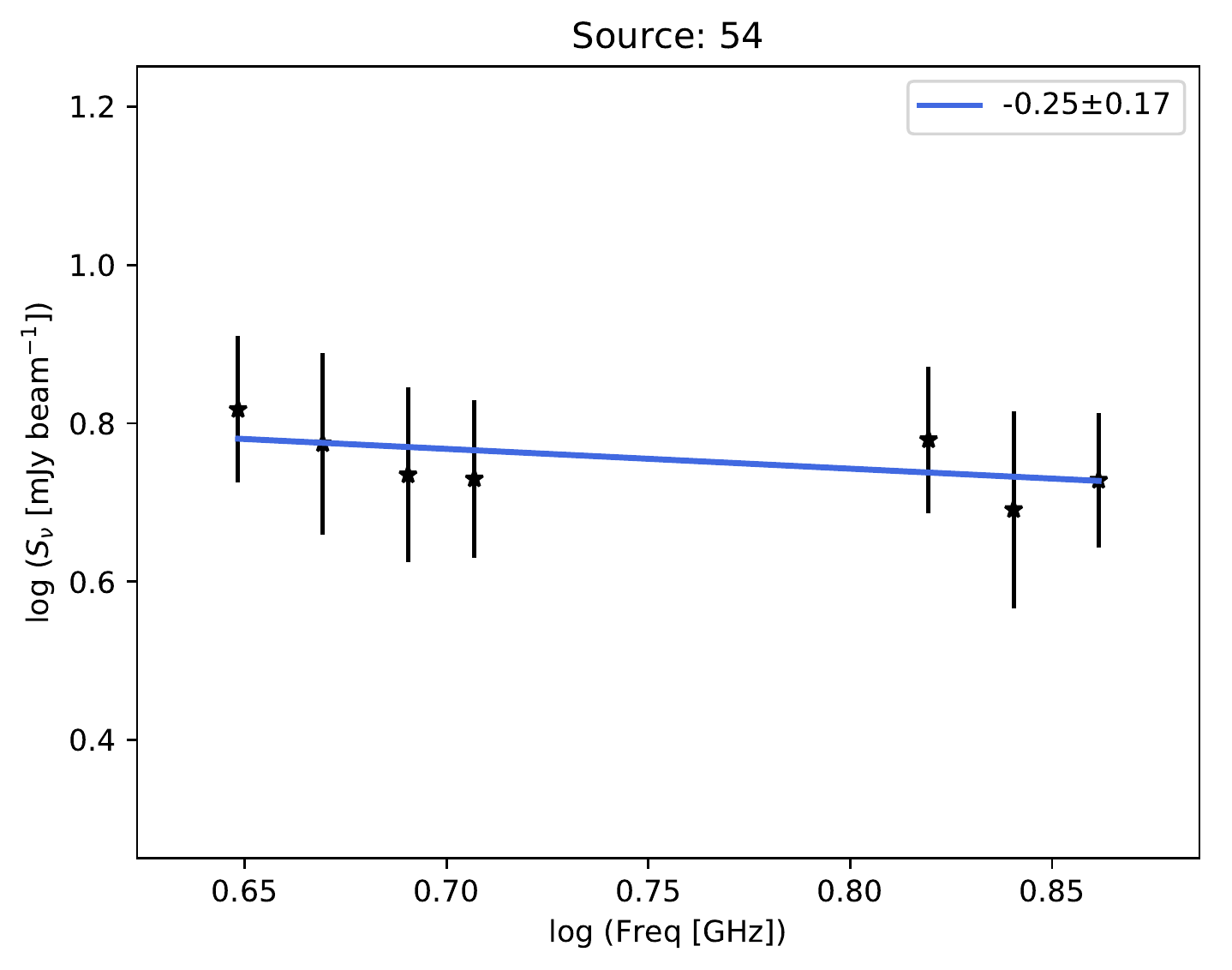}\\
\includegraphics[width=0.24\textwidth]{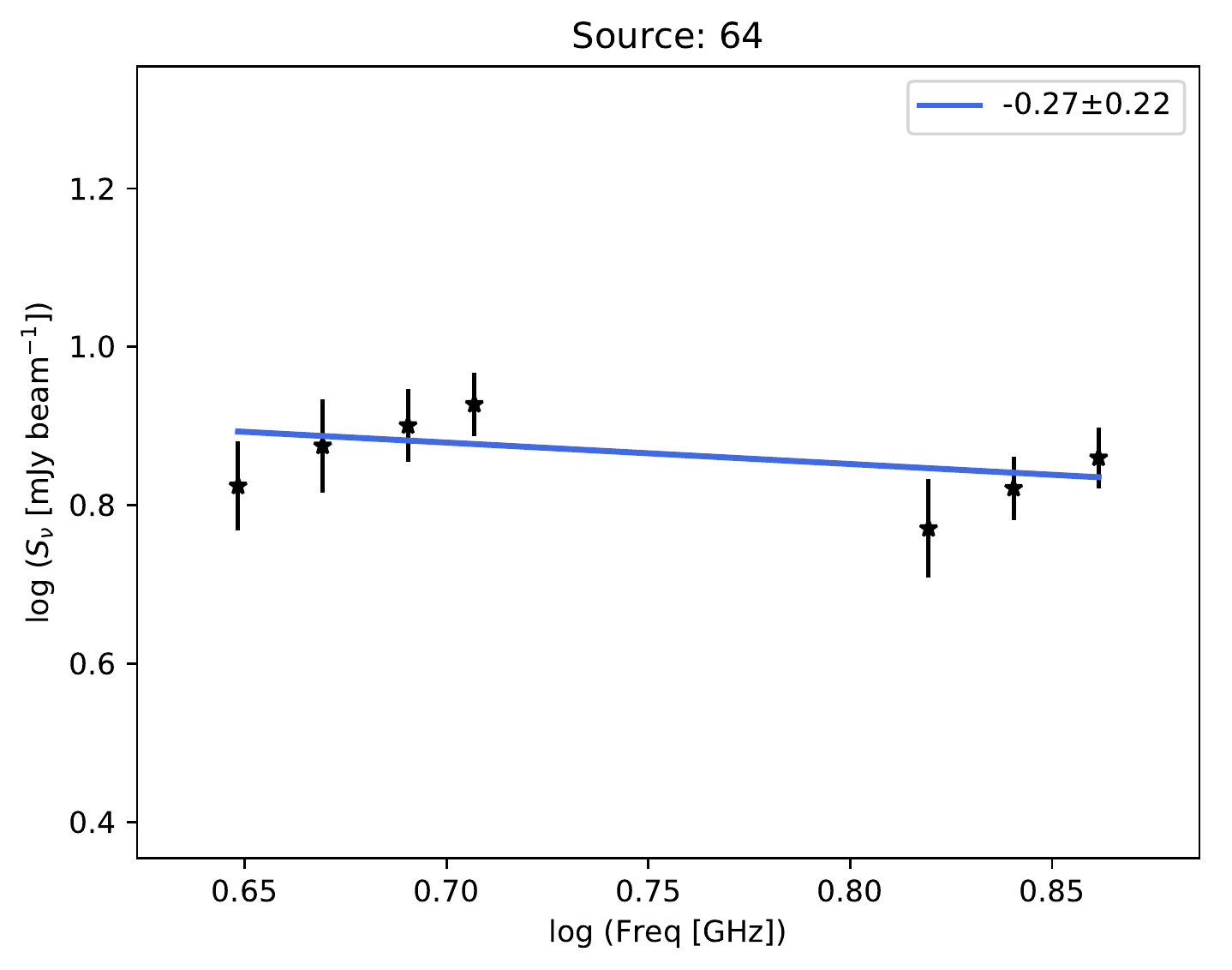}
\includegraphics[width=0.24\textwidth]{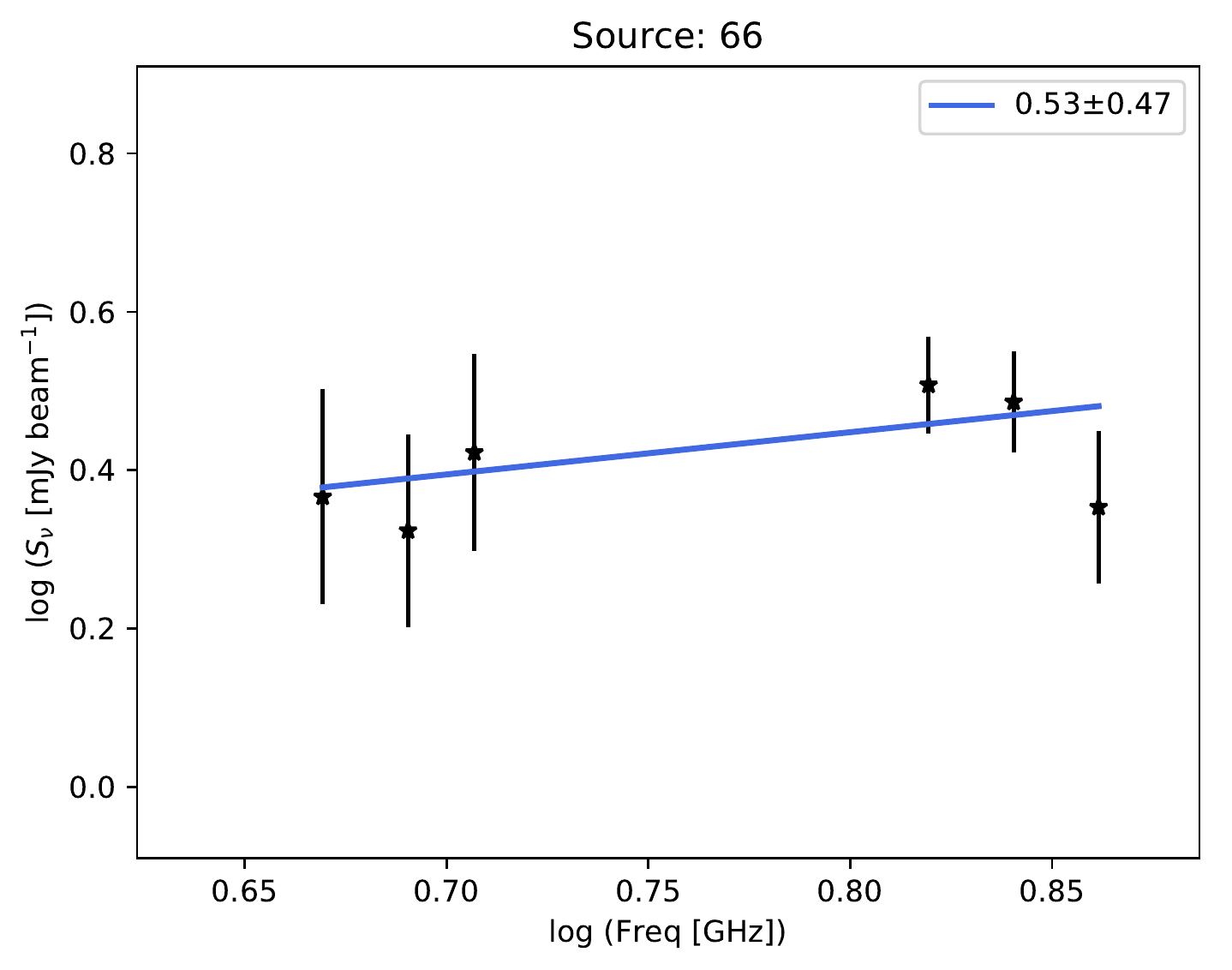}
\includegraphics[width=0.24\textwidth]{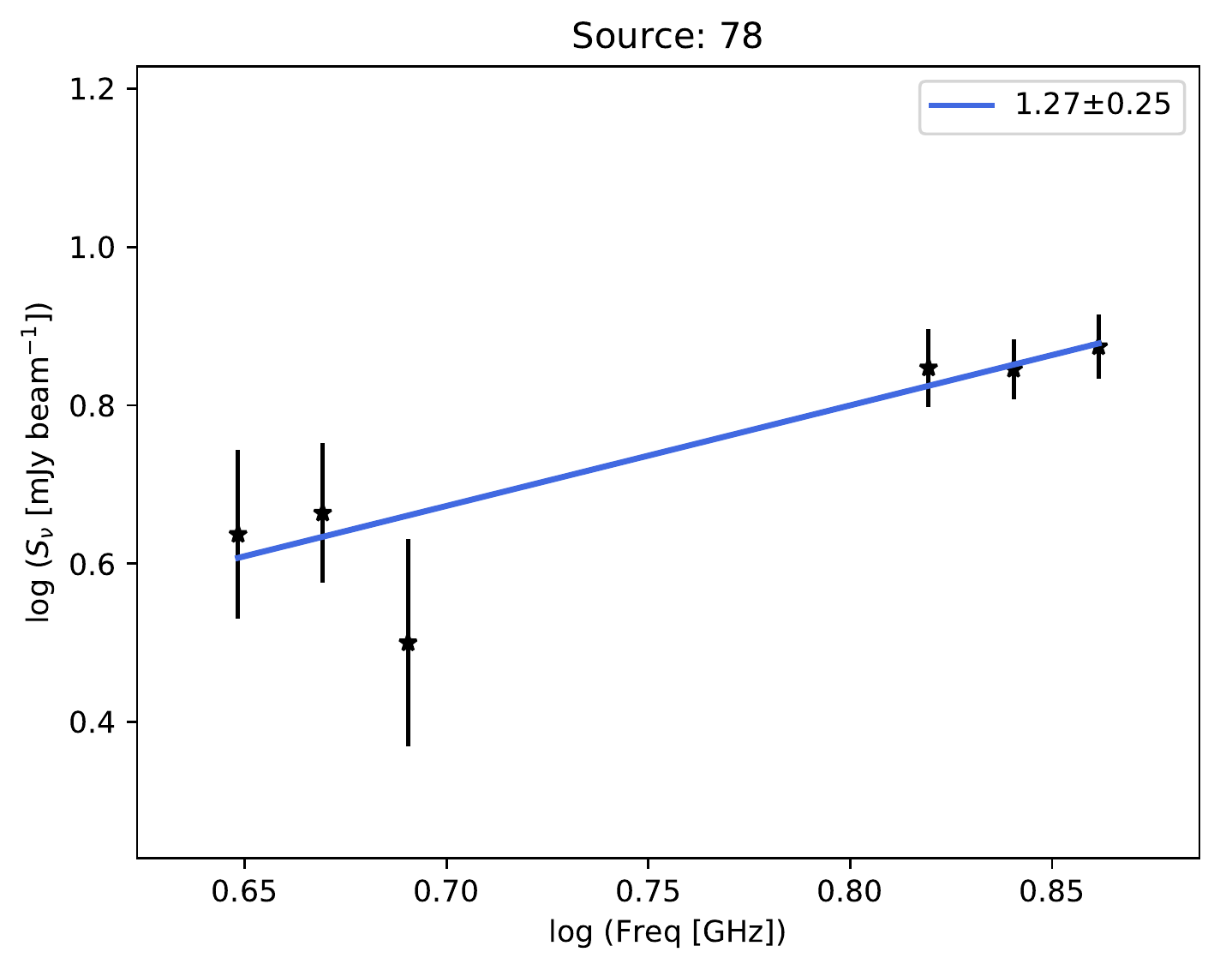}
\includegraphics[width=0.24\textwidth]{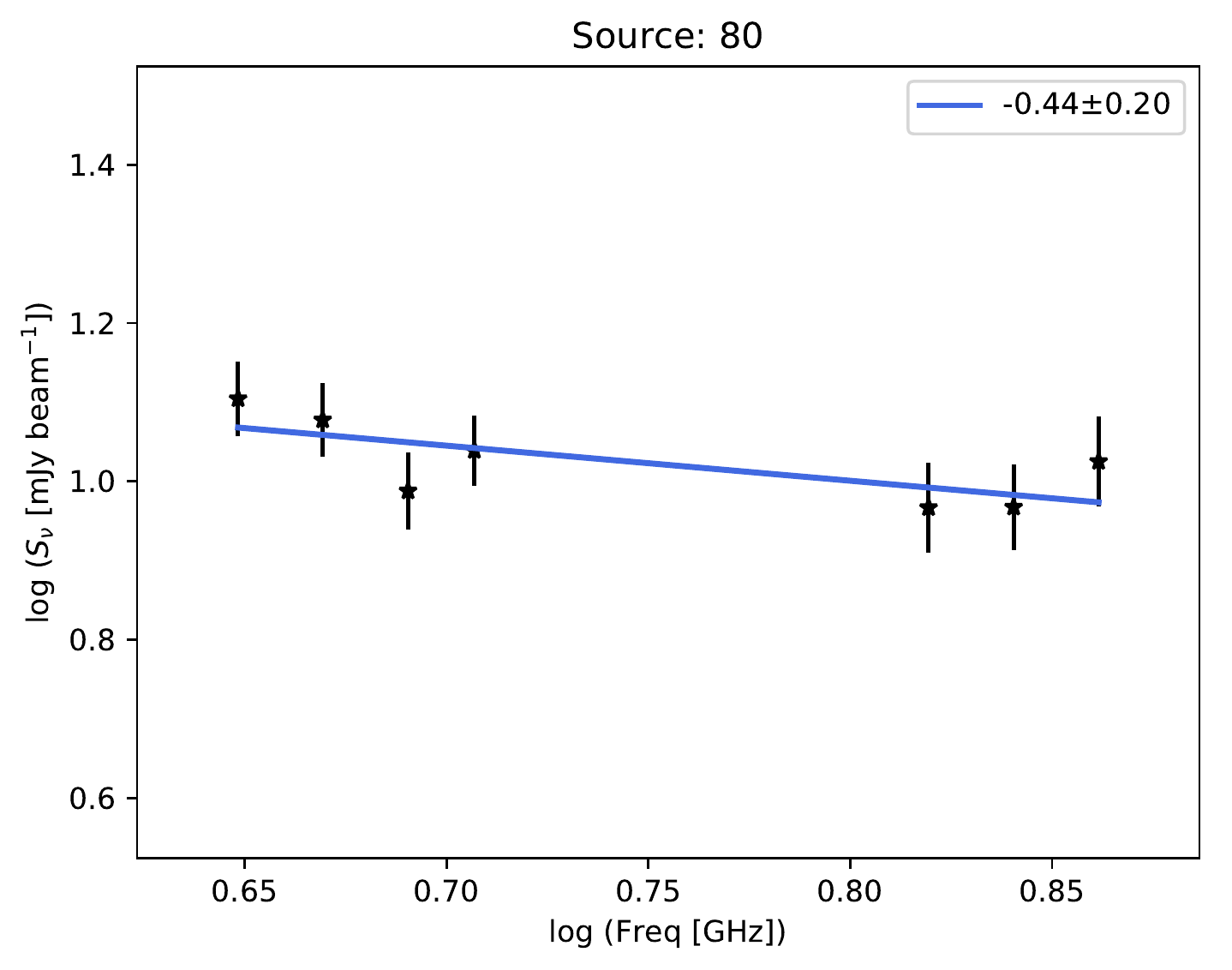}\\
\includegraphics[width=0.24\textwidth]{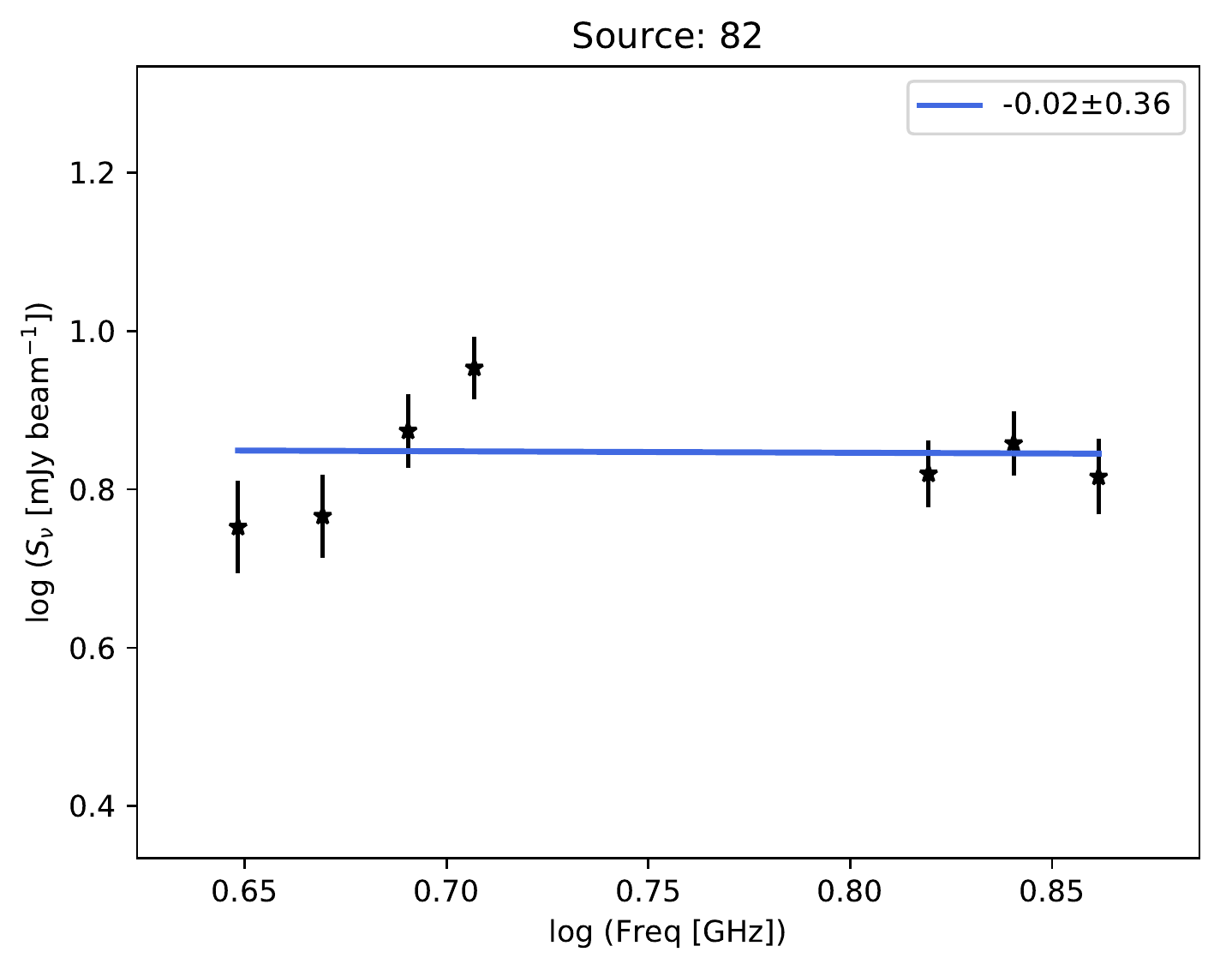}
\includegraphics[width=0.24\textwidth]{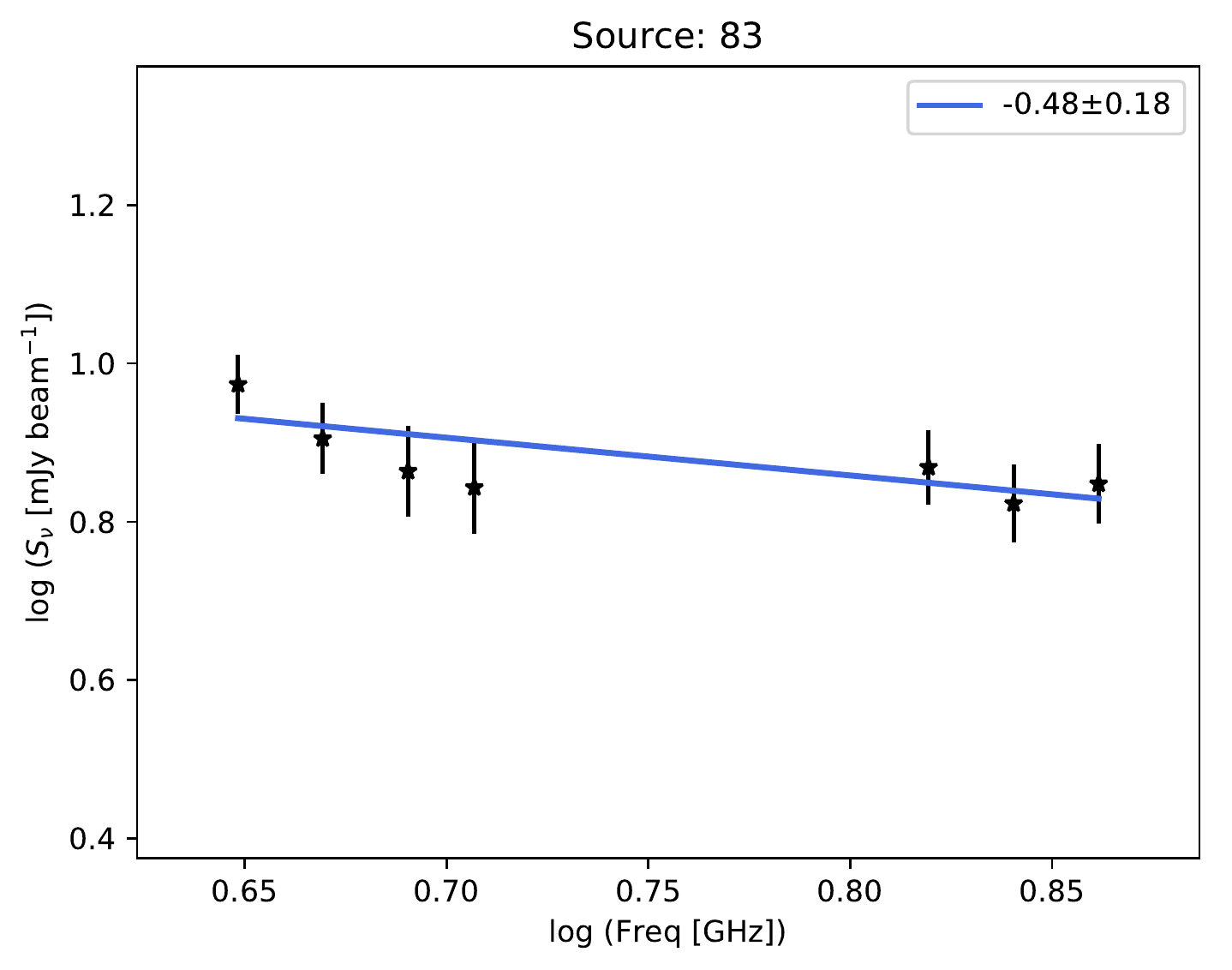}
\includegraphics[width=0.24\textwidth]{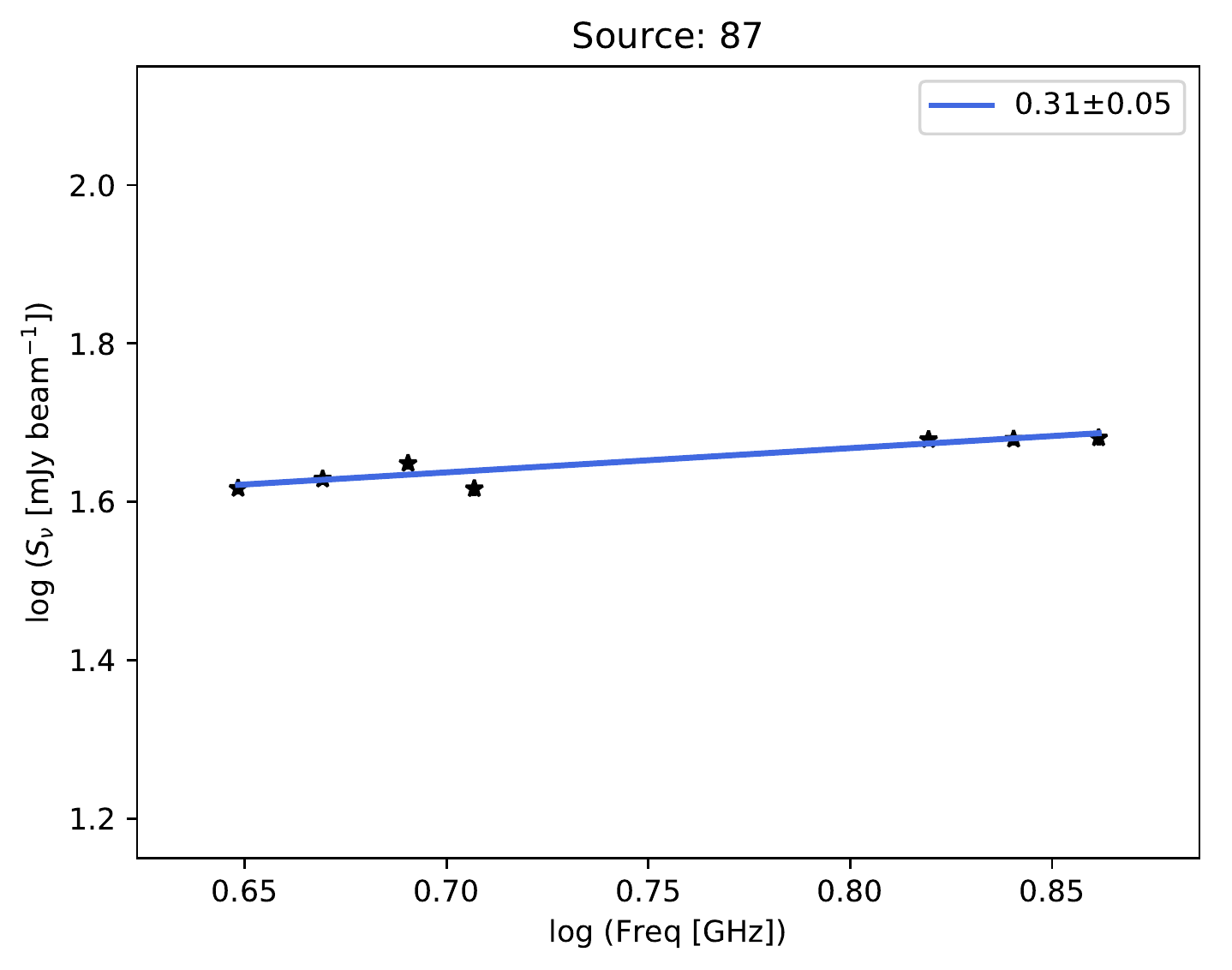}
\includegraphics[width=0.24\textwidth]{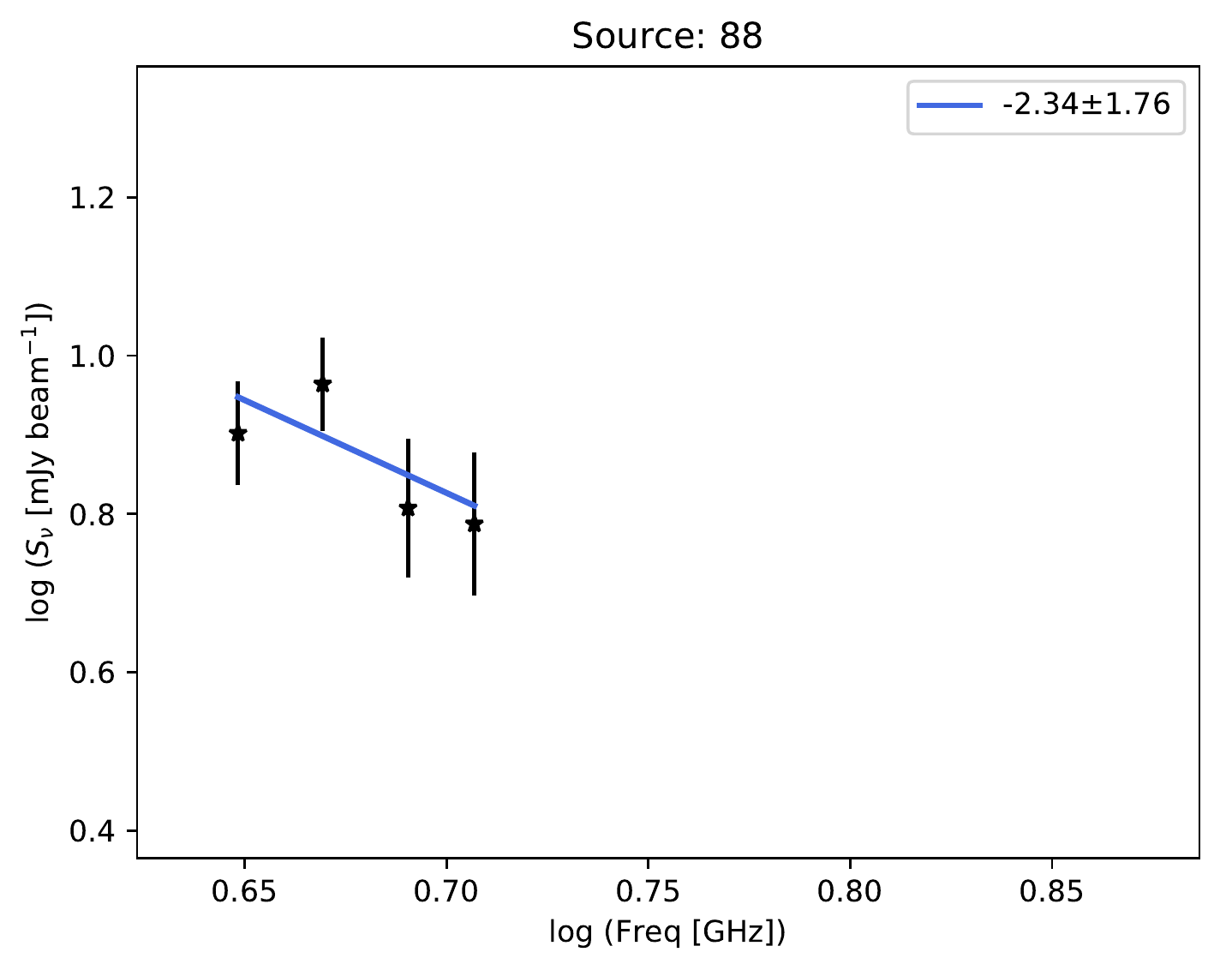}\\
\includegraphics[width=0.24\textwidth]{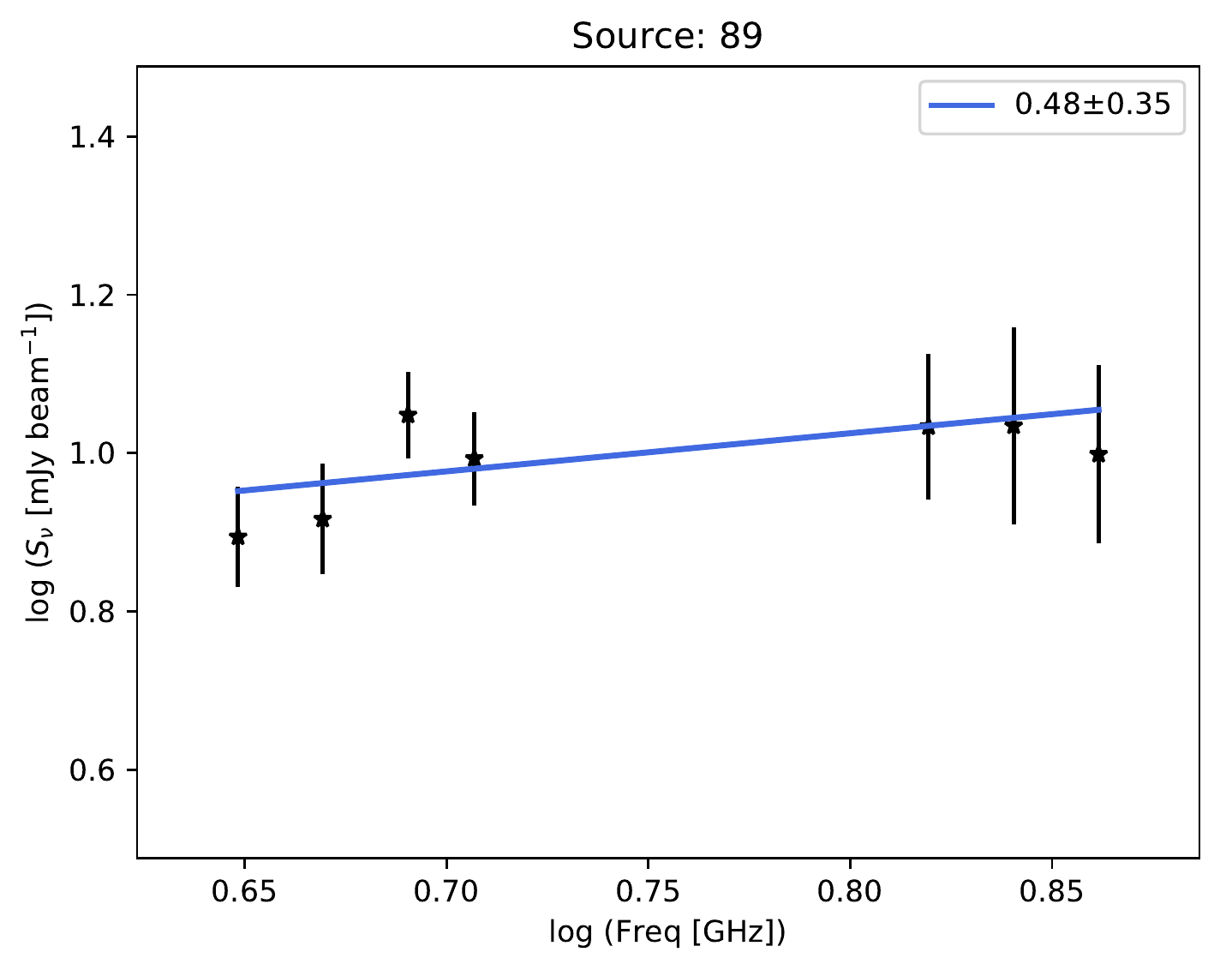}
\includegraphics[width=0.24\textwidth]{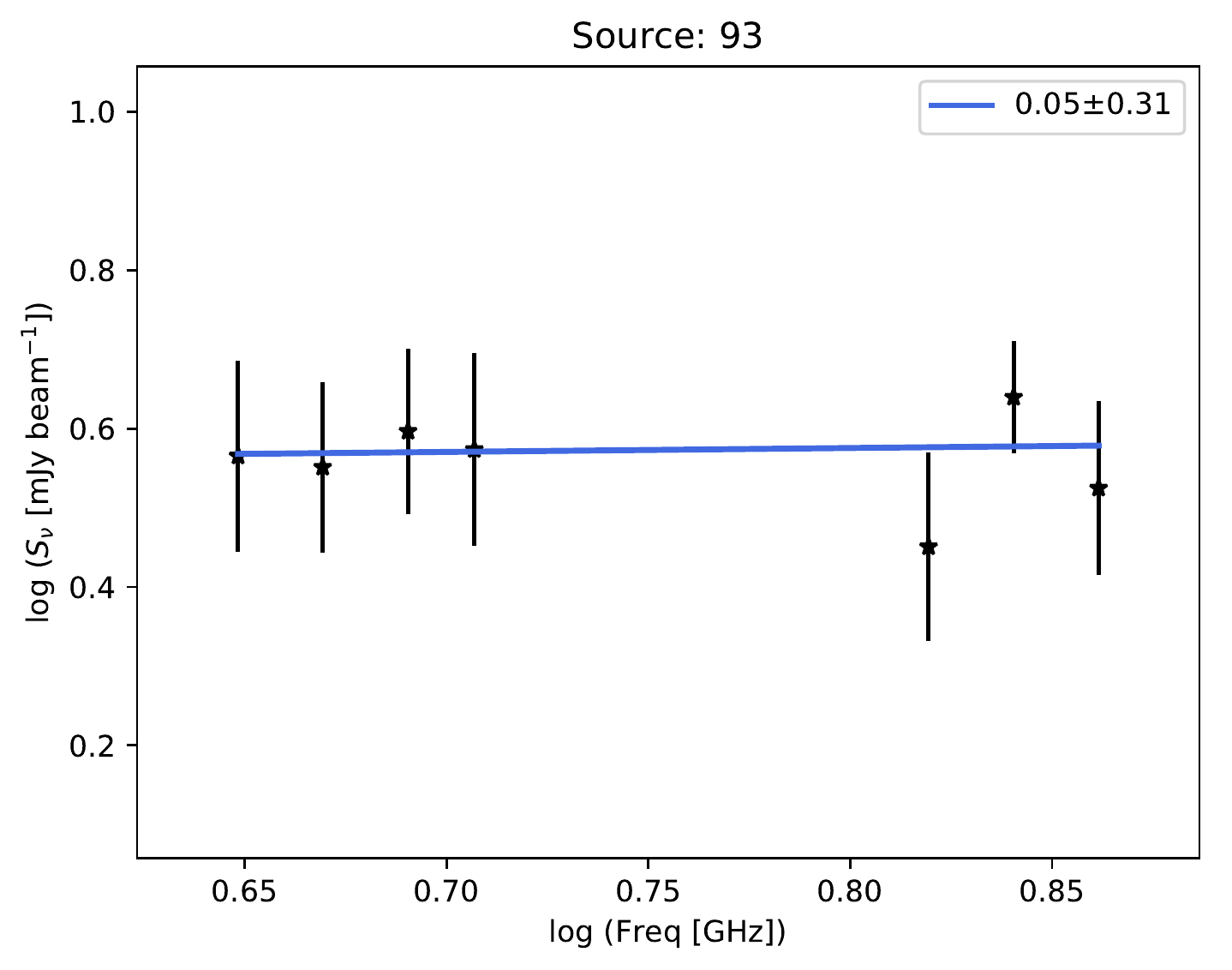}
\includegraphics[width=0.24\textwidth]{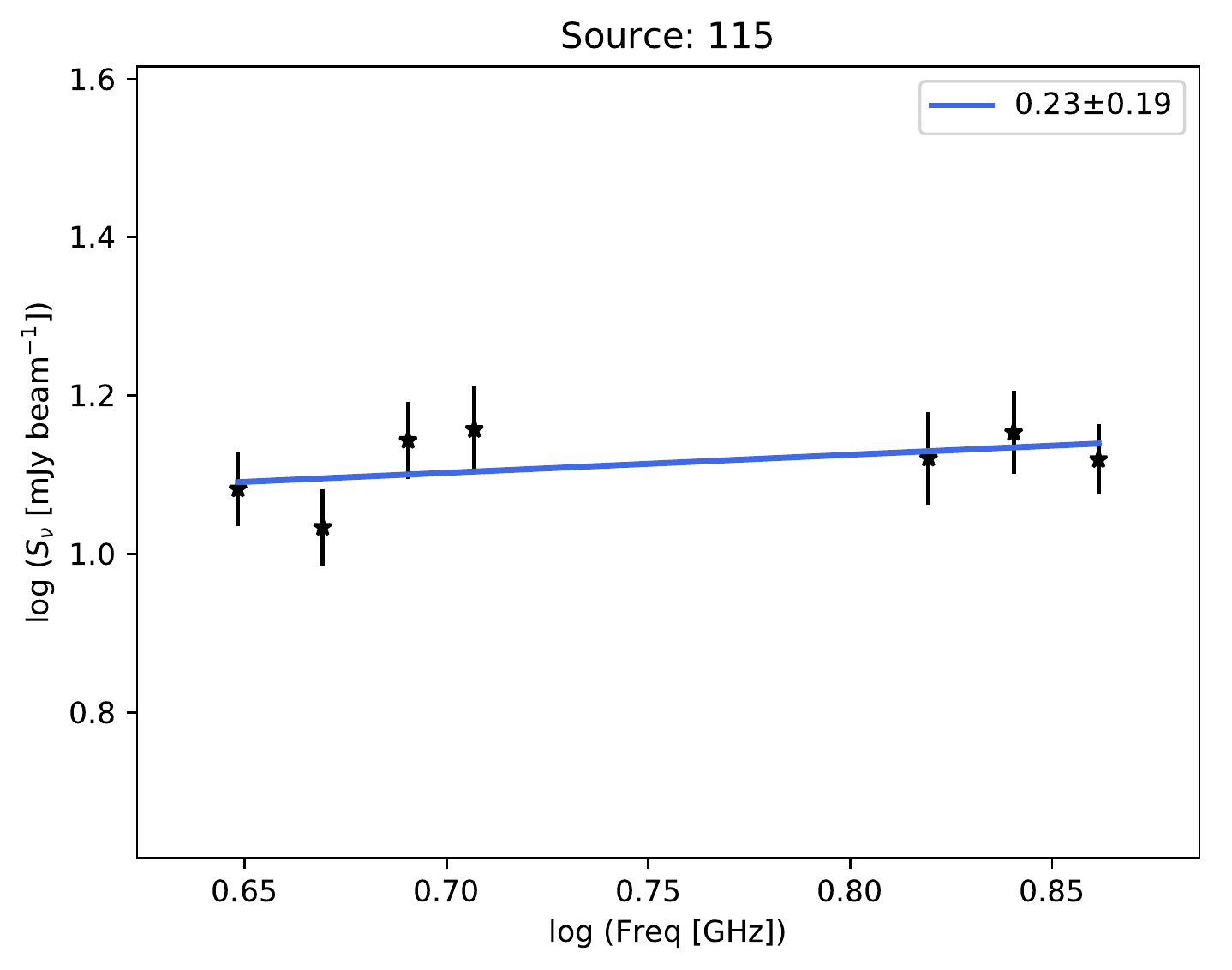}
\includegraphics[width=0.24\textwidth]{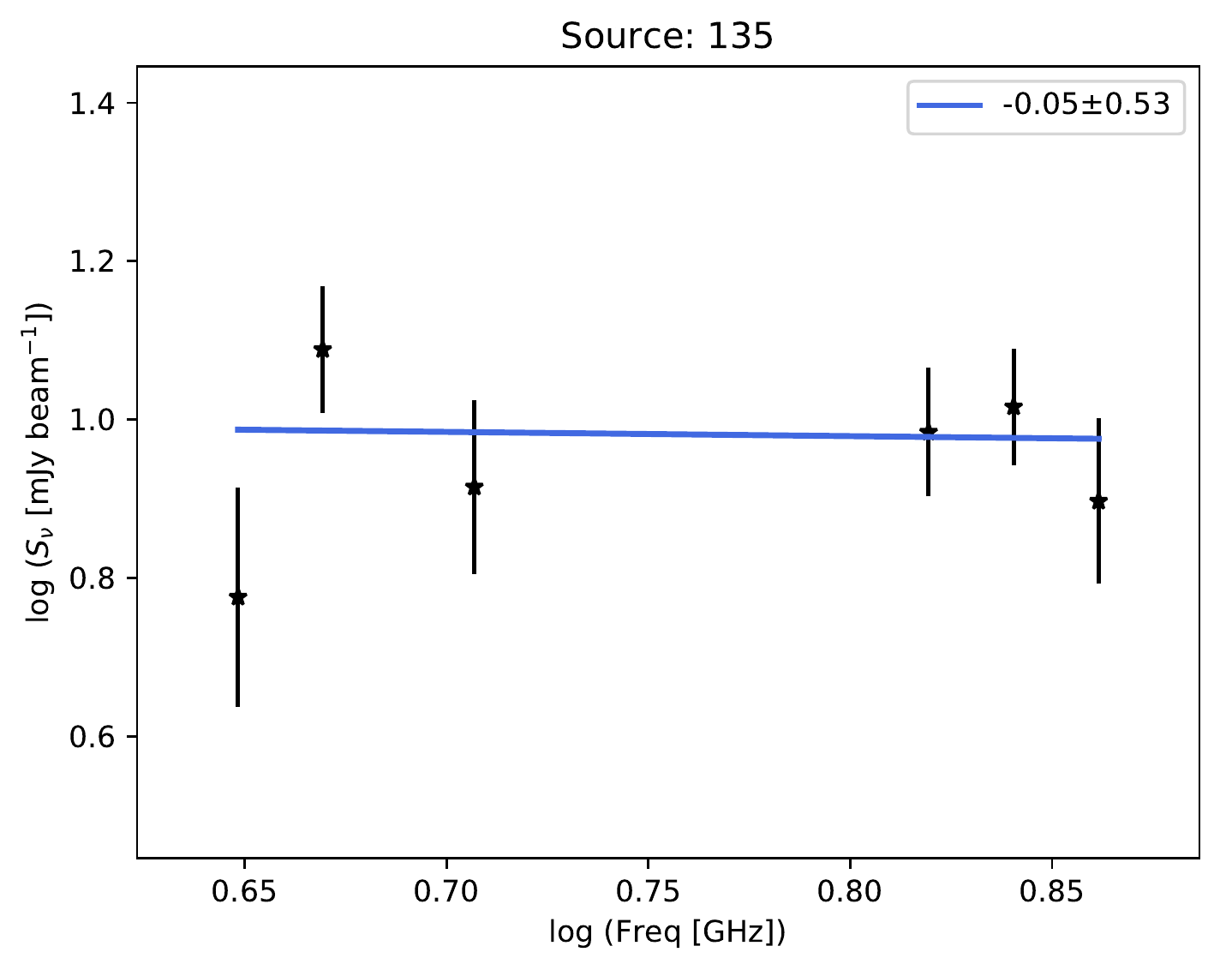}\\
\includegraphics[width=0.24\textwidth]{specIndex/LOG_spectral_index_fit_139_140.pdf}
\includegraphics[width=0.24\textwidth]{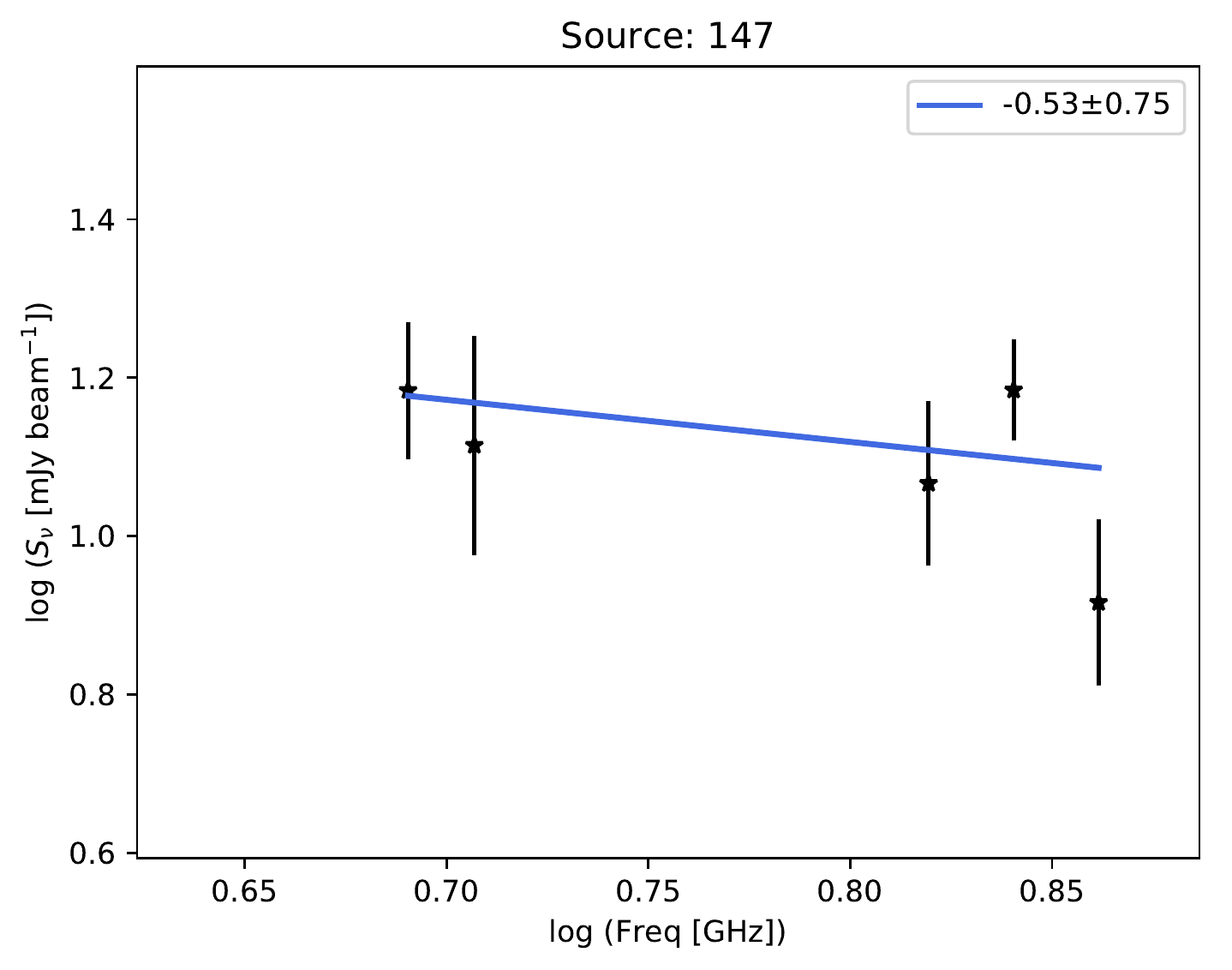}
\includegraphics[width=0.24\textwidth]{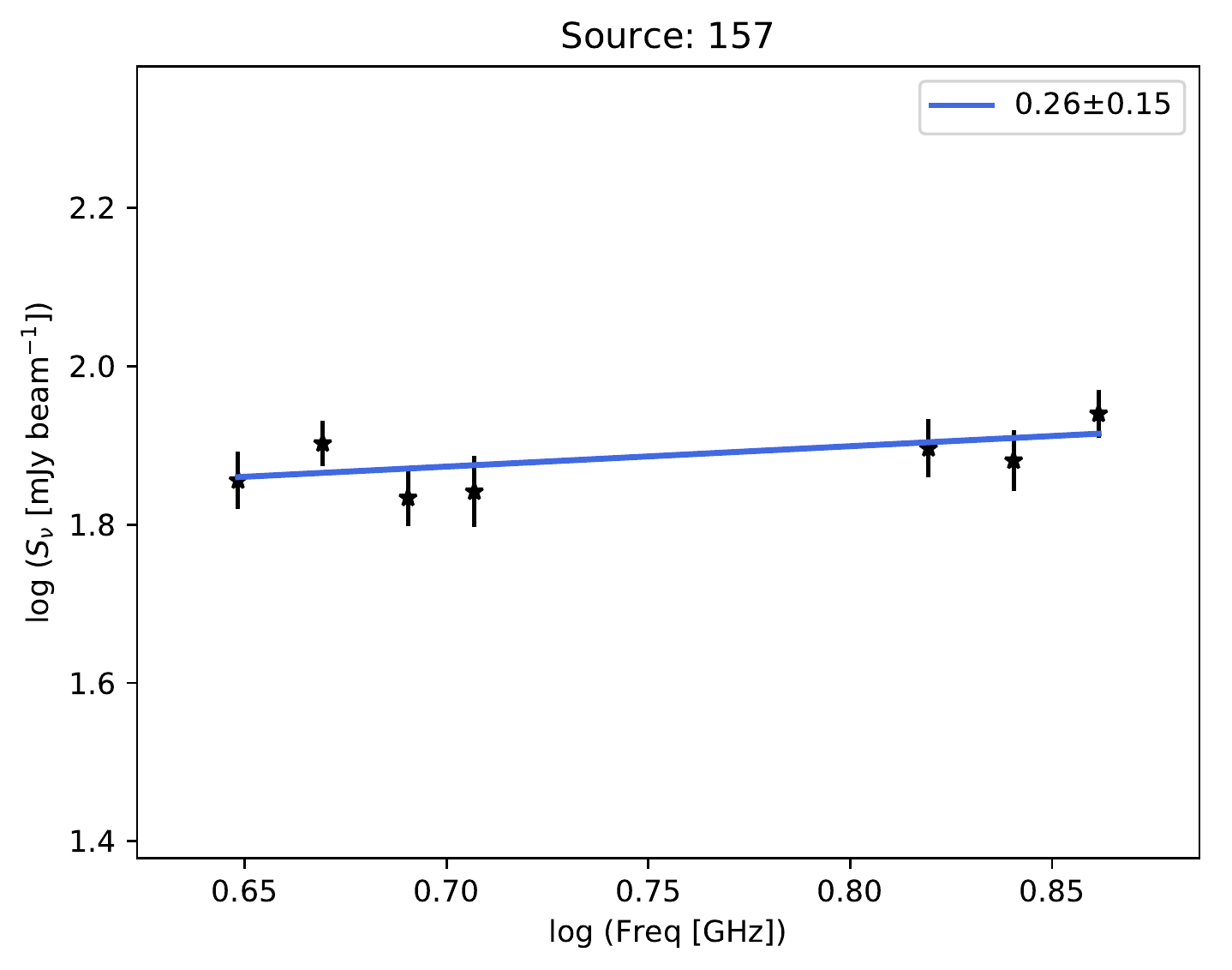}
\includegraphics[width=0.24\textwidth]{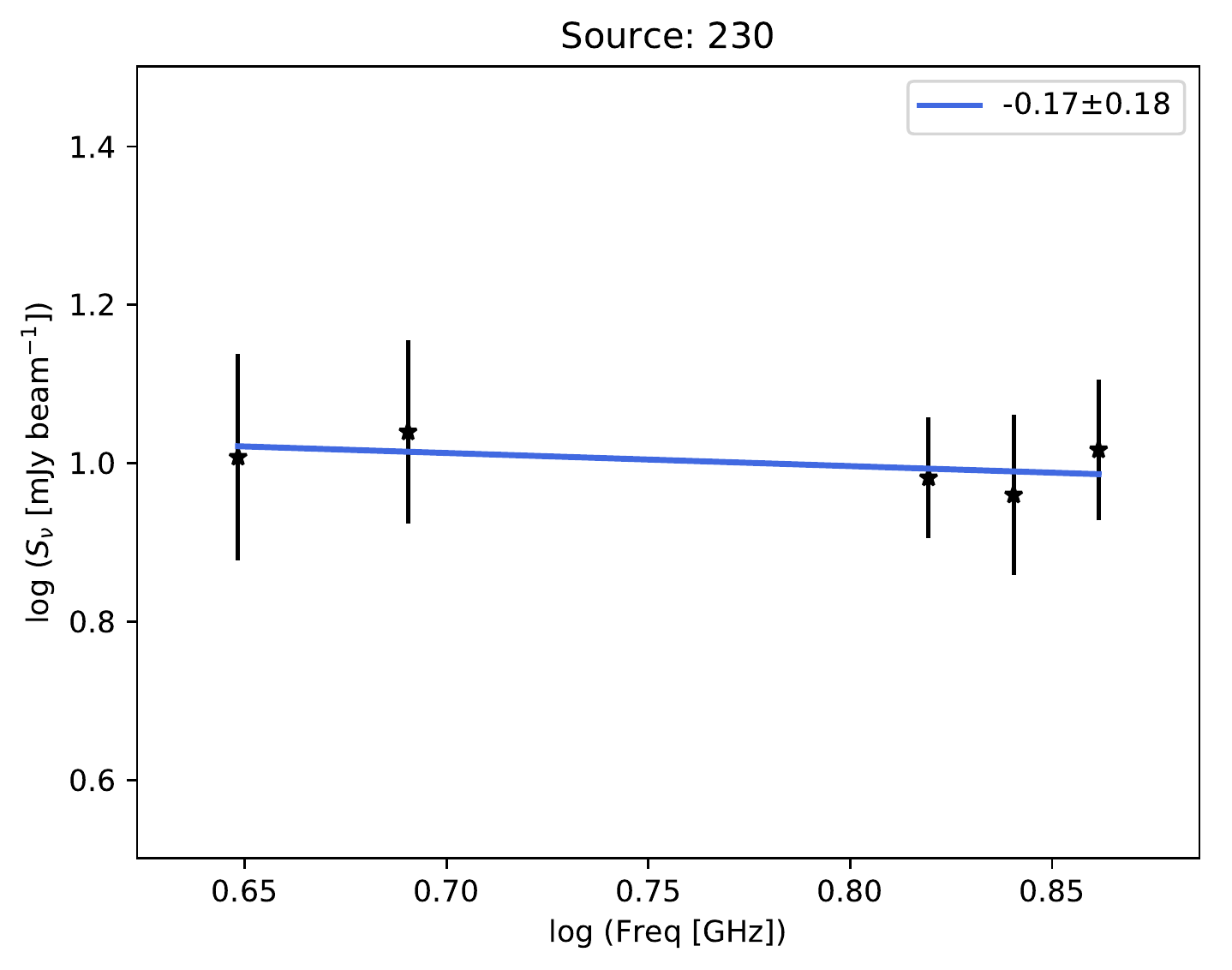}\\
\includegraphics[width=0.24\textwidth]{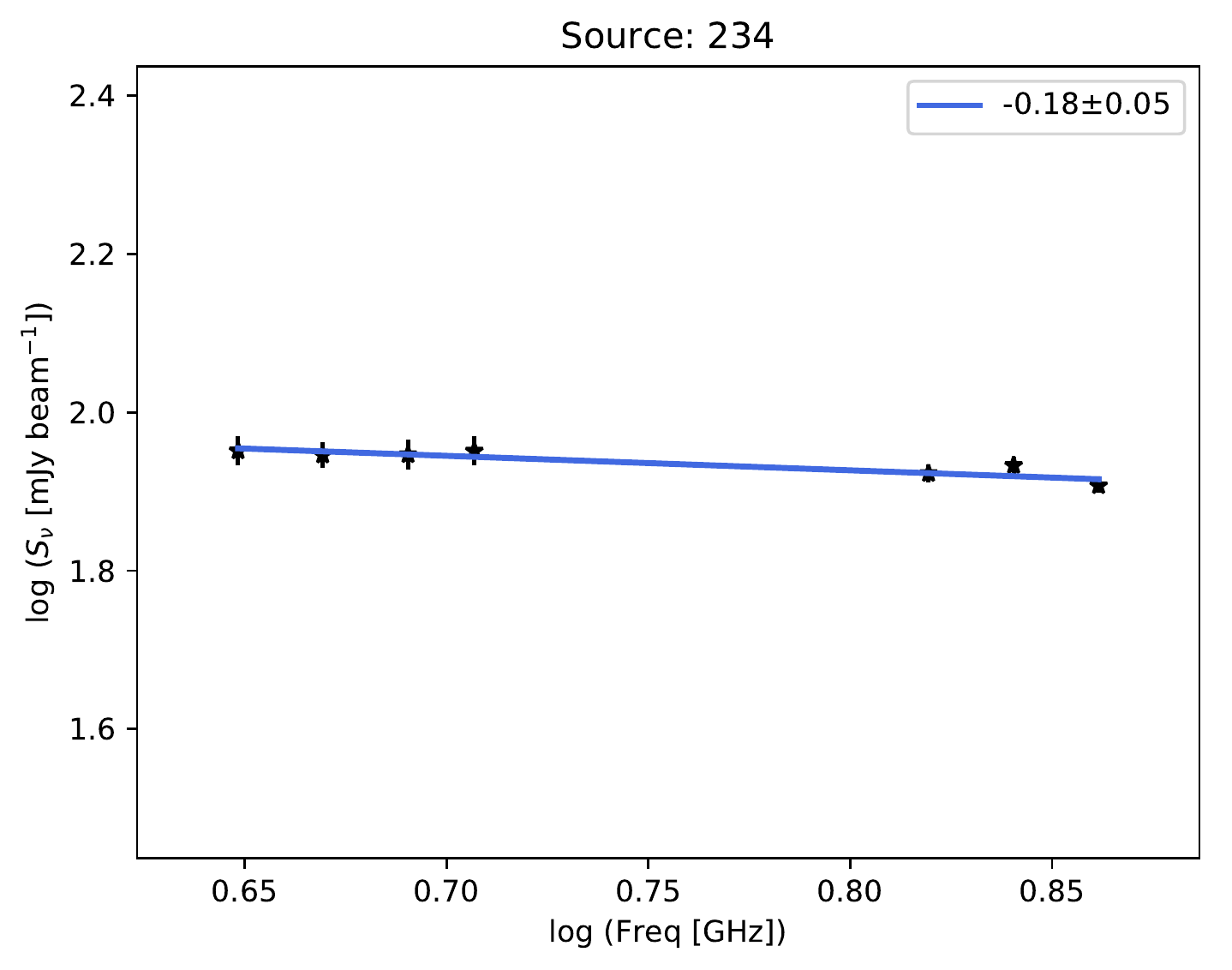}
\includegraphics[width=0.24\textwidth]{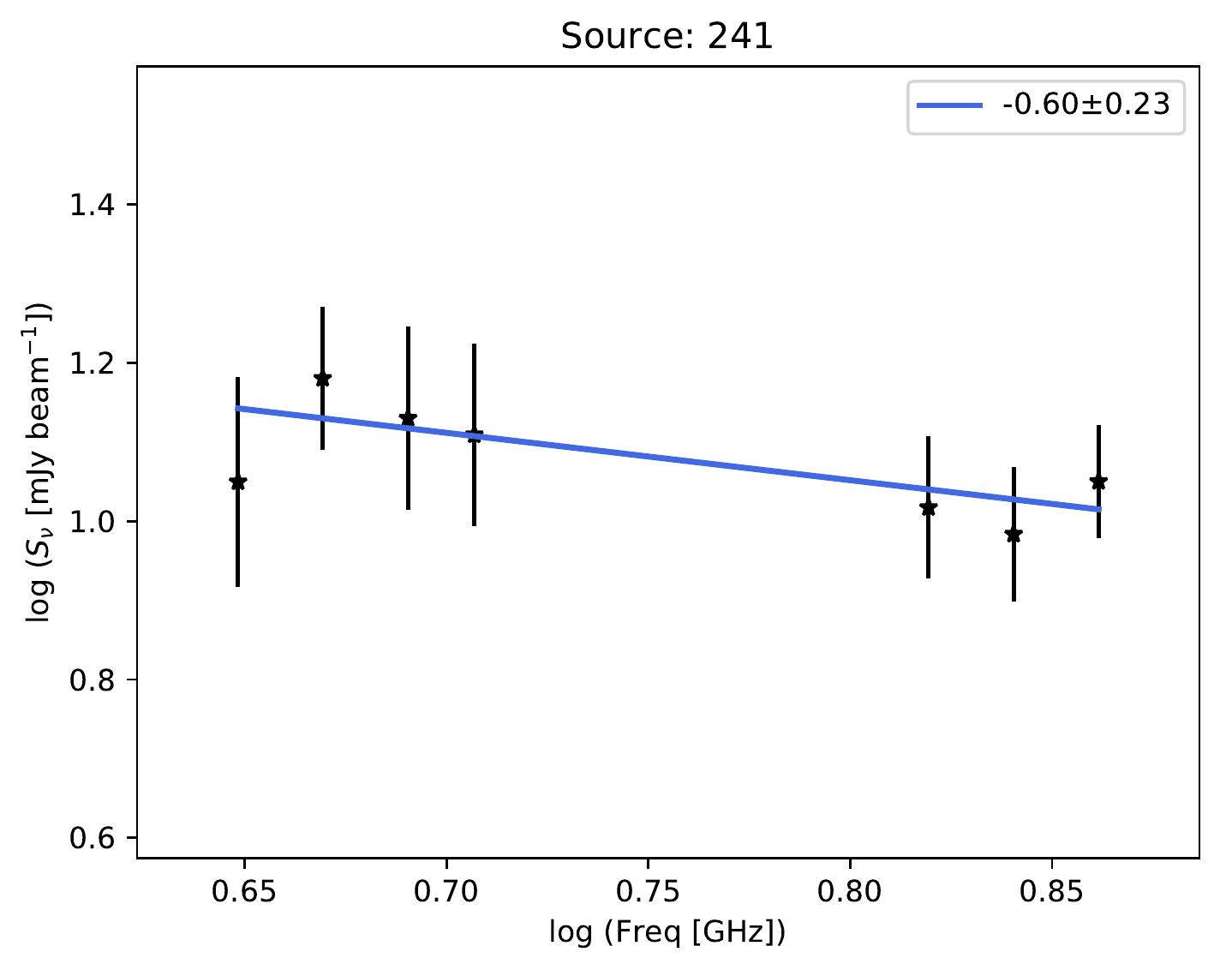}
\includegraphics[width=0.24\textwidth]{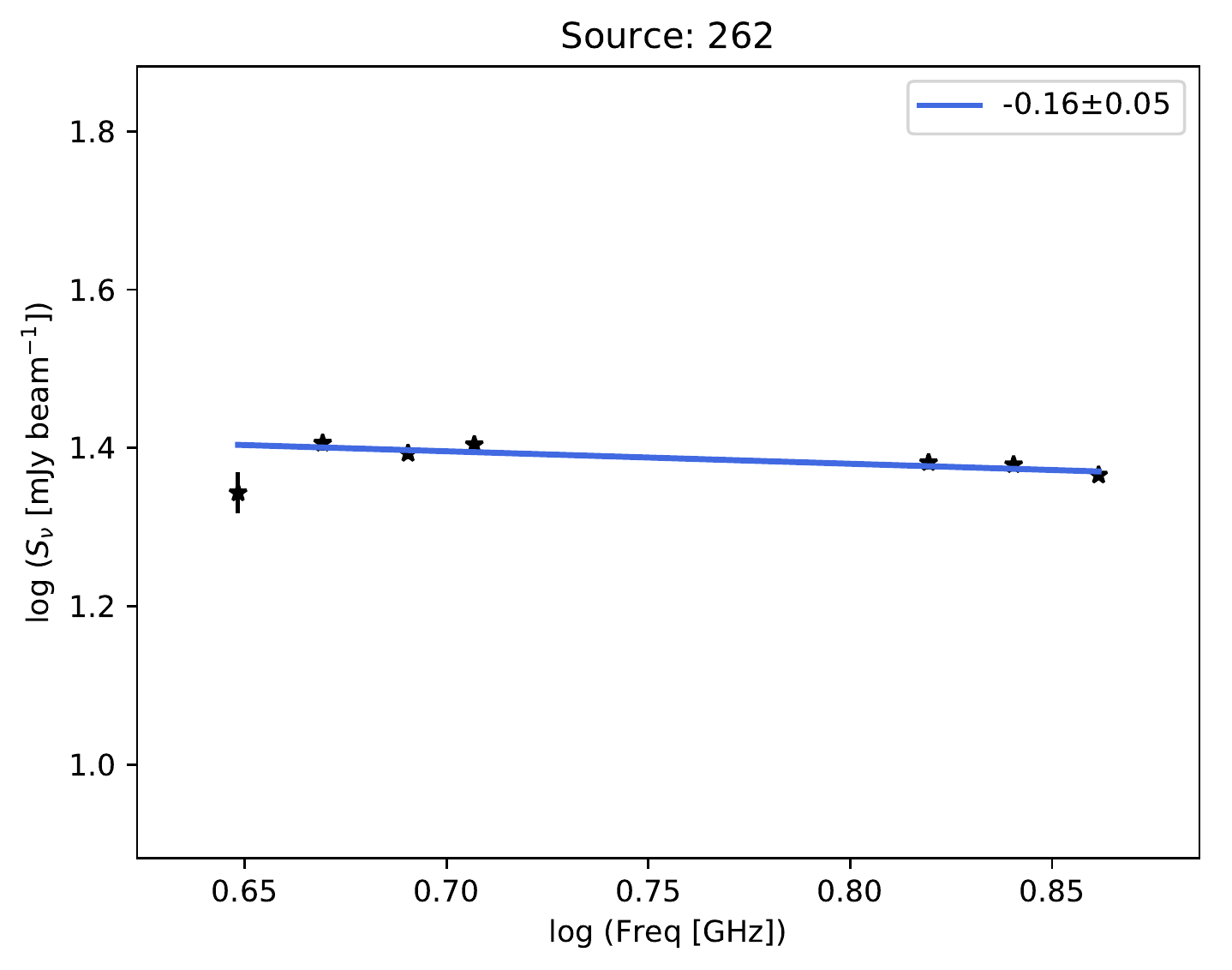}
\includegraphics[width=0.24\textwidth]{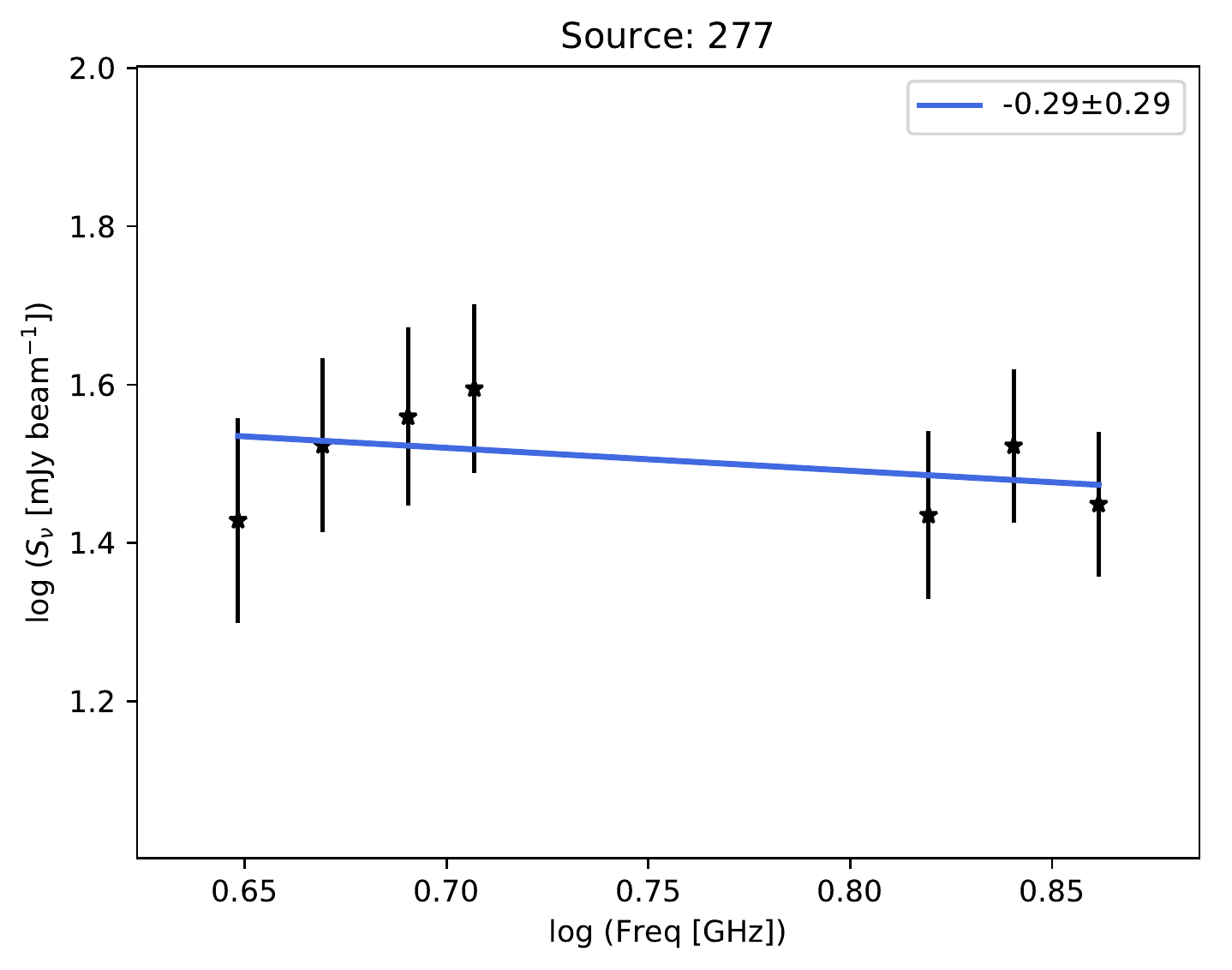}
\end{tabular}
\caption{Linear fits of the peak brightness against frequency in logarithmic scale to determine the spectral index of a source. To better appreciate the error bars, all panels have a total range of 1 on the y-axis, centred on the mean brightness of that panel. The frequencies range from 4.45\,GHz to 7.2\,GHz.}
\label{fig:specIndices}
\end{figure*}

\begin{figure*}[!tbhp]
\begin{tabular}{cccc}
\includegraphics[width=0.24\textwidth]{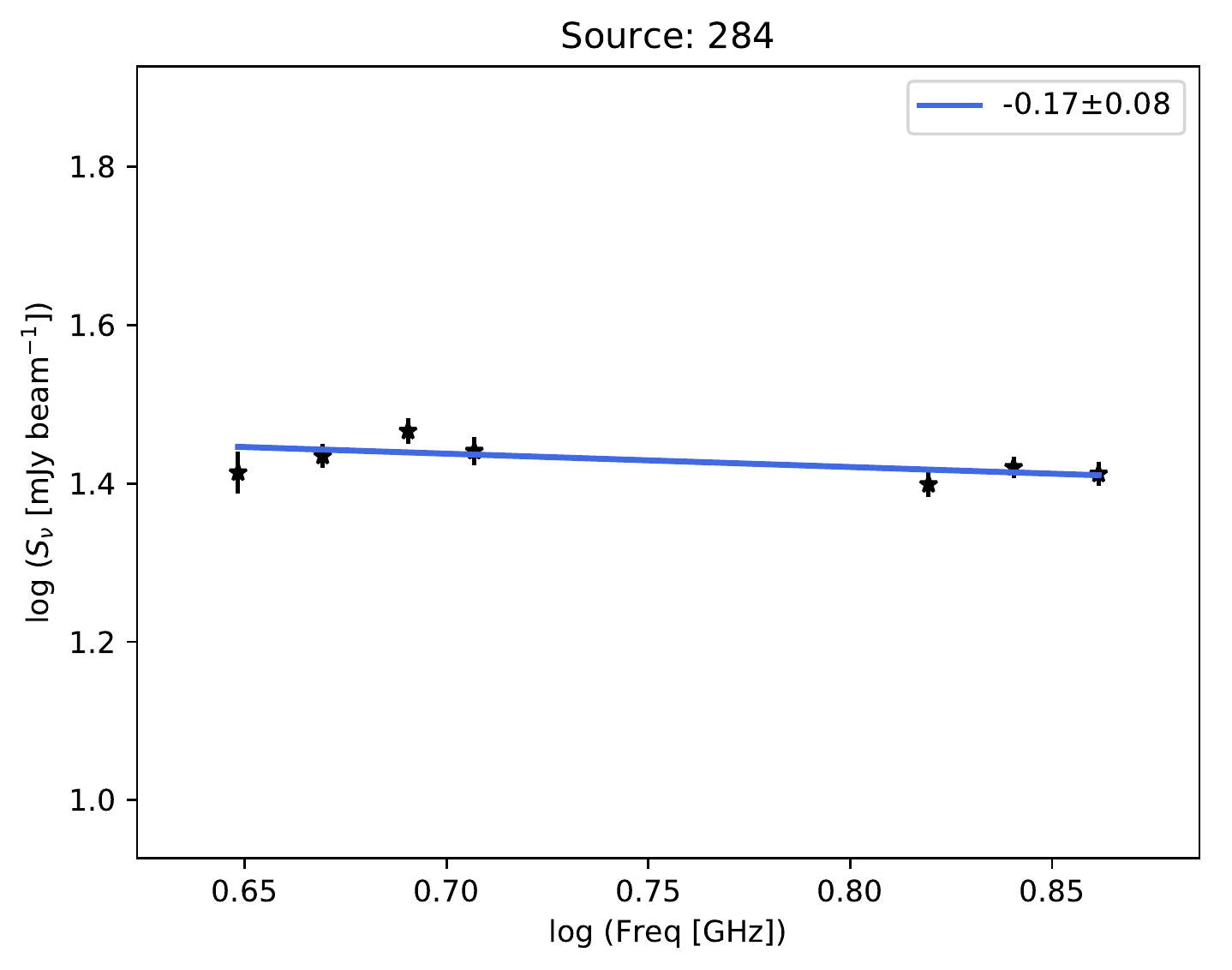}
\includegraphics[width=0.24\textwidth]{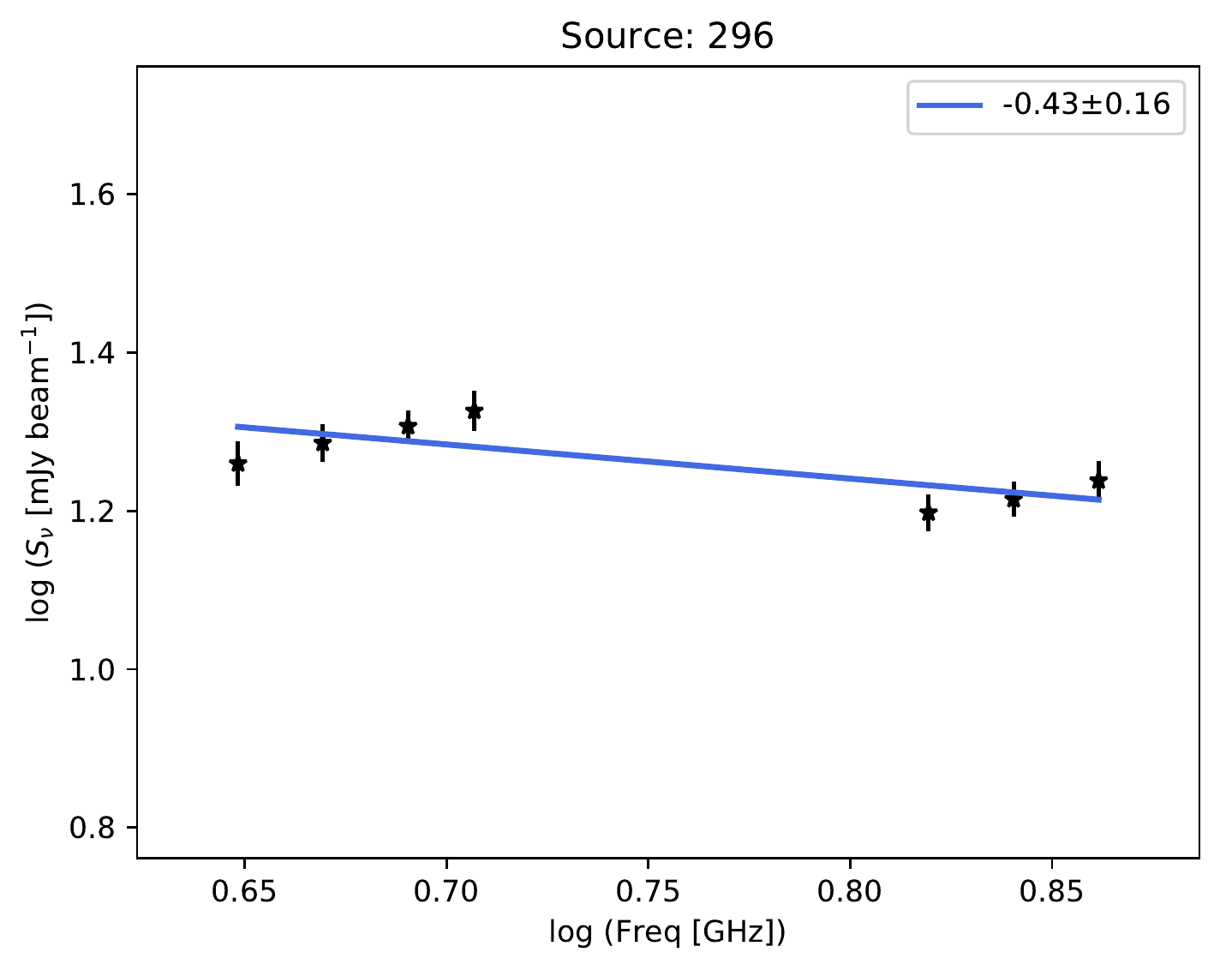}
\includegraphics[width=0.24\textwidth]{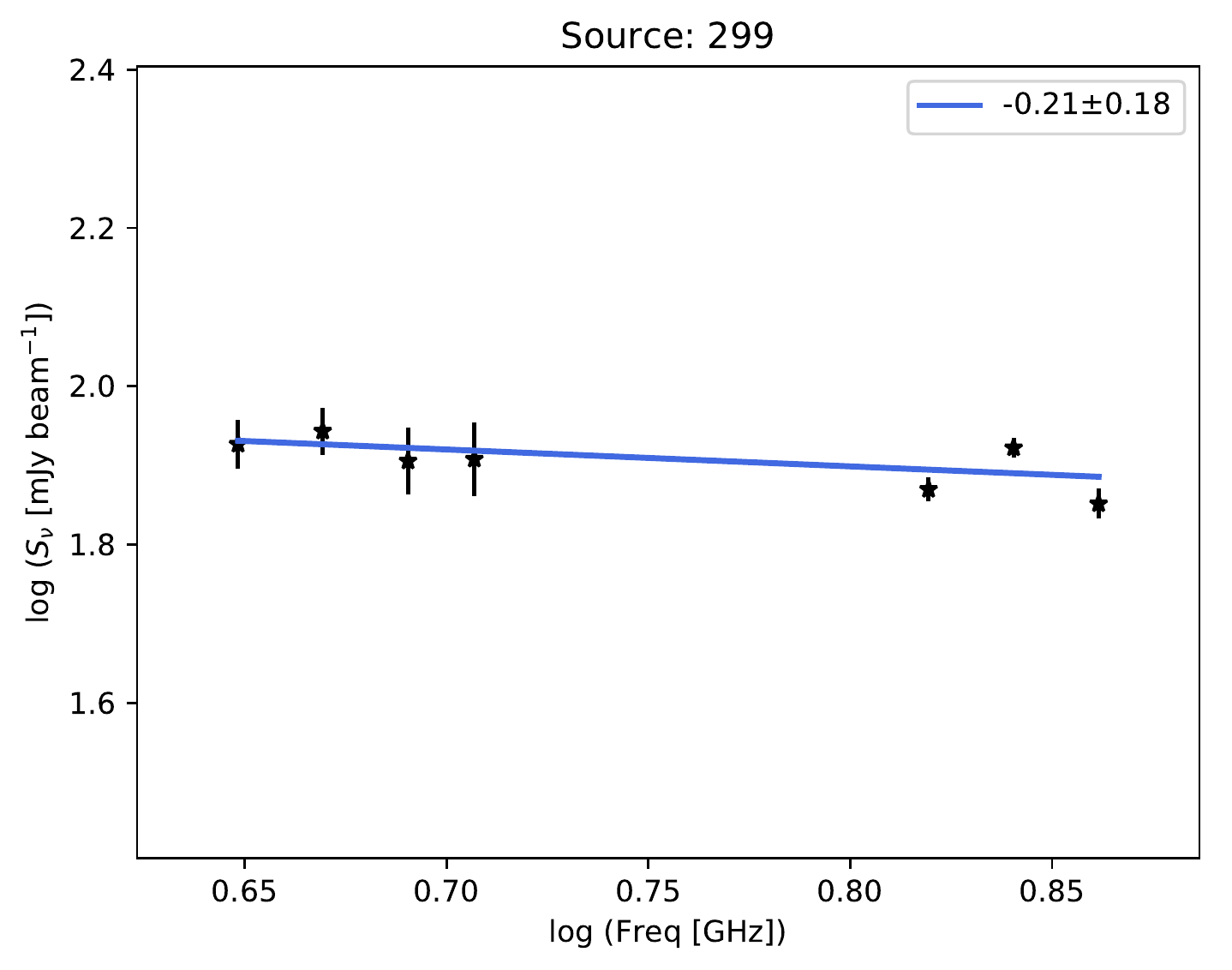}
\includegraphics[width=0.24\textwidth]{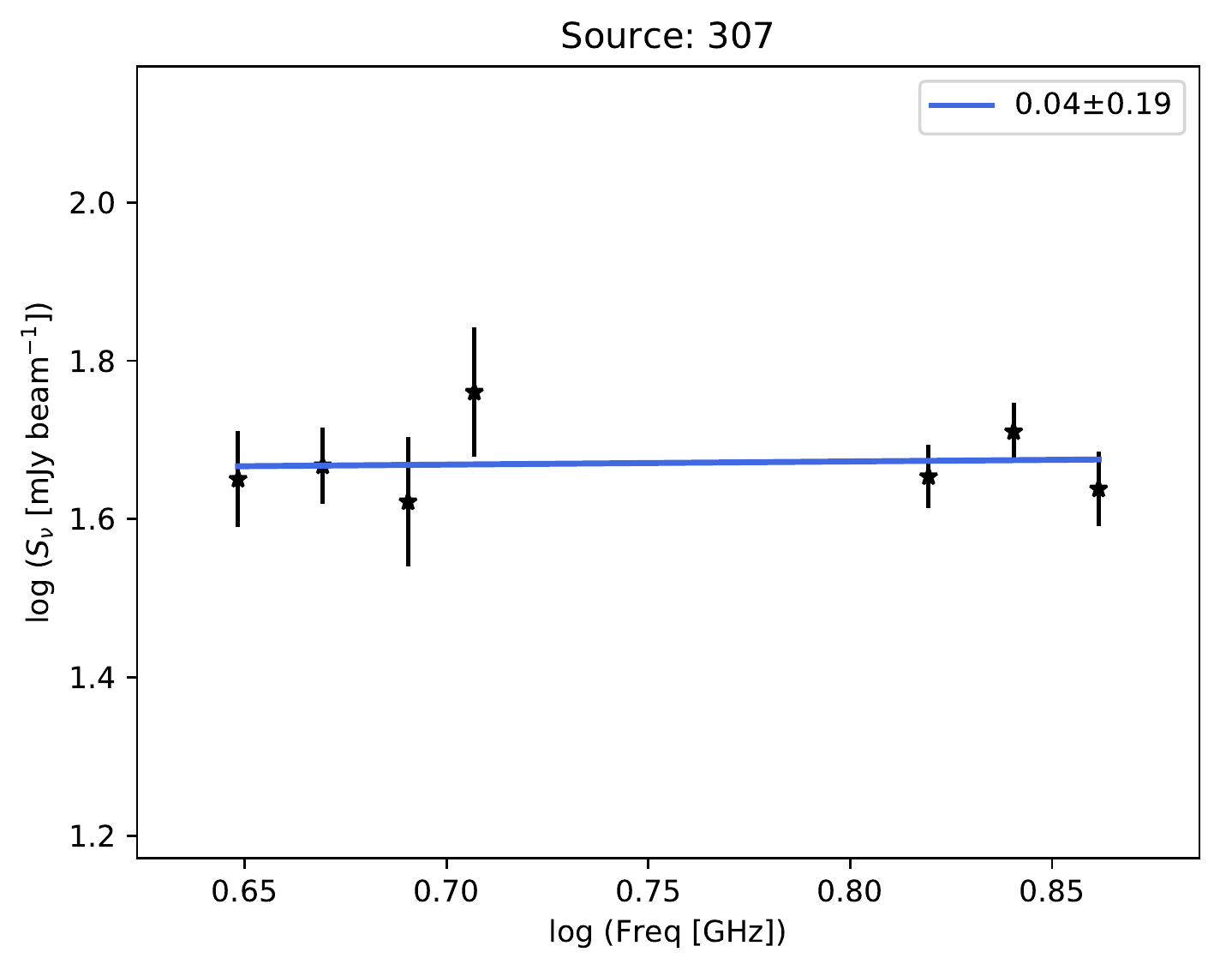}\\
\includegraphics[width=0.24\textwidth]{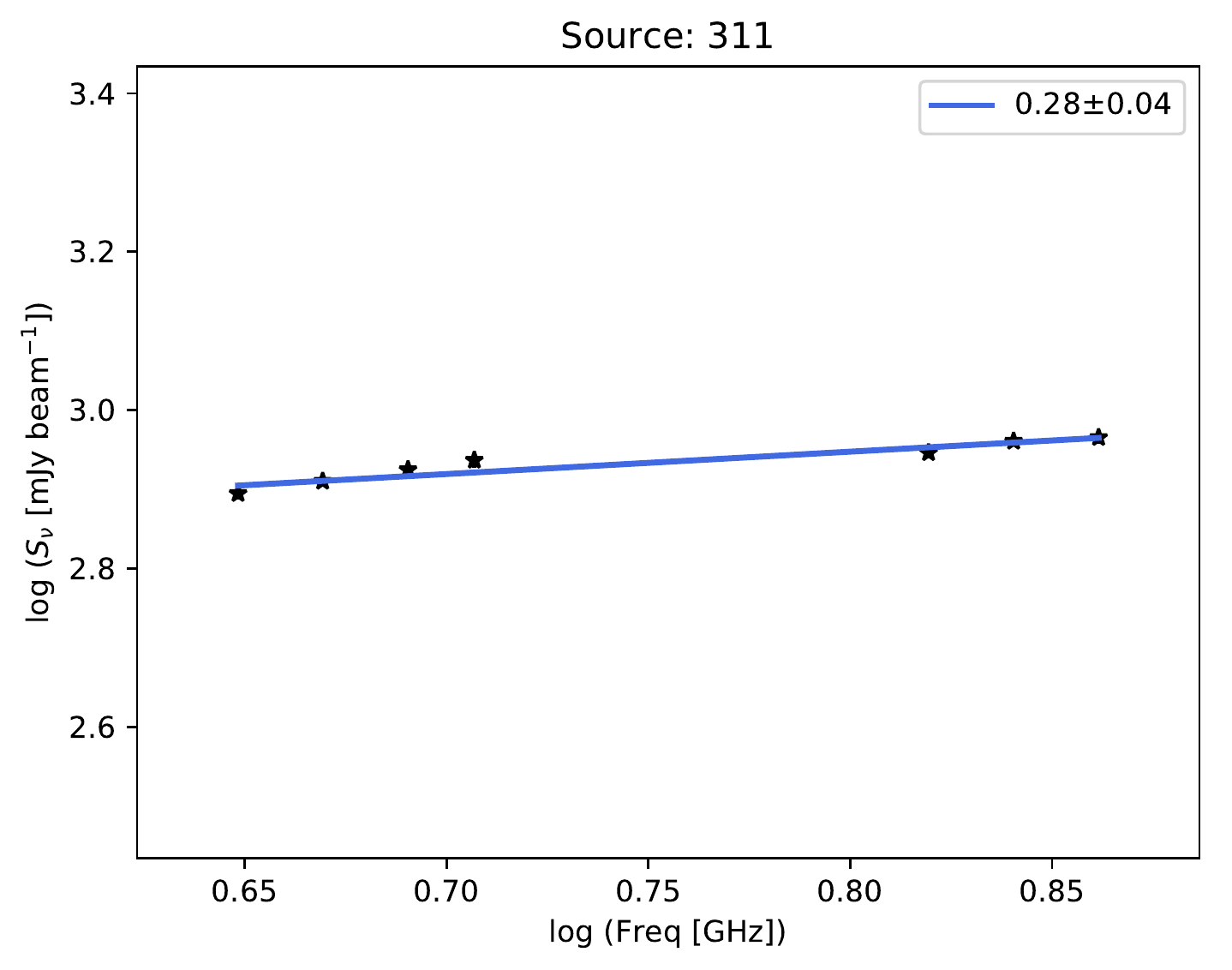}
\includegraphics[width=0.24\textwidth]{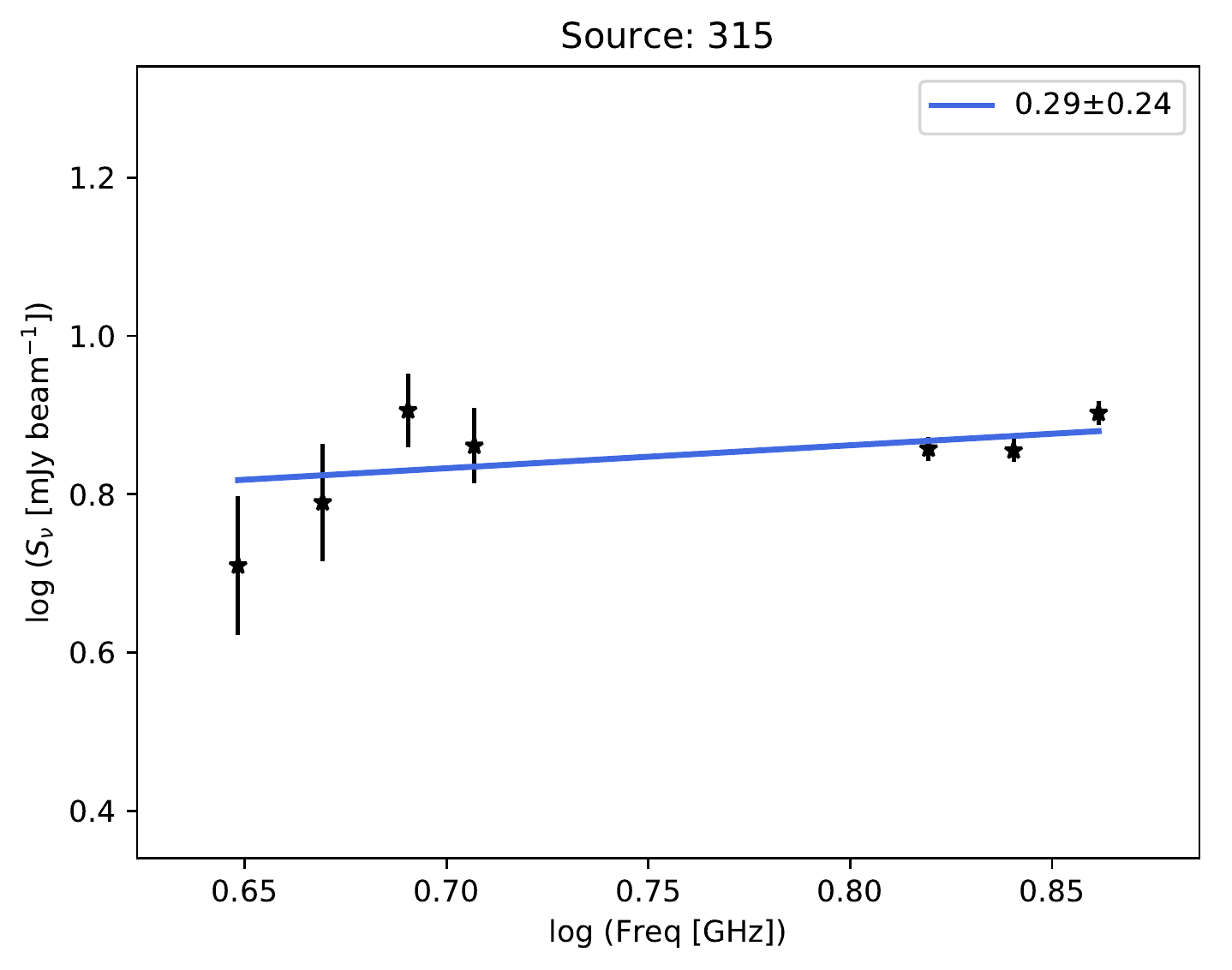}
\includegraphics[width=0.24\textwidth]{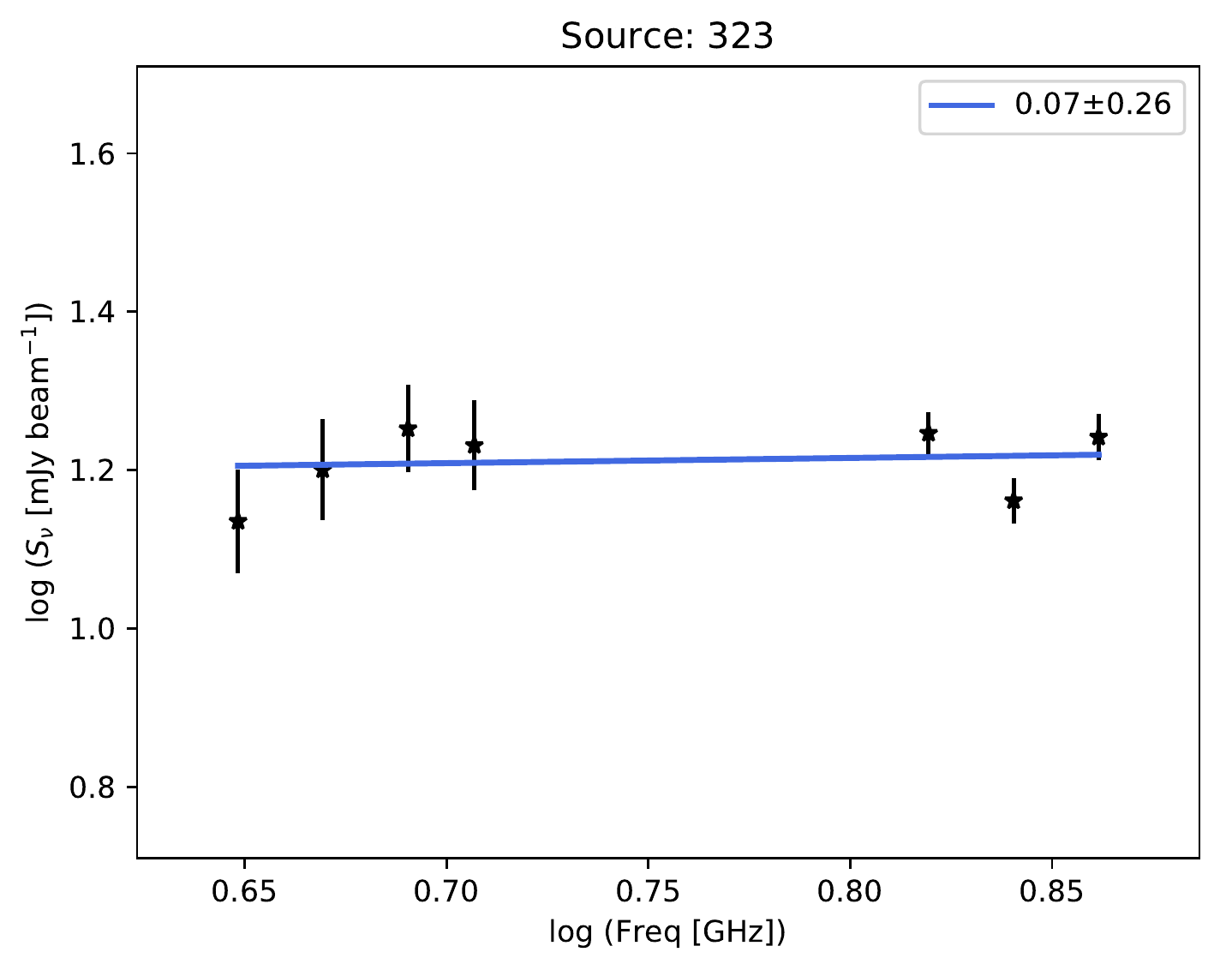}\end{tabular}

\caption{Continued from Fig.~\ref{fig:specIndices}.}
\label{fig:specIndices5}
\end{figure*}

\end{appendix}

\end{document}